\documentclass[a4paper,11pt]{article}
\pdfoutput=1 
\usepackage{jhep_like}

\usepackage{amsmath,amssymb,color,epsf}
\usepackage{graphicx}
\usepackage{subfloat}
\usepackage{subfig}
\usepackage{multirow}
\usepackage{ulem}

\newcommand{\be}{\begin{equation}}
\newcommand{\ee}{\end{equation}}
\newcommand{\bea}{\setlength\arraycolsep{2pt} \begin{eqnarray}}
\newcommand{\eea}{\end{eqnarray}}
\newcommand{\nn}{\nonumber}

\def\0{{\sst{(0)}}}
\def\1{{\sst{(1)}}}
\def\2{{\sst{(2)}}}
\def\3{{\sst{(3)}}}
\def\4{{\sst{(4)}}}
\def\5{{\sst{(5)}}}
\def\6{{\sst{(6)}}}
\def\7{{\sst{(7)}}}
\def\8{{\sst{(8)}}}
\def\sst#1{{\scriptscriptstyle #1}}

\def\avg#1{\left\langle#1\right\rangle}
\def\bra#1{\left\langle#1\right|}
\def\ket#1{\left|#1\right\rangle}

\def\abs#1{\left|#1\right|}
\def\kc#1{\left(#1\right)}
\def\kd#1{\left[#1\right]}
\def\ke#1{\left\{#1\right\}}
\def\Re{{\rm Re}}

\def\be{\begin{equation}}       \def\ee{\end{equation}}
\def\bea{\begin{eqnarray}}      \def\eea{\end{eqnarray}}
\def\ba{\begin{array}}
	\def\ea{\end{array}}
\def\bnum{\begin{enumerate} }
	\def\enum{\end{enumerate}}

\def\nn{\nonumber}

\def\=>{\Rightarrow}
\def\>{\rightarrow}

\def\eye2{Fathbb{I}}

\def\Tr{\mathrm{Tr}}

\def\TT{{T\bar T}}

\title{\boldmath Quantum chaos, scrambling and operator growth in $T\bar{T}$ deformed SYK models}

\author[a,b]{Song He,}
\author[c,d]{Pak Hang Chris Lau,}
\author[e]{Zhuo-Yu Xian,}
\author[a,f]{and Long Zhao}

\affiliation[a]{Center for Theoretical Physics and College of Physics, Jilin University,  \newline Changchun 130012, People's Republic of China}
\affiliation[b]{Max Planck Institute for Gravitational Physics (Albert Einstein Institute),  \newline Am M\"uhlenberg 1, 14476 Golm, Germany}
\affiliation[c]{National Center for Theoretical Sciences, National Tsing-Hua University, \newline Hsinchu 30013, Taiwan, R.O.C.}
\affiliation[d]{Department of Physics, Kobe University, Kobe-shi 657-8501, Hyogo, Japan}
\affiliation[e]{Institute for Theoretical Physics and Astrophysics and W\"urzburg-Dresden Cluster of Excellence ct.qmat,   Julius-Maximilians-Universität W\"urzburg,  \newline 97074 W\"urzburg, Germany.}
\affiliation[f]{Institute of Theoretical Physics, Chinese Academy of Science, \newline Beijing 100190, People's Republic of China}

\emailAdd{hesong@jlu.edu.cn}
\emailAdd{phcl2@panda.kobe-u.ac.jp}
\emailAdd{zhuo-yu.xian@physik.uni-wuerzburg.de}
\emailAdd{zhaolong@mail.itp.ac.cn}

\preprint{KOBE-COSMO-22-14}

\abstract{In this work, we investigate the quantum chaos in various $T\bar{T}$-deformed SYK models with finite $N$, including the SYK$_4$, the supersymmetric SYK$_4$, and the SYK$_2$ models. We numerically study the evolution of the spectral form factor (SFF), the out-of-time ordered correlator (OTOC), and the Krylov complexity. We find that the characteristic evolution of the SFF, OTOC and K-complexity of both the SYK$_4$ and SSYK$_4$ models remains unchanged under the deformation, which implies that the properties of quantum chaos is preserved. We also identify a many-body localization behavior in the deformed SYK$_2$ model.
}

\begin{document}

\maketitle

\section{Introduction}
\label{sec:intro}
The $T\bar{T}$ deformation of field theories has attracted much research interest in recent years, both from the perspective of field theory and holographic duality. The $T\bar{T}$ deformation of a 2D translation-invariant field theory is defined by \cite{Zamolodchikov:2004ce,Smirnov:2016lqw,Cavaglia:2016oda}. 
It is typically defined on a plane or cylinder by \cite{Smirnov:2016lqw,Cavaglia:2016oda}
\begin{equation}
\frac{\mathrm{d} \mathcal{L}^{\lambda}}{\mathrm{d} \lambda}=\frac{1}{2} \epsilon^{\mu \nu} \epsilon^{\rho \sigma} T_{\mu \rho}^{\lambda} T_{\nu \sigma}^{\lambda} \,,
\end{equation}
where $T^{\lambda}$ is a function of $\lambda$ and is the stress tensor of the theory with Lagrangian $\mathcal{L}^{\lambda}$. Although the RHS is a composite operator, it is quantum mechanically well-defined \cite{Zamolodchikov:2004ce}. 
This paper will study an analog of the $T\bar{T}$ deformation in the one-dimensional quantum mechanical theory proposed by \cite{Gross:2019ach, Gross:2019uxi}. For example, one can refer to \cite{He:2021dhr,Chakraborty:2020xwo} for recent developments. In particular, we focus on a particular realization of the $T\bar{T}$ deformation of the SYK$_4$ model in the form of $f(H)$, where $H$ is the Hamiltonian. These are a broad class of integrable deformations of quantum mechanics, which can be viewed as transformations of the Hamiltonian $H\to f(H)$. They can also be viewed as generated by the first order flows ${\partial H\over \partial \lambda}=f(H)$ or ${\partial L_E\over \partial \lambda}=f(T)$, where $L_E$ is a Euclidean Lagrangian and $T$ is the stress tensor.

The $\TT$ deformation is an integrable deformation, which means that the classical integrability of a system is preserved under this deformation. On the otherhand, there are also quantum chaotic theories which behave quite differently from integrable theories. In quantum chaotic theories, the energy level spacing satisfies a Wigner-Dyson distribution, which is one of the characteristic properties of quantum chaos \cite{Bohigas:1983er, Guhr:1997ve, Lau:2020qnl}. The spectral form factor (SFF) ~\cite{Garcia-Garcia:2016mno,Cotler:2016fpe,Lau:2018kpa} can also capture such behavior. The pattern of the SFF contains three regions: the ``slope region", the ``ramp region", and the ``plateau region". The pattern of the SFF in the ramp region manifests the quantum chaotic character of the model. The time scale of this region is approximately $e^{S/2}<t<e^S$ where $S$ is the system's entropy. Thus, the chaotic behavior of a system can be explored at a late stage by SFF.

Recently, inspired by advances in the study of holography and the black hole, another observable has been proposed to diagnose quantum chaos, namely scrambling \cite{Sekino:2008he}, which is characterized by the out-of-time ordered correlation function (OTOC) \cite{Shenker:2014cwa, Maldacena:2015waa, Roberts:2014ifa}. It is generally assumed that OTOC captures the universal early time ($\beta<t<t_*$, with $t_*\sim\log S$ the scrambling time) evolution of the initial boundary condition of the quantum chaotic system. Scrambling describes the growth of a local perturbation in operator space. It leads to an exponential growth of the OTOC which is characterized by the Lyapunov exponent $\lambda_L$ 
 \cite{Shenker:2013pqa, Roberts:2014isa, Shenker:2013yza}. 
Meanwhile, many examples show that OTOCs can grow exponentially even for a classical integrable system \cite{Rozenbaum:2019nwn, Xu:2019lhc, Hashimoto:2020xfr}. Motivated by the proposal in \cite{Xu:2019lhc}, we treat scrambling and chaos as two distinct concepts. Empirically, a system exhibiting an exponential decay of OTOC is in the thermal phase \cite{Goldstein:1506.07494}. A system in the thermal phase has a power-law increasing entanglement entropy and an exponentially decaying OTOC \cite{Nandkishore:2014kca}. In addition to the thermal phase, there is another phase, namely the many-body localization (MBL) phase \cite{Anderson:1958vr,PhysRevB.21.2366}. A system in the MBL phase has a logarithmically increasing entanglement entropy and an OTOC with power-law decay \cite{Nandkishore:2014kca, Huang:2016knw, Fan:2016ean}.

However, the exponential growth behavior of the OTOC is not necessarily present outside the large-$N$ limit, and then, the Lyapunov exponent is not well-defined outside the large-$N$ limit \cite{Khemani:2017nda, Xu:2018xfz, Xu:2018dfp}. Fortunately, the Krylov complexity (K-complexity), $C_K$, introduced in \cite{Parker:2018yvk,Barbon:2019wsy,Avdoshkin:2019trj,Jian:2020qpp,Rabinovici:2020ryf,Kar:2021nbm,Hornedal:2022pkc,Caputa:2021sib,Balasubramanian:2022tpr,Balasubramanian:2022dnj} provides us a more general definition of operator growth. The operator size, as measured by the K-complexity, grows exponentially before the scrambling time and then increases linearly until $t\sim e^S$. After this time, the K-complexity saturates the upper bound $C_K\sim \frac{1}{2}e^{2S}$. Similar to the OTOC, the exponential growth of the K-complexity at the early time has been observed in the integrable system \cite{Bhattacharjee:2022vlt,Baek:2022pkt}, so we distinguish K-complexity from the quantum chaos also. 

It was found that the behaviors of OTOC remain unchanged under the 1d $T\bar{T}$ deformation in the conformal limit \cite{Gross:2019ach,Gross:2019uxi} and the 2d $T\bar{T}$ deformation in \cite{He:2019vzf,He:2020qcs}. Motivated by the understanding of the universal nature of quantum chaos, $T\bar{T}$-deformed theories are a good testbed since the $T\bar{T}$ deformation preserves the integrability properties of the un-deformed theories. Moreover, it is interesting to analyze further the connection and difference between these concepts: quantum chaos, scrambling, and K-complexity.

In this paper, we focus mainly on whether and how the one-dimensional $T\bar{T}$ deformation of SYK$_4$ (and supersymmetric SYK, SSYK$_4$) and SYK$_2$ \cite{Maldacena:2016hyu} changes the chaotic behavior of the original theories. Since the SYK$_4$ (SSYK$_4$) and SYK$_2$ model are maximally chaotic \cite{Maldacena:2015waa} and non-chaotic models, respectively, we would like to turn on the $T\bar{T}$ deformation stepwise to check whether the deformation preserves the chaotic properties of the non-deformed theories. 
As mentioned in the last paragraph, the classical integrable systems also show exponential growth in OTOCs. 
Consequently, we characterize the quantum chaotic behavior only through the SFFs. Moreover, we are interested in studying the scrambling, and K-complexity behavior under the $T\bar{T}$ deformation \cite{Cotler:2016fpe, Shenker:2014cwa}. 
We numerically study the evolution on these $\TT$-deformed models at finite $N$ and extract the characteristic behaviors of SFFs, OTOCs, and K-complexities.

The structure of this paper is as follows. In section \ref{setup} we give an overview of the SYK$_4$ (SSYK$_4$) and SYK$_2$ models and list the useful definition of their $T\bar{T}$ deformation. In section \ref{section-SFF}, we study the SFF in deformed SYK models and discuss the late time chaos. The OTOC and K-complexity in deformed theories are presented in section \ref{section:OTOC} and section \ref{kcomplexity}. We end in section \ref{Sec:Summary} with a summary and an outlook.

\section{Set up}
\label{setup}

\subsection{The $T\bar{T}$-deformed 1D system}
\label{QMttbar}
In this section we would like to pursue the idea of the dynamical coordinate transformation given in \cite{He:2021dhr} for the $T\bar{T}$ deformation in the $1$-dimensional case, which is a generalization of the 2D $T\bar{T}$ deformation in terms of the dynamical coordinate transformation that has been extensively studied \cite{Dubovsky:2018bmo,Dubovsky:2017cnj,Cardy:2018sdv,Conti:2018tca,Aguilera-Damia:2019tpe,Tolley:2019nmm,Mazenc:2019cfg}. One can couple an action $S_{0}$ to a $1$-dimensional massive 
``gravity"\cite{He:2021dhr}
\begin{align}
S\left[e_{\mu}, v^{\mu}, \phi\right] &=S_{\text {grav }}\left[e_{\mu}, v^{\mu}\right]+S_{0}\left[e_{\mu}, \phi\right] \,, \\
S_{\text {grav }}\left[e_{\mu}, v^{\mu}\right] &=\frac{1}{\lambda} \int d t e_{t} B\left(e_{t} v^{t}\right),
\end{align}
where the 1-form $e_{\mu}$ is the dynamical tetrad and the vector $v^{\mu}$ is a fixed co-tetrad corresponding to the metric on which the deformed theory lives. One can take $v^{t}=1$ and then
\begin{align}\label{TCoordinate}
v^{T}=\frac{d T}{d t}, \quad e_{T}=1, \quad e_{t}=\frac{d T}{d t}.
\end{align}
We start with a seed scalar theory as follows
\begin{equation}\label{S0}
S_{0}=\int d t e_{t}\left(\frac{1}{2\left(e_{t}\right)^{2}} \partial_{t} \phi \partial_{t} \phi-V(\phi)\right),
\end{equation}
and it can be written as the first order formalism
\begin{align}
S_{0}=\int d t e_{t}\left(\frac{1}{e_{t}} p \partial_{t} \phi-H_{0}(\phi, p)\right),
\end{align}
where $p$ is the canonical momentum in the phase space and $H_{0}$ is the corresponding Hamiltonian of the undeformed theory.
The equation of motion of $e_{t}$ gives
\begin{equation}\label{EoMTetrad}
e_tv^{t} B^{\prime}\left(e_{t} v^{t}\right)+B\left(e_{t} v^{t}\right)-\lambda H_{0}=0.
\end{equation}
From (\ref{TCoordinate}), it becomes
\begin{align}
\frac{d T}{d t} B^{\prime}\left(\frac{d T}{d t}\right)+B\left(\frac{d T}{d t}\right)-\lambda H_{0}=0.
\end{align}
Using $d T=f'\left(H_{0}\right) d t,$ one can obtain a relation between $f$ and $B$
\begin{align}
f'\left(H_{0}\right) B^{\prime}\left(f'\left(H_{0}\right)\right)+B\left(f'\left(H_{0}\right)\right)-\lambda H_{0}=0.
\end{align}
The solution is
\begin{align}\label{Bf}
B\left(f'(H)\right)=\lambda H-\frac{\lambda f(H)}{f'(H)}+\frac{C}{f'\left(H\right)},
\end{align}
where $C$ is a constant. There is no 1D massive gravity action available in the literature. Thanks to $T\bar{T}$ deformation of the SYK model, one can apply the particular case to derive so-called 1D massive gravity. 

In $t$ coordinate, the solution $B$ of (\ref{EoMTetrad}) can take the form (\ref{Bf}) such that $e_{t}=f'\left(H_{0}\right).$  By integrating out $e_{t}$ in the action\footnote{We apply the equation of motion of $e_{t}$.}, the resulting action is
\begin{align}
S=\int d t\left(p \partial_{t} \phi-f\left(H_{0}\right)\right),
\end{align}
where the constant term $C/\lambda$ has been dropped.
For $T \bar{T}$-deformation \cite{Gross:2019uxi}, we have 
\begin{align}\label{flowsyk}
f(H)=\frac{1-\sqrt{1-8 H \lambda}}{4 \lambda}.
\end{align}
The deformed Hamiltonian (\ref{flowsyk}) should satisfy Eq.(\ref{Bf})
and then the 1D massive gravity can be regarded as follow
\begin{align}
B(x)=\frac{(x-1)^{2}}{8 x^{2}}.
\end{align}
Finally, one can check that the deformed Hamiltonian satisfies the flow equation
\be \label{floweq}
2\partial_\lambda H=\frac{H^2}{4-2\lambda H},
\ee
which is consistent with \cite{Gross:2019uxi}.

We close this section by summarizing the result. We follow the dynamical coordinate transformation proposed by \cite{Dubovsky:2018bmo,Dubovsky:2017cnj} in 2D quantum field theories to realize the $T\bar{T}$ flow equation. Based on Ref.~\cite{Gross:2019ach}, We work out the generic framework to do $T\bar{T}$ deformation of the undeformed theory by coupling with the 1D massive gravity $B$. The undetermined function $B$ to characterize the massive gravity in 1D is fixed by the case offered by Ref.~\cite{Gross:2019ach}. Our approach gives the same results as Ref.~\cite{Gross:2019ach}. It is an important ingredient to deform generic one-dimensional quantum systems. One can set up a 1D quantum mechanical system coupled to a 1D massive gravity to realize the $T\bar{T}$ deformation satisfying the flow equation Eq.~(\ref{floweq}). Since there is no well-defined stress momentum tensor, it is not easy to generalize the 2D $T\bar{T}$ deformation to a 1D system. Here we apply the analogous 1D flow equation to define the 1D deformation, which is a generalization of the standard $T\bar{T}$ deformation. In the following sections, we would like to investigate the behavior of quantum chaos when the deformation is turned on.

\subsection{The Majorana SYK models}
\label{sec:Majorana-SYK}

The Hamiltonian of the SYK$_q$ model with $N$ Majorana fermions is \cite{Maldacena:2016hyu,Kit.KITP.1,Kit.KITP}
\begin{align}
    H=\frac{i^{q/2}}{q!}\sum_{j_1j_2\cdots j_q}^N J_{j_1j_2\cdots j_q} \psi_{j_1}\psi_{j_2}\cdots\psi_{j_q}\,,
\end{align}
where the fermions satisfy the anticommutation relation $\ke{\psi_i,\psi_j}=2\delta_{ij}$, the coupling tensor $J_{j_1j_2\cdots j_q}$ is totally antisymmetric, and each independent element is randomly drawn from a Gaussian distribution with zero mean and variance $\avg{J_{j_1j_2\cdots j_q}^2}=\frac{(q-1)!}{N^{q-1}}J_0^2=\frac{2^{q-1} (q-1)! }{q N^{q-1}}\mathcal{J}^2$.

For even $N$, one can define $N_d=N/2$ Dirac fermions with annihilation and creation operators $c_i$ and $\bar{c}_i$ from the $N$ Majorana fermions
\begin{align}
	\psi_{2i}=\frac{c_i+\bar{c}_i}{\sqrt{2}},\quad 	\psi_{2i-1}=\frac{i(c_i-\bar{c}_i)}{\sqrt{2}} \,.
\end{align}
The fermion number charge is given by $Q=\sum_{i=1}^{N_d}\bar{c}_ic_i$ and the charge parity $\tilde{Q}=Q$ mod 2 is conserved. For the SYK$_4$ model, the theory has a particle-hole symmetry under the operator \cite{Cotler:2016fpe,Fu:2016yrv,You:2016ldz}
\begin{align}
	P=K\prod_{i=1}^{N_d}(\bar{c}_i+c_i),
	\label{particle-hole}
\end{align}
where $K$ is an anti-linear operator. This operator $P$ is a symmetry of the SYK$_4$ model but not of the SYK$_2$ model, namely
\begin{align}
	[H_{\text{SYK}_4},P]=0,\quad [H_{\text{SYK}_2},P]=2H_{\text{SYK}_2}P.
\end{align}
As a result, it can be shown that the energy level of the SYK$_4$ model is doubly degenerate for $N$ mod $8=2,4,6$ and non-degenerate for $N$ mod $8=0$, while the energy level of the SYK$_2$ model is non-degenerate for all $N$.

The SYK$_2$ Hamiltonian can be diagonalized and expressed as the Hamiltonian of $N/2$ free Dirac fermions $\chi_a$, namely
\begin{align}\label{SYK2-diagonal}
H_{\text{SYK}_2}
=\sum_{a=1}^{N/2} \varepsilon_a \kc{\hat n_a -\frac12},
\end{align}
where $\hat n_a=\chi_a^\dagger\chi_a/2$. We have diagonalized the anti-symmetric random couplings as $J_{ij}=-i \sum_k^N U_{ki} J_{k} U_{kj}^*$, where $U_{2a,i}=U_{2a-1,i}^*$, $\sum_i U_{2a,i}^*U_{2b,i}=\delta_{ab}$, and $J_{2a}=-J_{2a-1}\geq0$. The Dirac fermions are defined as $\chi_a=\sum_i U_{2a,i}^*\psi_i$ with the anti-commutation relations $\ke{\chi_a,\chi_b^\dagger}=2\delta_{ab},\,\ke{\chi_a,\chi_b}=\ke{\chi_a^\dagger,\chi_b^\dagger}=0$ and the energy band $\varepsilon_a=2J_{2a}$. 
Thus, the eigenstates are labeled by the occupation numbers $\ket{\vec n}=\ket{n_1,n_2,\cdots,n_{N/2}}$ with energy $E_0+\sum_{a}\varepsilon_a n_a$, where $n_a=0,1$ and the ground state energy $E_0=-\frac12\sum_a\varepsilon_a$.
For the $\TT$-deformed SYK$_2$ model, the eigenstates are unchanged, namely,
$f(H-E_0)\ket{\vec n}=f\kc{\sum_a n_a\varepsilon_a} \ket{\vec n}$.
The Hamiltonian can be expanded as
\begin{align}\label{TTSYK2H}
    f(H-E_0)=\sum_a \varepsilon_a \hat n_a +2\lambda \sum_{ab}\varepsilon_a\varepsilon_b\hat n_a\hat n_b+\cdots,
\end{align}
which coincides with the phenomenological model of many-body localization (MBL) \cite{Maksym:2013}. Based on this observation, we investigate the OTOC of this model in Sec.~\ref{section:OTOC}.

The $T\bar{T}$-deformed partition function of the SYK model is
\begin{align}
Z(\beta)_\lambda
=\int_{-\infty}^\infty \mathrm{d}Ee^{-\beta E}\rho_\lambda(E)
=\int_{-\infty}^\infty \mathrm{d}Ee^{-\beta f(E)}\rho(E)\,,
\end{align}
which can be written as an integral transform of the undeformed partition function by introducing a kernel defined as
\begin{align}
K_f(\beta,\beta')=\frac{1}{2\pi i}\int_\mathcal{C}\mathrm{d}Ee^{-\beta f(E)+\beta' E}\,.
\label{TTbarkernal}
\end{align}
For the deformation Eq.~(\ref{flowsyk}), the contour $\mathcal{C}$ of the integral runs from $0$ to $\infty$. The eigenvectors $|E\rangle$ are unchanged under the deformation $H\rightarrow f(H-E_0)$. Consequently, the deformed $n$-point thermal correlators with arguments $\tau_1>\cdots>\tau_{n-1} > 0$ can be expressed as follows
\begin{align}
G_\lambda(\beta,\{\tau_i\})
=\int\left(\prod_{i=0}^{n-1}\mathrm{d}\beta'_iK_f(\beta_i,\beta_i')\right)G_0(\beta,\{\tau_i\})\,.
\label{TTbarCorrelator}
\end{align}
According to equations (\ref{TTbarkernal}) and (\ref{TTbarCorrelator}), it is easy to show that the conformal 2-point function is unchanged at finite temperature under $T\bar{T}$ deformation.

In the limit of large $N$, the effect of $T\bar{T}$ deformation on the SYK model takes a simple form. Let us consider the SYK$_q$ model averaged over the disorder after applying a $T\bar{T}$ deformation, the expression is
\begin{align}
S_E(\lambda,e)
=&\int \mathrm{d}\tau\left(\psi_i\partial_\tau\psi_i-\frac{e}{8\lambda}\left(1-e^{-1}\right)^2-eE_0\right)\nn\\
&-\frac{N}{2q}\int\mathrm{d}\tau\mathrm{d}\tau'J^2e(\tau)e(\tau')G(\tau,\tau')^q\,,
\label{TTbarEffectriveAction}
\end{align}
where $e$ is a Lagrange multiplier. The Dyson-Schwinger equations of this effective action for $G$ and $\Sigma$ have the same forms as the undeformed one with $J^2$ replaced by $J^2e(\tau)e(\tau')$. The equation of motion of $e$ is
\begin{align}
\frac{e^{-2}-1}{8N\lambda}-\frac{J^2}{q}\int\mathrm{d}\tau e(\tau)G(\tau,\tau')^q-\frac{E_0}{N}=0\,.
\label{EoMe}
\end{align}
Since the $T\bar{T}$ deformation preserves most of symmetries of the original theory, the effective action (\ref{TTbarEffectriveAction}) is time translation invariant and the auxiliary field $e(\tau)$ should be a constant. Its value can be determined by the equation of motion (\ref{EoMe}), and the result is
\begin{align}
e^{-1}=\sqrt{1+8\lambda\left(E_0+\frac{cJ}{q}\right)}\,,
\end{align}
where the constant $c$ is the integral of $G^q$. The solutions for $G$ and $\Sigma$ remain the same, but now $J$ is rescaled to $Je$. The two point function in Eq.~(\ref{EoMe}) satisfies the $T\bar{T}$-deformed D-S equation and is related to the undeformed one by $G(\tau,\tau')^q\sim G_0(\tau,\tau')^q/e$. The Hamiltonian $H_0$ of the undeformed SYK model is 
\begin{align}
    H_0=-\frac{NJ^2}{2}\int d\tau G_0(\tau,\tau')^q \,,
\end{align}
where $G_0$ is the original two-point function. So we can replace the integral of $G(\tau,\tau')$ in Eq.~(\ref{EoMe}) by $H_0/Ne$. The solution of Eq.~(\ref{EoMe}) is
\begin{align}\label{Jeff}
    e^{-1}=\sqrt{1-8\lambda\left(H_0-E_0\right)}\,.
\end{align}
We can estimate $H_0-E_0$ by $(E_{\rm max}-E_0)/2$ at infinite temperature. For the SYK model with finite $N$, we can also consider the effect of the $T\bar{T}$ deformation as a redefinition of the effective coupling constant $J_{\rm eff}=J_0e$.

\subsection{The supersymmetric SYK model}
\label{sec:ssyk}
This section studies the $T\bar{T}$-deformed supersymmetric SYK$_4$ model (SSYK$_4$). The SSYK$_q$ model can be constructed through the supercharge \cite{Fu:2016vas,Li:2017hdt,Stanford:2019vob}
\begin{align}
    Q &= i^{\frac{\hat{q}-1}{2}} \sum_{a_1a_2\cdots a_{\hat{q}}} C_{a_1 a_2\cdots {\hat{q}}} \psi_{a_1} \psi_{a_2}\cdots \psi_{a_{\hat{q}}}\,, \quad q=2\hat{q}-2\,,\\
    H &= Q^2 \,,
\end{align}
where $\hat{q}\geq 3$ and $C_{a_1a_2\cdots a_{\hat{q}}}$ is the random coupling with mean 
\begin{align}
	\left\langle C_{a_1a_2\cdots a_{\hat{q}}} \right\rangle =0\,,\quad
	\left\langle C_{a_1a_2\cdots a_{\hat{q}}} C_{a_1a_2\cdots a_{\hat{q}}} \right\rangle = \frac{(\hat{q}-1)!J_0}{N^{\hat{q}-1}}. 
\end{align}
The Hamiltonian is positive semi-definite by construction. By definition, the supersymmetric extension of the SYK$_2$ model is absent, so we only focus on the SSYK$_4$ model in this work.

Besides the particle-hole symmetry generator $P$ defined in Eq.~(\ref{particle-hole}), the Hilbert space is divided by the Witten parity operator $\hat{P}$. More precisely, the spectrum of the SSYK$_4$ model is divided into the two parity sectors with $\hat{P}=\pm 1$. The Witten parity operator is defined by the fermion number mod 2, $\hat{P}= (-1)^{\hat{N}}$ with the fermion number operator $\hat{N}= \sum_{i=1}^{N/2} \bar{c}_i c_i$ defined by the Dirac fermion creation and annihilation operators or $(-i)^{\frac{N}{2}}\prod_{i=1}^{N} \psi_i$ equivalently.

The $T\bar{T}$-deformed SSYK model follows the same level statistic classification as the undeformed model. It is due to the fact that the deformed spectrum $E(\lambda)$ is generated from the un-deformed spectrum $E_0$ by Eq.~(\ref{flowsyk}). The deformed spectrum retains the degeneracy of the original spectrum. We have also confirmed this by examining the spectrum explicitly. For $N\mod 8 = 0, 6$, the two parity sectors are doubly degenerate. For $N \mod 8 = 2, 4$, there are double degeneracies within each parity sector and between the parity sectors. As a result, the total spectrum is quadruply degenerate.

The different classifications of the SSYK with different $N$ can also be observed from the plot of the SFF. The shape of the ramp region with $N\mod 8=2,4$ reflects that they follow the same GSE class. The SFF curves of $N=18, 20$ are more prominent when compared to $N=10, 12$ due to the finite number of samples used. The other cases with $N\mod 8=0, 6$ follow the GOE classification, and the SFF curves behave differently from the GSE class.

\section{The SFF in the $T\bar{T}$-deformed SYK models}
\label{section-SFF}
\subsection{The energy level spacing distributions}
\begin{figure}[htbp]
	\centering
	\captionsetup[subfloat]{farskip=10pt,captionskip=1pt}
	\subfloat[t][\centering{SYK$_4$, $N=18$}]{\label{PlotSYK4LevelSpacingLN18}
		\includegraphics[height =0.2\linewidth]{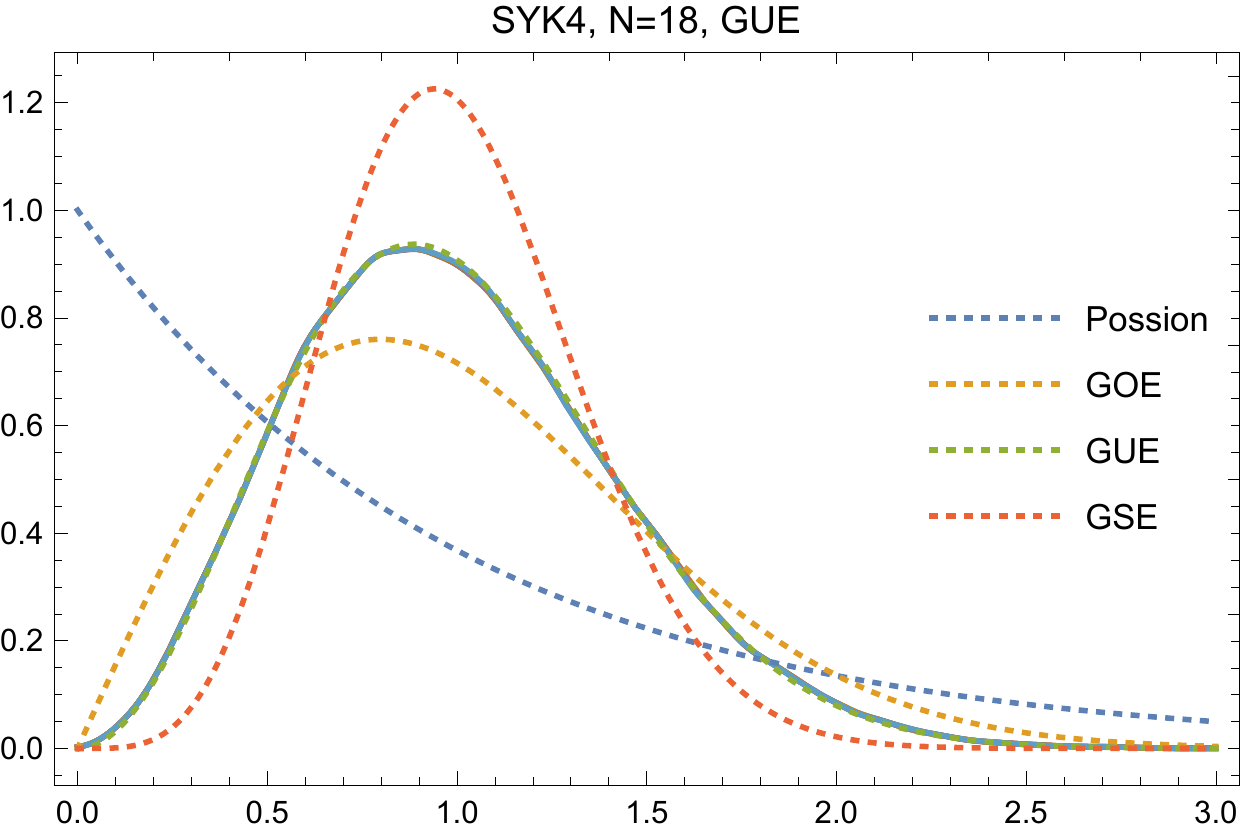}}
	\subfloat[t][\centering{SSYK$_4$, $N=18$}]{\label{PlotSSYK4LevelSpacingLN18}
		\includegraphics[height =0.2\linewidth]{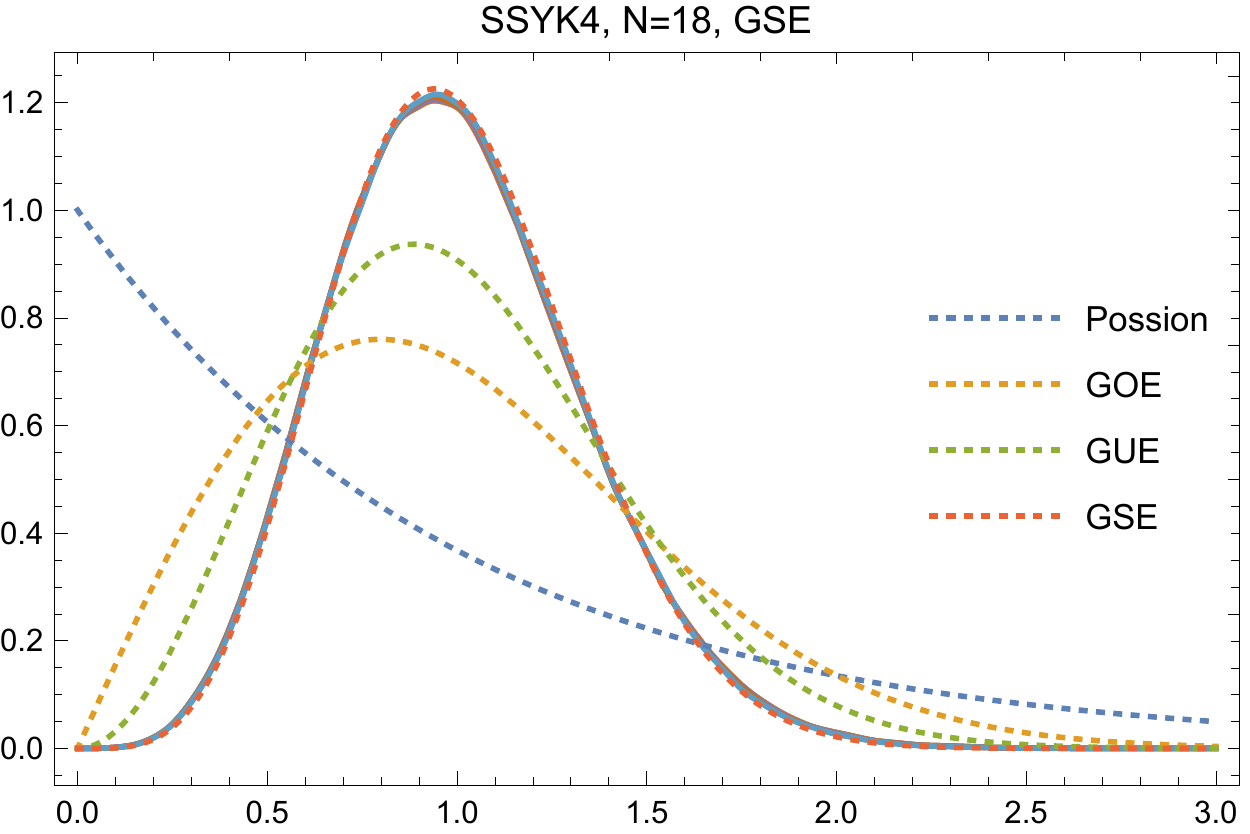}}
	\subfloat[t][\centering{SYK$_2$, $N=18$}]{\label{PlotSYK2LevelSpacingLN18}
		\includegraphics[height =0.2\linewidth]{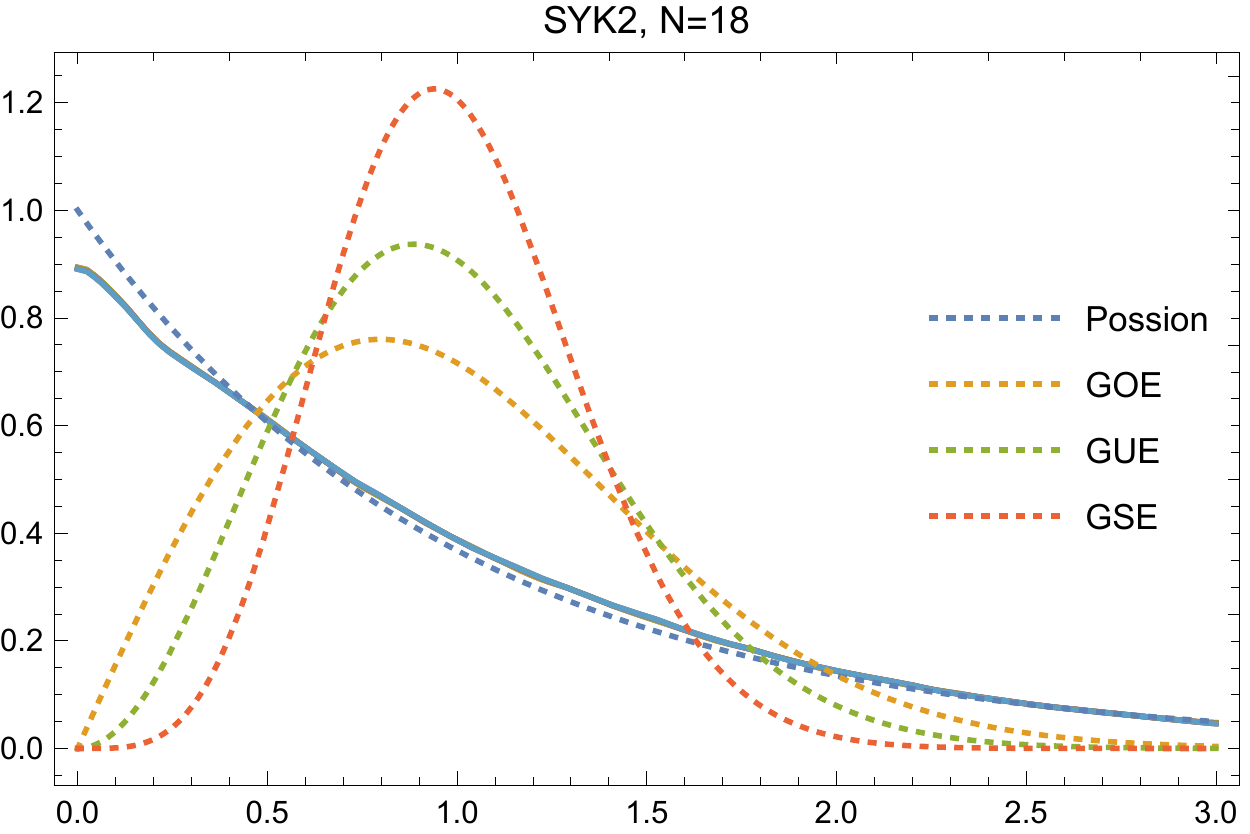}}\\
	\subfloat[t][\centering{SYK$_4$, $N=20$}]{\label{PlotSYK4LevelSpacingLN20}
		\includegraphics[height =0.2\linewidth]{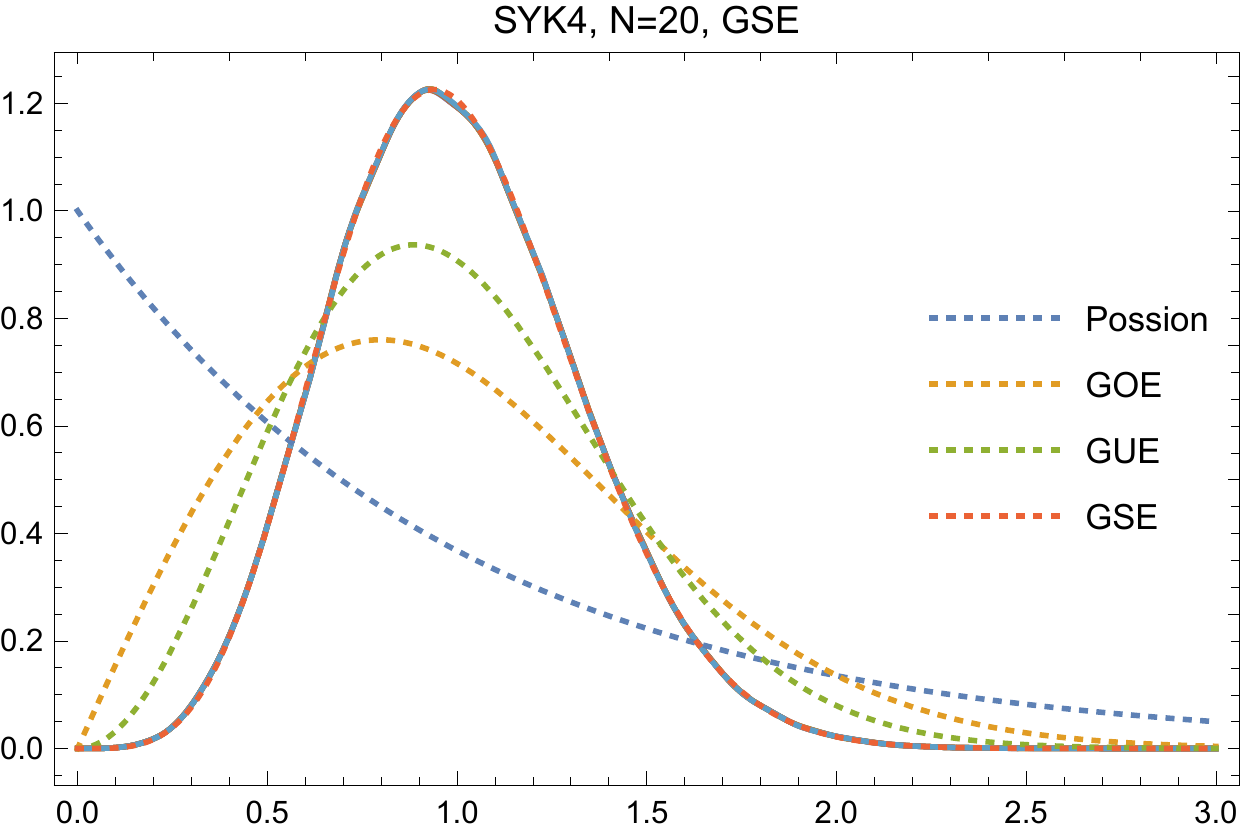}}
	\subfloat[t][\centering{SSYK$_4$, $N=20$}]{\label{PlotSSYK4LevelSpacingLN20}
		\includegraphics[height =0.2\linewidth]{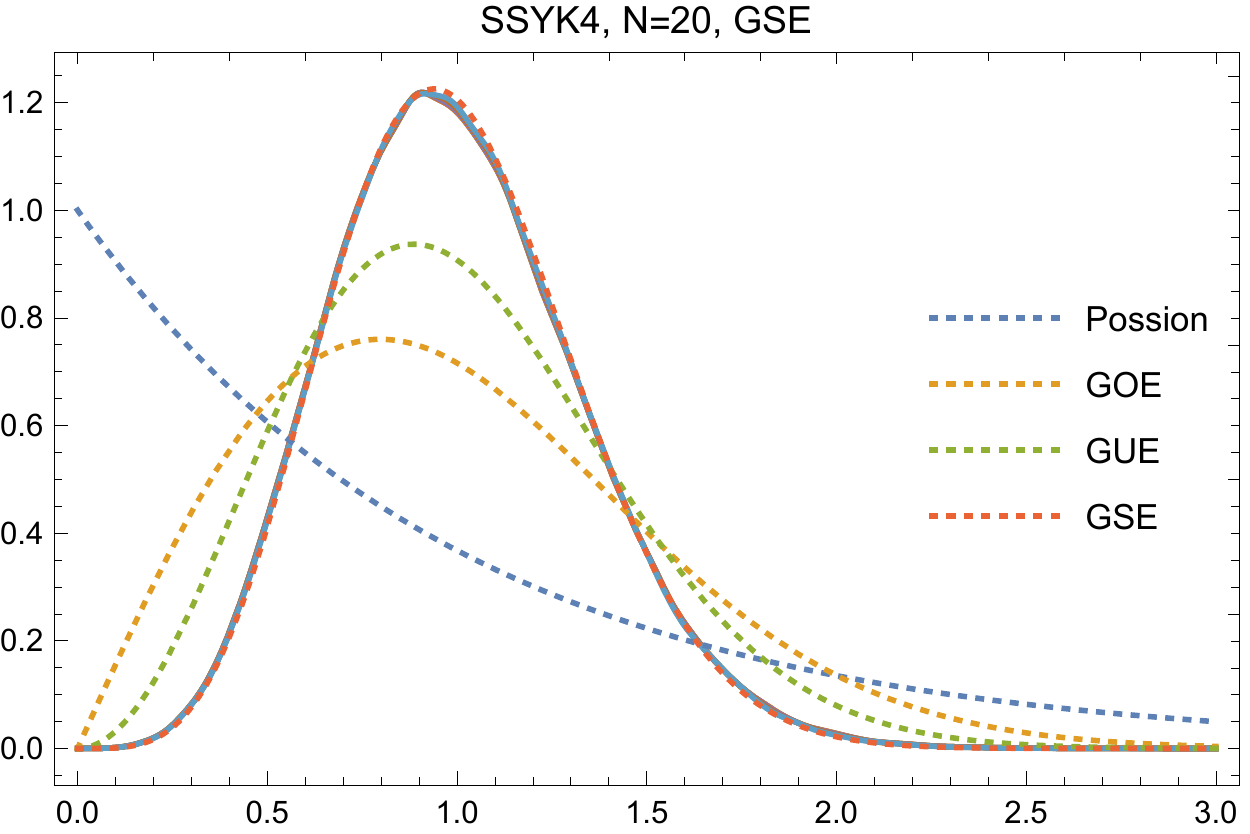}}
	\subfloat[t][\centering{SYK$_2$, $N=20$}]{\label{PlotSYK2LevelSpacingLN20}
		\includegraphics[height =0.2\linewidth]{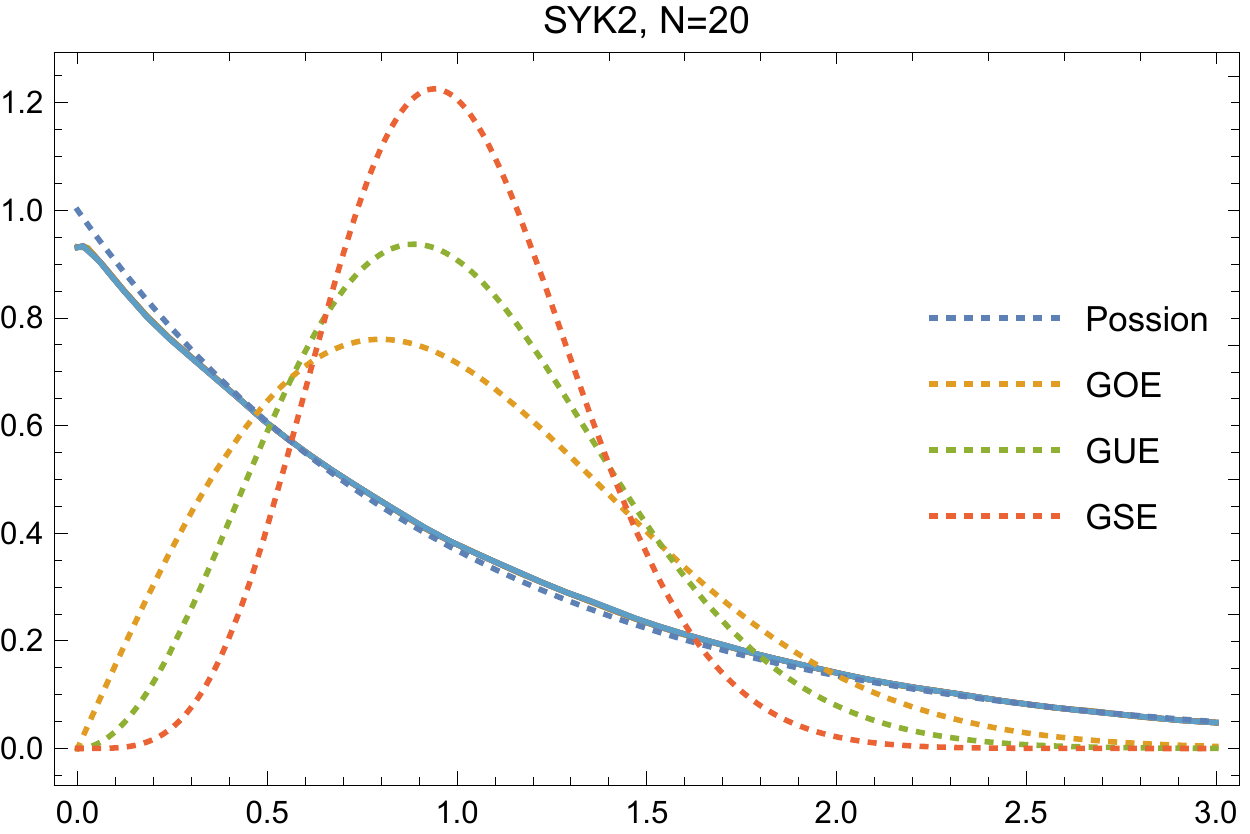}}\\
	\subfloat[t][\centering{SYK$_4$, $N=22$}]{\label{PlotSYK4LevelSpacingLN22}
		\includegraphics[height =0.2\linewidth]{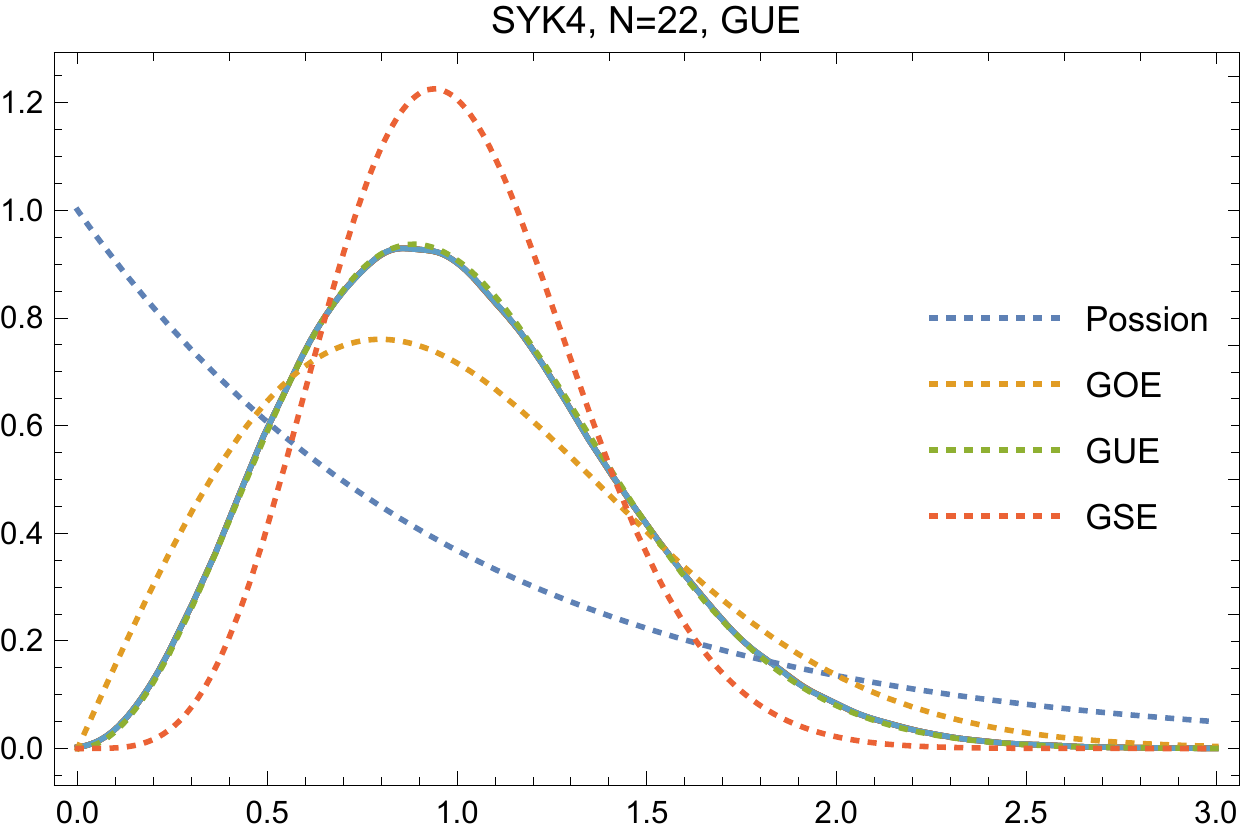}}
	\subfloat[t][\centering{SSYK$_4$, $N=22$}]{\label{PlotSSYK4LevelSpacingLN22}
		\includegraphics[height =0.2\linewidth]{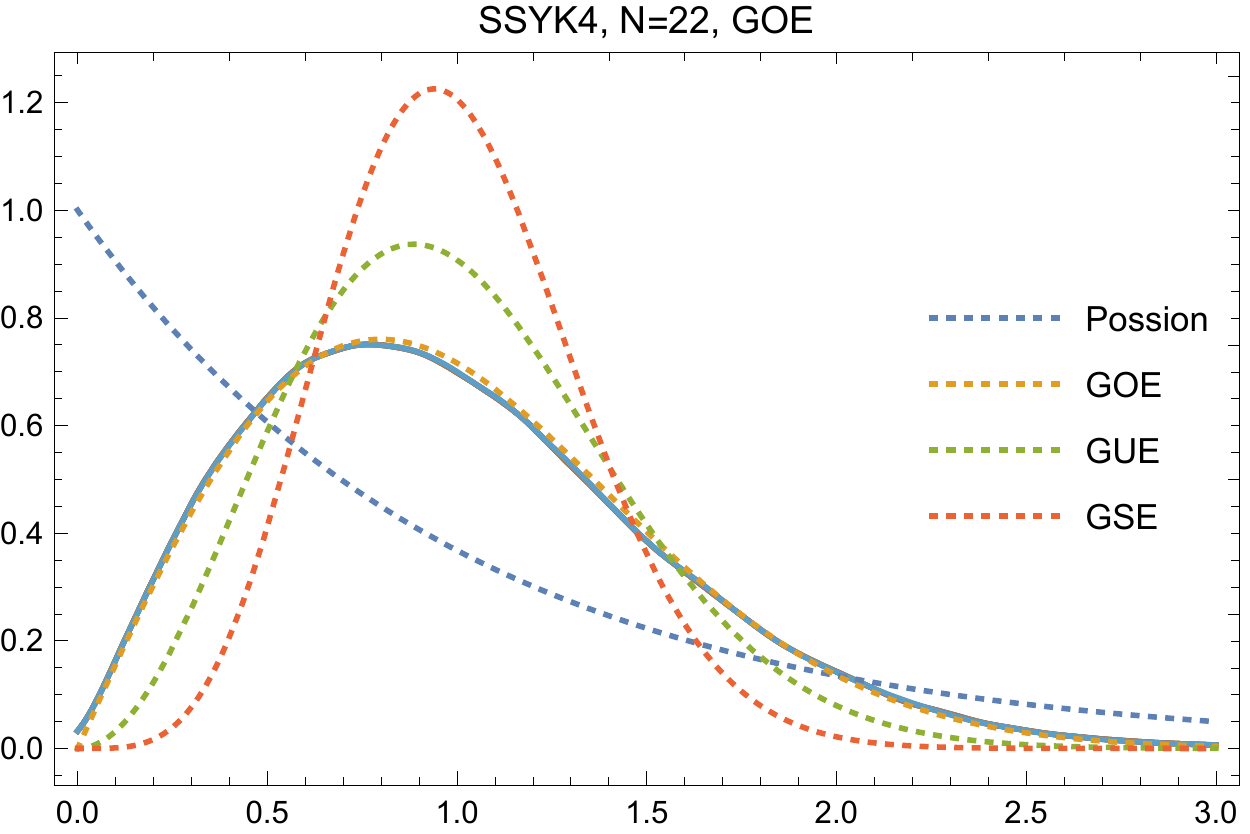}}
	\subfloat[t][\centering{SYK$_2$, $N=22$}]{\label{PlotSYK2LevelSpacingLN22}
		\includegraphics[height =0.2\linewidth]{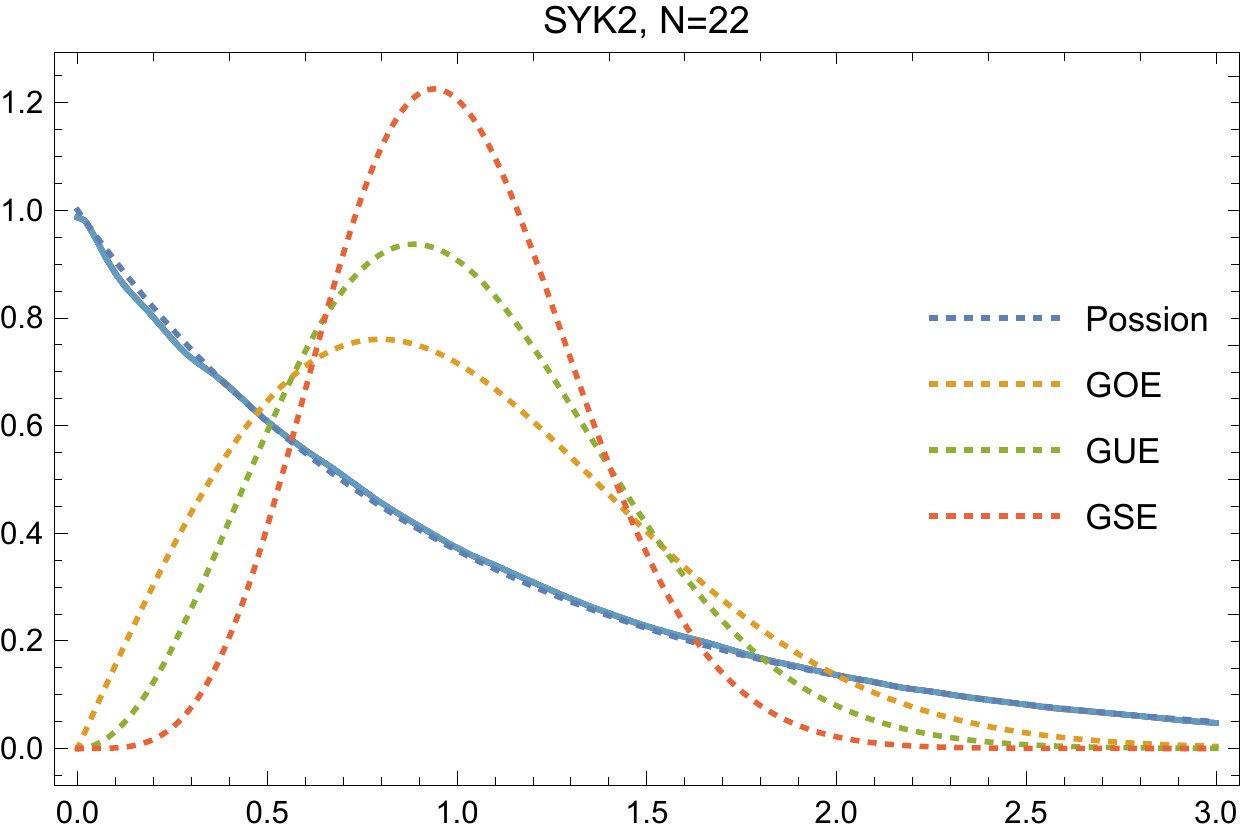}}\\
	\subfloat[t][\centering{SYK$_4$, $N=24$}]{\label{PlotSYK4LevelSpacingLN24}
		\includegraphics[height =0.2\linewidth]{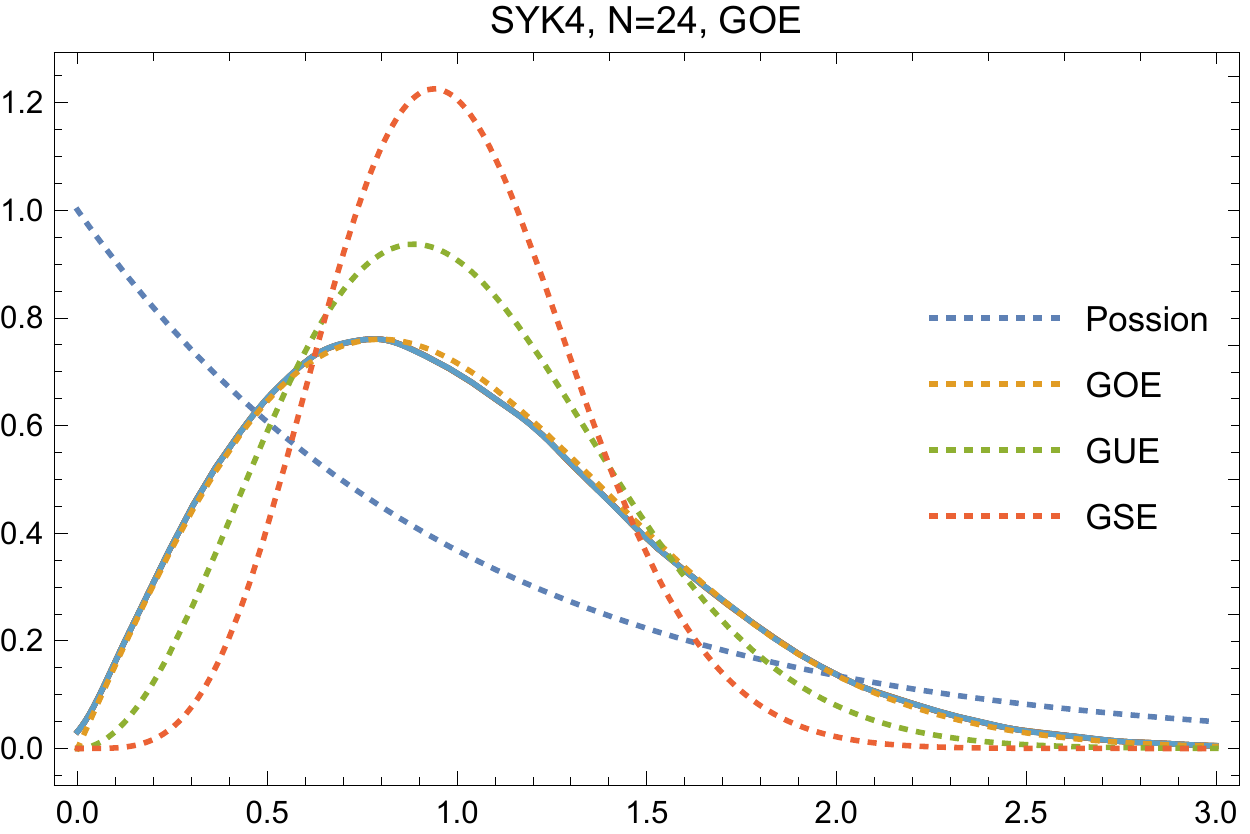}}
	\subfloat[t][\centering{SSYK$_4$, $N=24$}]{\label{PlotSSYK4LevelSpacingLN24}
		\includegraphics[height =0.2\linewidth]{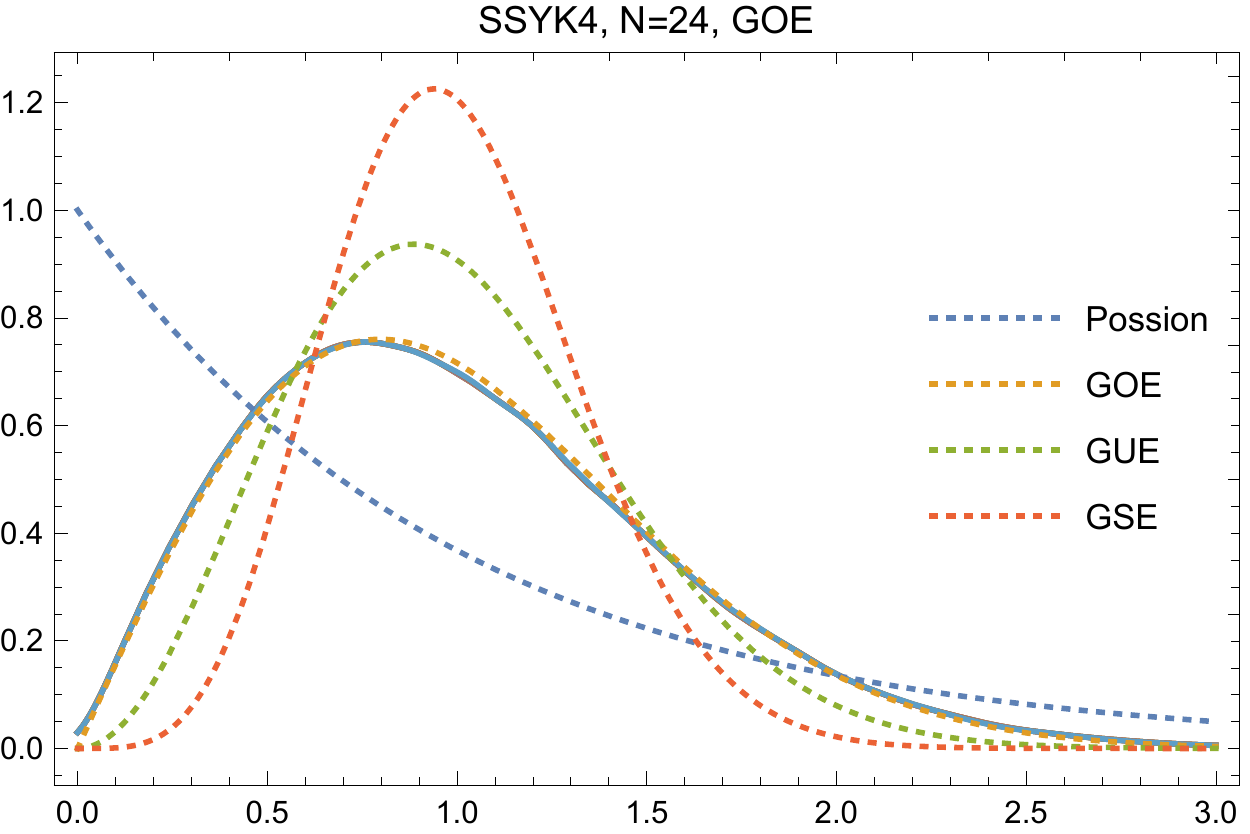}}
	\subfloat[t][\centering{SYK$_2$, $N=24$}]{\label{PlotSYK2LevelSpacingLN24}
		\includegraphics[height =0.2\linewidth]{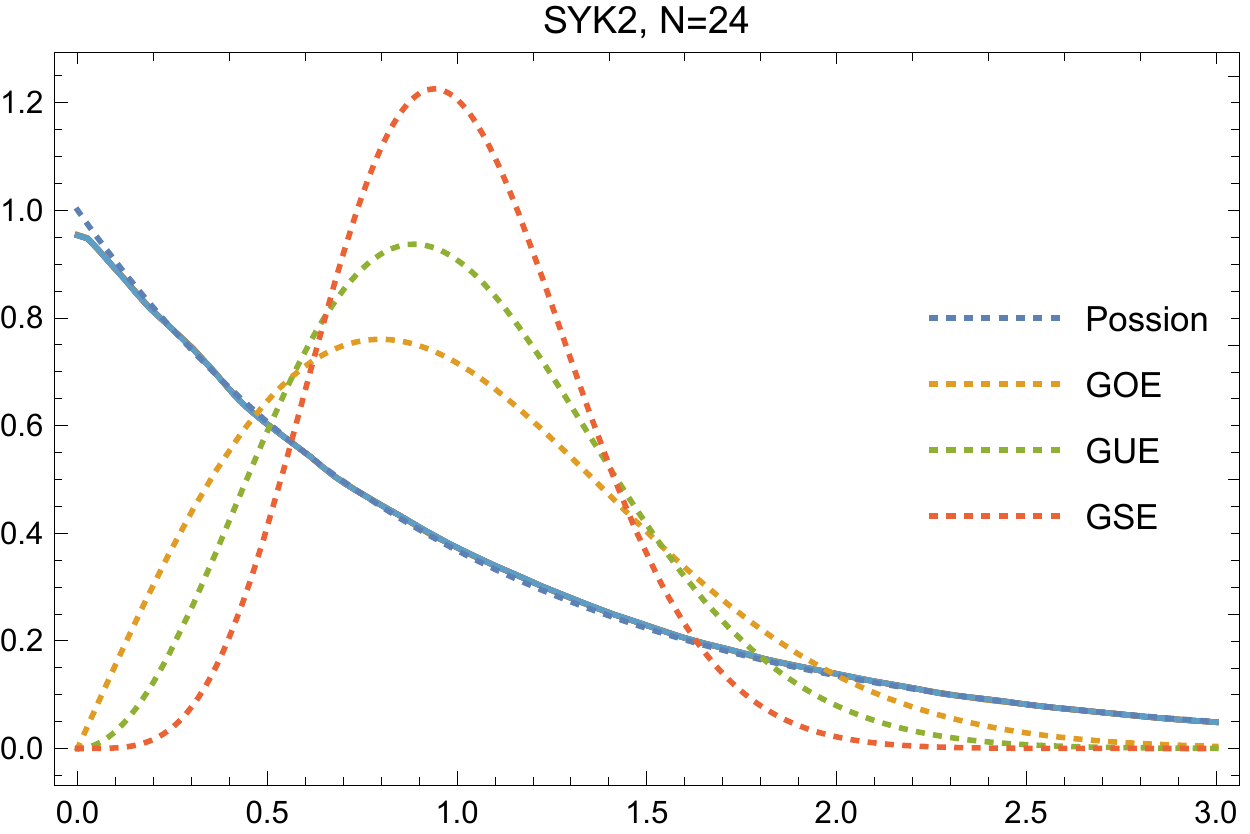}}
	\caption{Plot of the nearest neighbor energy level spacing distributions of the SYK$_4$, SSYK$_4$ and  SYK$_2$ models with $\lambda=0$, $-0.001$, $-0.1$, $-1$, $-10$, $-100$, $-1000$. The curves with different $\lambda$ overlap each other exactly.	}
	\label{level-spacing}
	\vspace{-0.5em}
\end{figure}

In this section, we investigate the quantum chaos signals of the $\TT$-deformed theory. The quantum chaotic behavior can be characterized by the nearest neighbor energy level spacing distribution \cite{Bohigas:1983er,Guhr:1997ve,Lau:2020qnl}. For the Gaussian random matrices \cite{Guhr:1997ve,mehta:2004}, it is easy to deduce that the level spacing distributions are unchanged under the $\TT$ deformation. We will display it below. For Gaussian random matrices, the Hamiltonian can be expressed as 
\begin{align}
H=U\Theta U^{-1}
\end{align}
where $U$ is an element of some symmetric group and $\Theta$ is a diagonal matrix with the diagonal elements $x_1$, $x_2$,$\cdots$, $x_N$. The $\TT$ deformation only changes the diagonal elements by $x_i\rightarrow f_i\equiv f(x_i-E_0)$ where $E_0$ is the minimum value of $x_i$. One can derive the joint probability density function for the eigenvalues of the Gaussian random matrices under the $\TT$ deformation
\begin{align}
P^\lambda_{N\beta}\left(f_1,\cdots,f_N\right)
=\left(\prod_{i=1}^N {f'_i}^{-1}\right)P^0_{N\beta}(x_1,\cdots,x_N)\,,
\label{equ:joint-probability-density}
\end{align}
where $x_i$ is the $i$th eigenvalue of the original Hamiltonian and $\beta=1$, $2$, $4$ for GOE, GUE and GSE respectively. The superscripts $0$ and $\lambda$ label the original and the deformed quantity, respectively. The probability density $A^\lambda_n$ for $n$ levels $f_i$ lying within an interval $\mathcal{I}=[f(-\theta-E_0),f(\theta-E_0)]$ can be obtained by integrating out the last $N-n$ arguments outside an interval $\mathcal{I}$, we can express the probability density $A^\lambda_n$ for several levels $f_i$ inside $\mathcal{I}$ by 
\begin{align}
A^\lambda_n(\mathcal{I},f_1,\cdots,f_n)=
\prod_{i=1}^n {f'_i}^{-1}A^0_n\left(\theta;x_1,\cdots,x_n\right)\,,
\label{eq:n-probability-density}
\end{align}
where $A^0_n$ is the probability density without the $\TT$ deformation. The $n$-point correlation function with the $\TT$ deformation is 
\begin{align}\label{npointfunction}
R^\lambda_n(f_1,\cdots,f_n)=
\prod_{i=1}^n {f'_i}^{-1}R^0_n\left(x_1,\cdots,x_n\right)\,,
\end{align}
where $R^0_n$ is the undeformed quantity as before. In order to eliminate the dependence on the mean level density $R_1^\lambda(f)$, one should introduce the dimensionless scaled variables
\begin{align}
\xi_i=\int_0^{f_i}R^\lambda_1(g_i)dg_i\
=\int_{-\infty}^{x_i}R_i^0(x'_i)dx'_i,.
\label{eq:mean-level-spacing}
\end{align}
The unfolded version of $A^\lambda_n$ is defined by
\begin{align}
B^\lambda_n(\mathcal{I}';\xi_1,\cdots,\xi_n)d\xi_1\cdots d\xi_n
&=A_n^\lambda(\mathcal{I};f_1,\cdots,f_n)df_1\cdots df_n  \nn\\
&=A_n^0\left(\theta;x_1,\cdots,x_n\right)dx_1\cdots dx_n\,,
\label{eq:unfolded-probability-density}
\end{align}
where $\mathcal{I}'$ is the image of $\mathcal{I}$ under the map (\ref{eq:mean-level-spacing}). In the second line, we have used Eq.~(\ref{eq:n-probability-density}). From Eq.~(\ref{eq:mean-level-spacing}) and (\ref{eq:unfolded-probability-density}), we find that the $\TT$ deformation doesn't affect the unfolded probability density $B_n$. The nearest neighbor energy level spacing \cite{mehta:2004} is defined by
\begin{align}
p(s)=2B^\lambda_2(\mathcal{I}';\mathcal{I}'_-,\mathcal{I}'_+)\,,
\end{align}
where $\mathcal{I}'_\pm$ are the right/left endpoint of $\mathcal{I}'$ and $s=\mathcal{I}'_+-\mathcal{I}'_-$. Based on the derivation from Eq.~(\ref{equ:joint-probability-density}) to (\ref{eq:unfolded-probability-density}), we find that the conclusion doesn't depend on the concrete expression of the undeformed probability density $P^0_{N\beta}(x_1,\cdots, x_N)$ so we expect that the level spacing distributions of the SYK model are unchanged under the $\TT$ deformation. We display the numerical results of the SYK models in Figure \ref{level-spacing}. We find that the level spacing distributions are unchanged under the $\TT$ deformation as claimed.

\subsection{The SFF of the chaotic models}
\label{section-SFF}

In the following paragraph, we will investigate the SFF of the deformed SYK models. The SFF for a disordered system is \cite{Cotler:2016fpe}:
\begin{equation}
	\begin{aligned} 
	g(t, \beta) &=\frac{\left\langle Z(\beta, t) Z^{*}(\beta, t)\right\rangle_{J}}{\langle Z(\beta)\rangle_{J}^{2}} \,, \\ 
	g_{d}(t, \beta) &=\frac{\langle Z(\beta, t)\rangle_{J} \cdot\left\langle Z^{*}(\beta, t)\right\rangle_{J}}{\langle Z(\beta)\rangle_{J}^{2}} \,, \\ 
	g_{c}(t, \beta) &=g(t, \beta)-g_{d}(t, \beta) \,,
	\end{aligned}
	\label{def-SFF}
\end{equation}
where the disorder average is taken separately in the numerator and denominator. The late-time behavior of this quantity is complicated but can be extracted by taking the long-time average. In this process, the terms with oscillating phases vanish, and only terms with $E_m=E_n$ survive. The result is
\begin{align}
     \lim_{T\rightarrow\infty}\frac{1}{T}\int_0^Tdt
     \left|\frac{Z(\beta,t)}{Z(\beta)}\right|^2
     =\frac{1}{Z(\beta)^2}\sum_EN_E^2e^{-2\beta E}\,,
 \end{align} 
where $N_E$ is the degeneracy of the energy level $E$. The degeneracy of the Majorana SYK$_4$ model and the SSYK$_4$ model have been reviewed in Sec.~\ref{setup}. For the $N$ Majorana SYK$_4$ model, the energy eigenvalues are doubly degenerate for $N$ mod $8=2$, $4$, $6$ and nondegenerate for $N$ mod $8=0$. For the SSYK$_4$ model, the energy eigenvalues are quadruply degenerate for $N$ mod $8=2$, $4$ and doubly degenerate for $N$ mod $8=0$, $6$. As a result, the late-time behavior of the SFF of these two types of models at infinite temperature can be summarized as follows
\begin{align}
    &\lim_{t\rightarrow\infty}\left|\frac{Z(it)}{Z(0)}\right|^2=
    \begin{cases}
    	\frac{1}{L}\quad N\text{ mod } 8=0\\
    	\frac{2}{L}\quad N\text{ mod } 8=2,4,6
    \end{cases} &&\text{for Majorana SYK$_4$} \,,\nonumber\\
    &\lim_{t\rightarrow\infty}\left|\frac{Z(it)}{Z(0)}\right|^2=
    \begin{cases}
    	\frac{2}{L}\quad N\text{ mod } 8=0,6\\
    	\frac{4}{L}\quad N\text{ mod } 8=2,4
    \end{cases}&&\text{for SSYK$_4$} \,,
    \label{degeneracy}
\end{align}
where $L=2^{N/2}$ is the dimension of the Hilbert space. Recall that the $T\bar{T}$ deformation $H\rightarrow f(H-E_0)$ preserves the classical integrability of the theory, so the late time behavior of the SFF is unchanged under the $T\bar{T}$ deformation.  

For the RMT, $g_d(t,0)$ and $g_c(t,0)$ behavior as
\begin{align}
g_d(t,0)\sim\frac{1}{t^{3/2}}
\quad\text{and}\quad
g_{\rm c}(t,0)=
\begin{cases}
t/(2\pi L^2),\quad t<2L\\
1/(\pi L),\quad t\geq2L
\end{cases}\,,
\label{SFFofRMT}
\end{align}
where $L$ is the dimension of the matrix model and corresponds to the dimension of the Hilbert space of the SYK model. According to Eq.~(\ref{SFFofRMT}), one can get the ``dip time" $t_d$ and the ``plateau time" $t_p$ as $t_d\sim L^{1/2}\sim e^{S/2}$ and $t_p\sim 2L\sim e^S$. Explicitly, the SFF of the original SYK model in the conformal limit is \cite{Cotler:2016fpe}
\begin{align}
&g_d(t,\beta)=\frac{\beta^3}{(\beta^2+t^2)^{3/2}}\exp\left(-\frac{cNt^2}{\beta(\beta^2+t^2)}\right)\,, \nn \\
&g_{\rm ramp}(t,\beta)\sim
\begin{cases}
\frac{t}{2\pi}\exp\left[-2Ns_0-\frac{cN}{\beta}\right],\quad &t<2\pi e^{Ns_0} \,, \\
\frac{t}{2\pi}\exp\left[-2Ns_0-\frac{cN}{\beta}-\frac{\beta}{cN}\log^2\left(\frac{t/(2\pi)}{e^{Ns_0}}\right)\right],
\quad &e^{Ns_0}<t<t_p \\
\exp\left[-Ns_0-\frac{3cN}{4\beta}\right],\quad &t>t_p
\end{cases}\,,
\label{SFFofSYK}
\end{align}
where $t_p=2\pi e^{Ns_0+\frac{cN}{2\beta}}=2\pi e^{S(\beta)}$ and $s_0$ is the entropy of the ground state. From Eq.~(\ref{SFFofSYK}), one can extract the ``dip time" as $t_d\sim e^{Ns_0/2}$. On the contrary, the time scale of $t_p$ for the original SYK$_2$ model is about $t_p\sim N$ which is much smaller than $L\sim 2^{N/2}\sim e^S$.

\subsection{The SFF in the $T\bar{T}$-deformed SYK$_4$ models}

The SFF defined in Eq.~(\ref{def-SFF}) of the Majorana SYK$_4$ model was first discussed in \cite{Cotler:2016fpe} and the curves of $g(t,\beta)$ are in agreement with the RMT results. The SFF of the RMT contains a slope region before the dip time $t_d$, a ramp region between $t_d$ and the plateau time $t_p$, and a plateau region at the late time. The ramp region configuration characterizes the system's quantum chaotic behavior, and the late time behavior demonstrates the discreteness of the spectrum. As discussed in Eq.~(\ref{degeneracy}), the value of the SFF at the late time is determined completely by the system's symmetry. The discussion on the similarity between the SYK model and the RMT was extended to the supersymmetric version \cite{Li:2017hdt}. In this case, the systems can be further classified according to the matrix type of the supercharge $Q$. The special structures of the Hamiltonian $H$ are modified from Gaussian to Wishart-Laguerre \cite{Dumitriu2002, Edelman2005} and the RMT ensembles to which they belong are called LOE and LSE. However, the SFFs of these systems belong to the class of GOE, and GSE \cite{Li:2017hdt}. As discussed in \cite{Hunter-Jones:2017crg}, the density of states $\rho(E)$ of SSYK models satisfy the Mar\v cenko-Pastur distribution \cite{Marcenko:1967} in the large $N$ limit. The slope region of SFF is calculated by the Fourier transformation of the square of $\rho(E)$. The result is 
\begin{align}
    g_d(t,0)=J_0^2(2t)+J_1^2(2t)\,
\end{align}
which decay slower than the RMT result in Eq.~(\ref{SFFofRMT}). So the ramp region is obscured. However, the ramp region is exposed by subtracting off the disconnected contribution from the SFF. So we display the connected SFF (cSSF) for the SSYK case. 
\begin{figure}[!htb]
		\centering
		\includegraphics[width=0.6\linewidth]{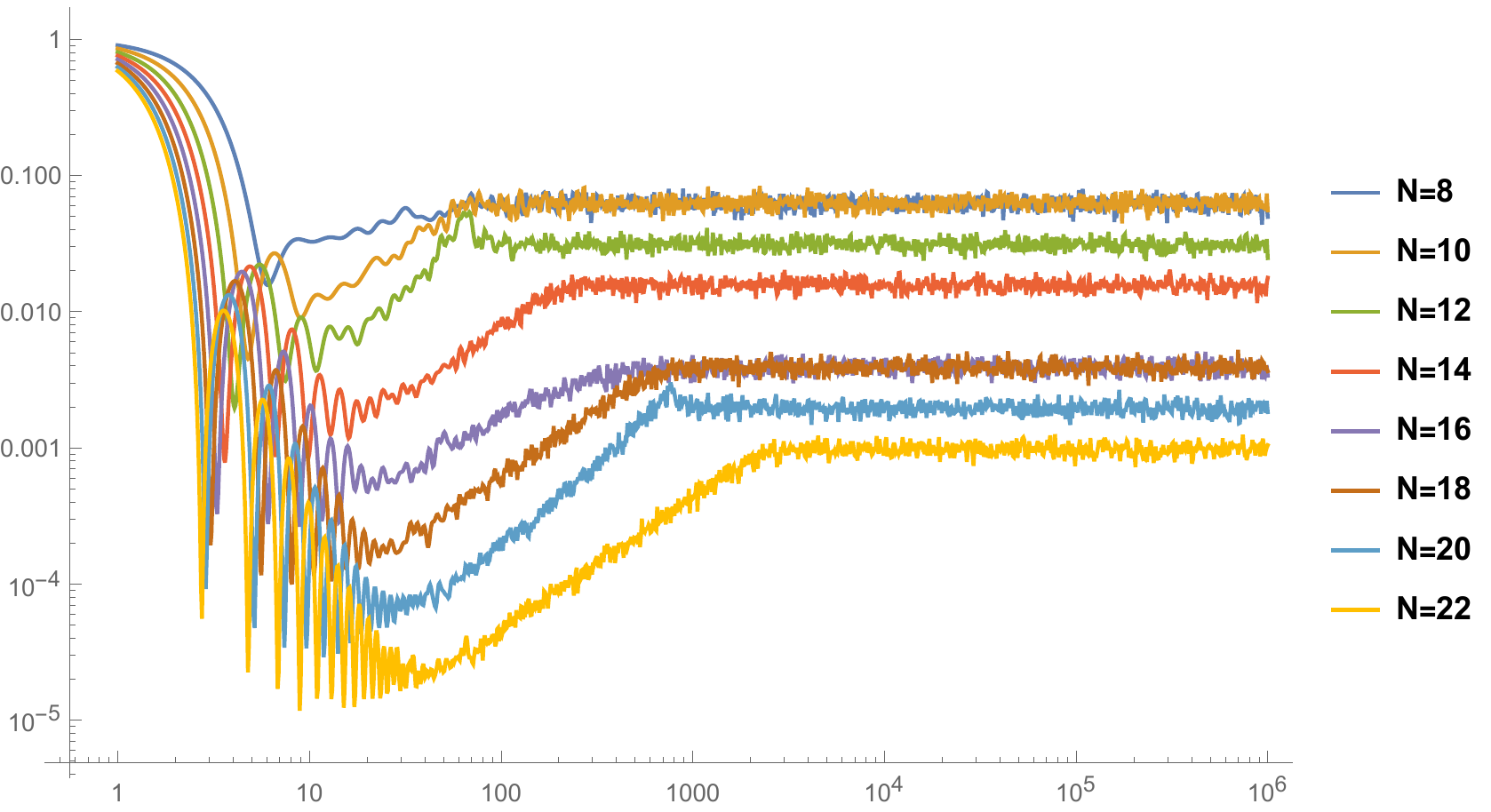}
	\caption{Plot of SFF of the undeformed SYK$_4$ model with various values of $N$.}
	\label{fig:syk4_L0_SFF_Nscan}
\end{figure}

\begin{figure}[htbp]
	\centering
	\captionsetup[subfloat]{farskip=10pt,captionskip=1pt}
	\subfloat[t][\centering{SFF for the SYK$_4$}]{\label{SFFSYK4lambda}
		\includegraphics[height =0.275\linewidth]
		{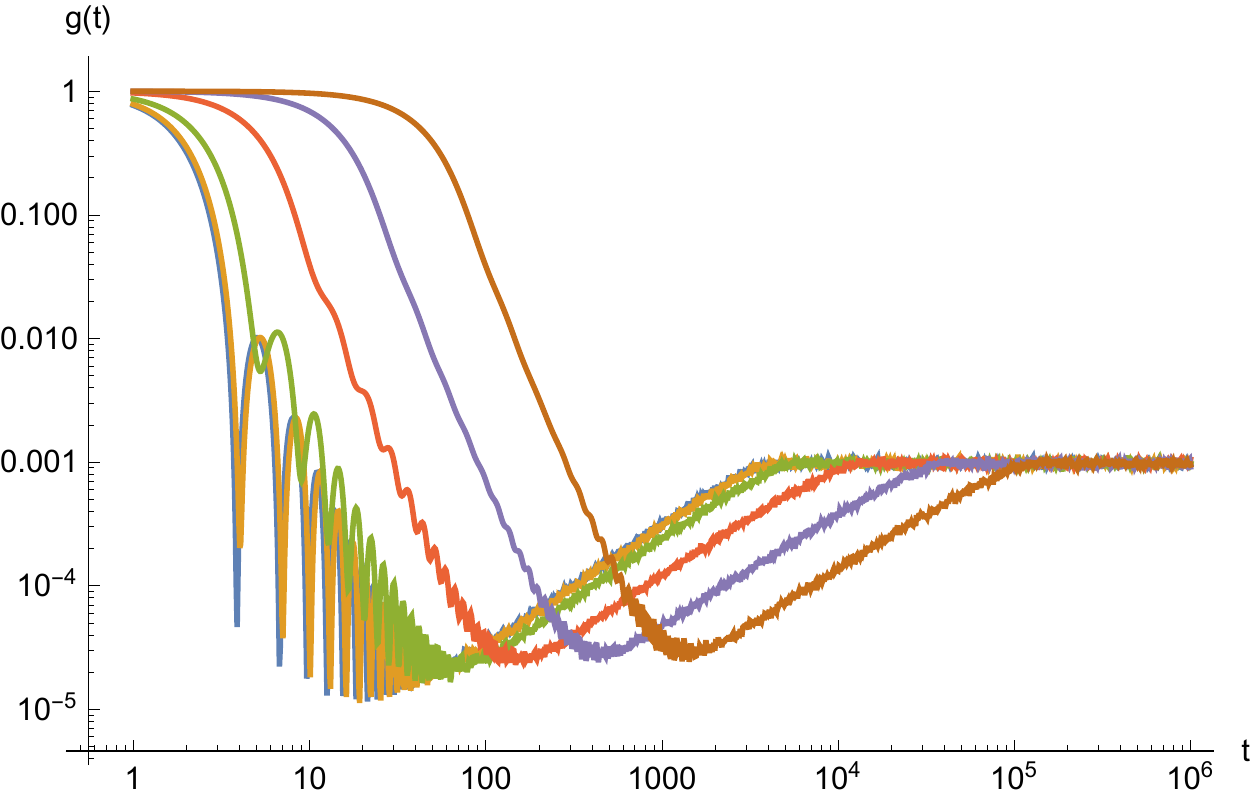}}
	\subfloat[t][\centering{cSFF for the SSYK$_4$}]{\label{fig:ssyk4_SFF_Lscan_N22}
		\includegraphics[height =0.275\linewidth]
		{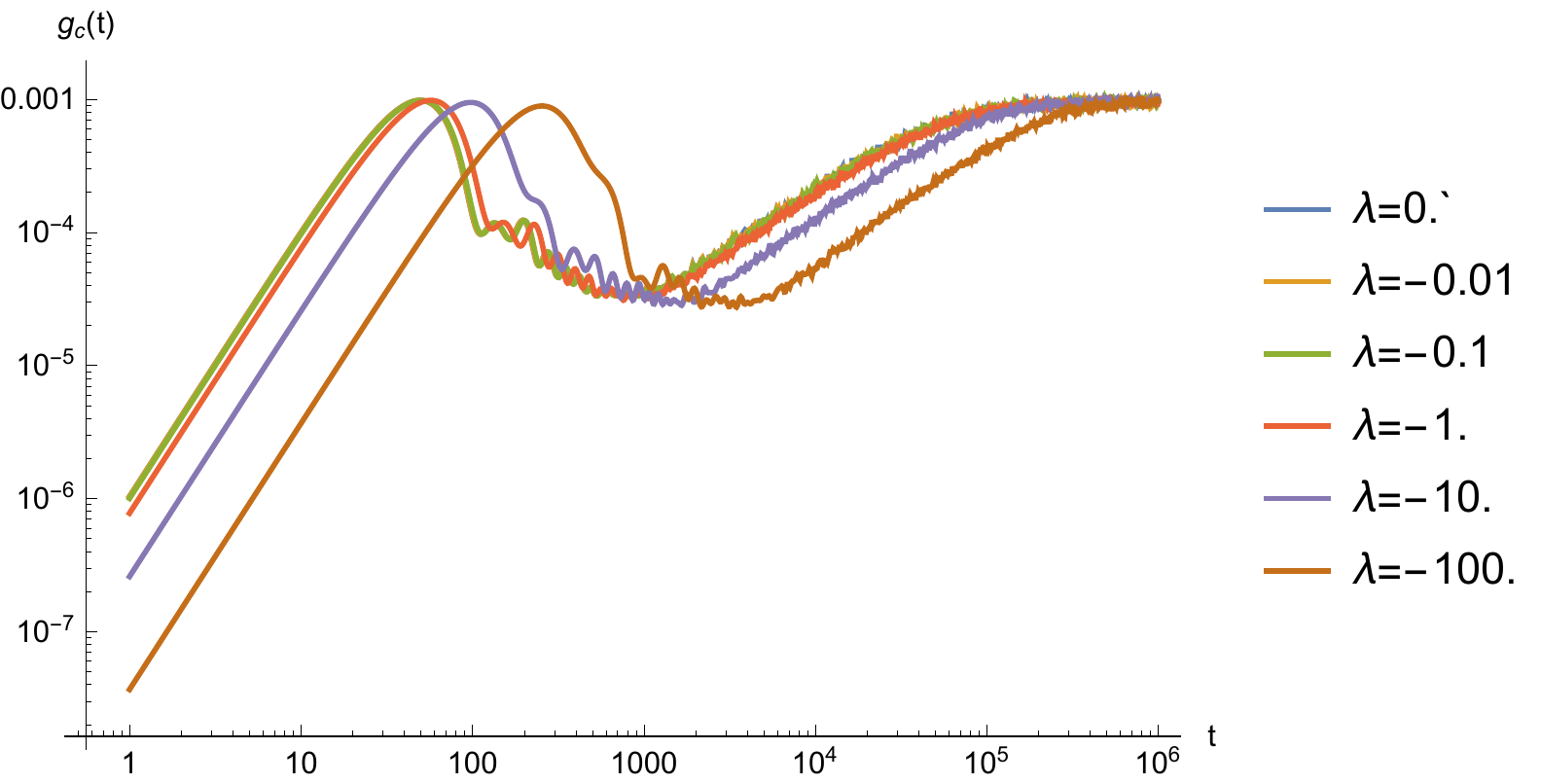}}\\
	\subfloat[t][\centering{SFF for the SYK$_4$ with respect to $J_{\rm eff}t$}]{\label{SFFSYK4JL}
		\includegraphics[height =0.275\linewidth]{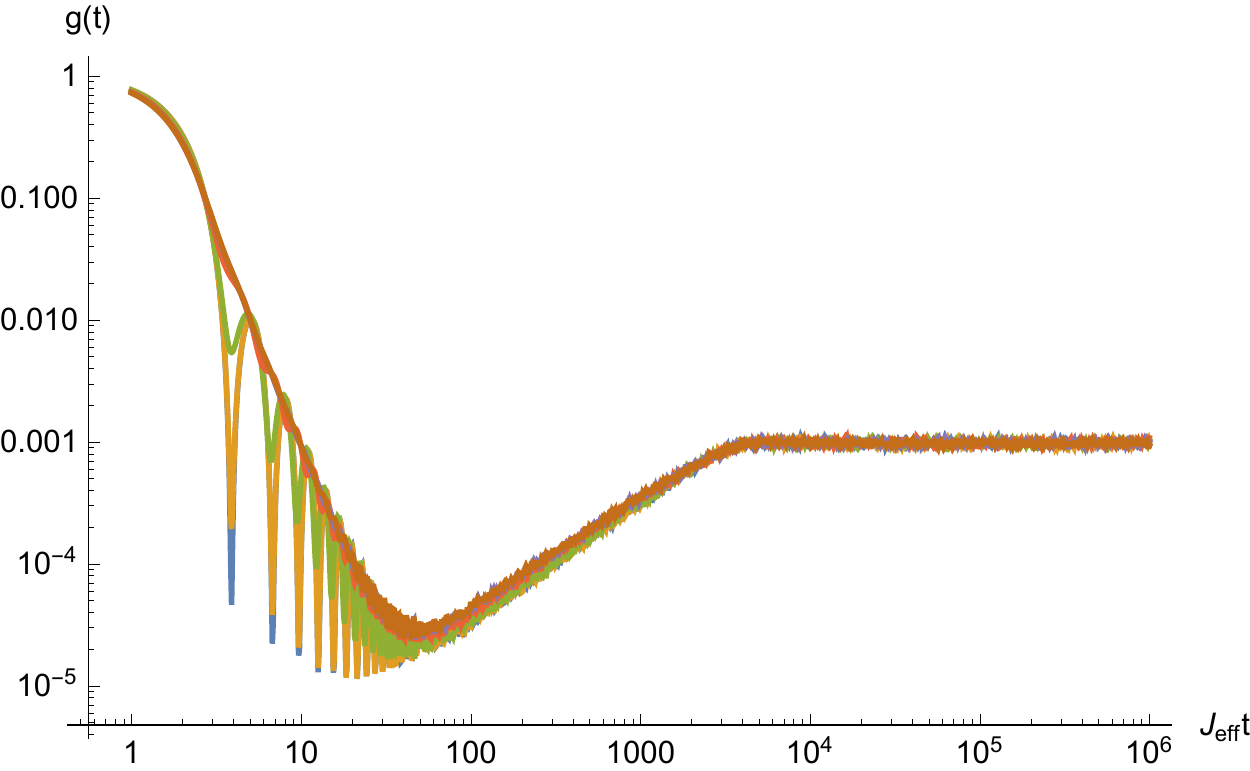}}
	\subfloat[t][\centering{cSFF for the SSYK$_4$ with respect to $J_{\rm eff}t$}]{\label{PlotSSYK4SFFcJ0L}
		\includegraphics[height =0.275\linewidth]{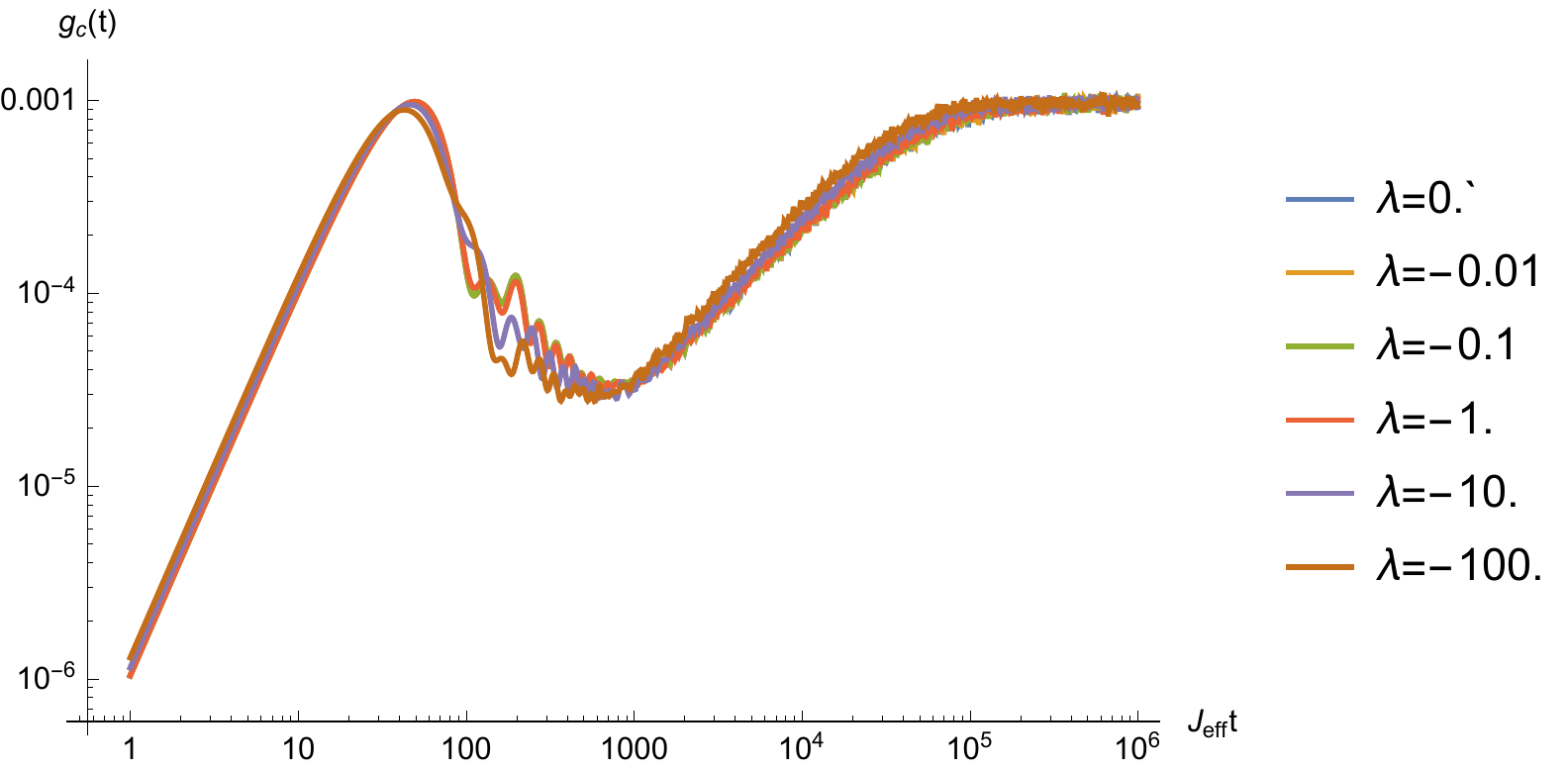}}
	\caption{(a): SSF of the $T\bar{T}$-deformed SYK$_4$ model and (b): cSSF of the $T\bar{T}$-deformed SSYK$_4$ model with various value of $\lambda$. 
	(c): SSF of the $T\bar{T}$-deformed SYK$_4$ model and (d): cSSF of the $T\bar{T}$-deformed Majorana SSYK$_4$ model with respect to $J_{\rm eff}t$ and various value of $\lambda$. We choose $N=22$, $\beta=0$ and $\lambda=0$, $-0.01$, $-0.1$, $-1$, $-10$, $-100$.
	}
	\vspace{-0.5em}
\end{figure}

In a chaotic model, the late-time behavior of the SFF is controlled by the small level spacing $s$, which approximately obeys the statistic of Gaussian ensembles. Here, we choose GUE statistics for simplicity. The two-point function has the sine kernel \cite{Cotler:2016fpe}
\begin{align}
    R_2^0(x_1,x_2)\approx \avg{\rho(x)}\delta(s)+\avg{\rho(x_1)}\avg{\rho(x_2)}\kc{1-\frac{\sin^2(\pi\avg{\rho(x)}s)}{(\pi\avg{\rho(x)}s)^2}} \,,
\end{align}
where $x=(x_1+x_2)/2$ and $s=x_1-x_2$. We will take $\rho(x)$ as the one-fold spectral density of the SYK$_4$ model with $\int\rho(x)dx=2^{N/2-1}$. The SFF of the $\TT$-deformed chaotic theory becomes
\begin{align}
\avg{\abs{Z_\lambda(\beta,t)}^2} 
=&\int_{-\infty}^\infty R^0_2(x_1,x_2)e^{-\beta(f_1+f_2)+it(f_1-f_2)}dx_1dx_2\nn\\
\approx&\abs{\avg{Z_\lambda(\beta,t)}}^2+\int dx e^{-2\beta f(x-E_0)}\min\ke{\frac{tf'(x-E_0)}{2\pi},\avg{\rho(x)}} \,.
\end{align}
Here, we have used \eqref{npointfunction} in the first step and approximated the exponent as $-2\beta f(x-E_0)+itf'(x-E_0)s$ in the off-diagonal part of the integral in the second step, since the sine kernel is localized around $s=0$.
To estimate the integral in the large $N$ limit, we consider the one-fold spectral density of the SYK$_4$ model near the edge of the spectrum and in the interior respectively as \cite{Maldacena:2016hyu,Cotler:2016fpe,Garcia-Garcia:2016mno,Feng:2018zsx}
\begin{align}\label{SpectralDensity}
    \avg{\rho(x)}=\begin{cases}
        \sqrt{\frac{2\pi}{c J_0^3}}e^{S_0} \sinh(\sqrt{2c (\abs{E_0}-\abs{x})}),& \abs{x}\to \abs{E_0}\\
        2^{N/2-1}\frac1{\sqrt{2\pi \sigma}}\exp\kc{-\frac{x^2}{2\sigma}},& \abs{x}\ll \abs{E_0}
    \end{cases},
\end{align}
where $c/2$ is the specific heat, $E_0$ is the ground state energy and the variant $\sigma\approx (0.6J_0)^2$ is determined by fitting the numerical spectrum of $N=22$.
According to the shape of the spectral density, the cSFF $g_c(t)$ with $\beta=0$ has a ramp
\begin{align}
    g_c(t)\approx
        \frac{1-\sqrt{1-16 \lambda \abs{E_0} }}{4 \lambda }2^{2-N} \frac{t}{2\pi}
        \sim 2^{2-N} \frac{\abs{E_0}J_{\rm eff}t}{\pi J_0}
        \label{equ:ramp-analysis}
\end{align}
initially and slows down gradually, where $J_{\rm eff}$ is given in \eqref{Jeff}. The ramp surpasses the slope at the dip time $t_d\approx 2^{(N-1)/3} \exp\kc{-E_0 {}^2/3 \sigma }\kc{\sigma\left| E_0\right| }^{-1/3}J_0/J_{\rm eff}$. The ramp stops at the plateau time $t_p\approx  \sqrt{2\pi}2^{N/2-1}/(0.6J_{\rm eff})$.
So the ramps of different $\lambda$ will overlap with each other as shown in Figure \ref{SFFSYK4JL}. 

\begin{figure}[htbp]
	\centering
	\captionsetup[subfloat]{farskip=10pt,captionskip=1pt}
	\subfloat[t][\centering{SFF for the SYK$_4$}]{\label{fig:syk4-L10-SFF-Nscan}
		\includegraphics[height =0.275\linewidth]{
		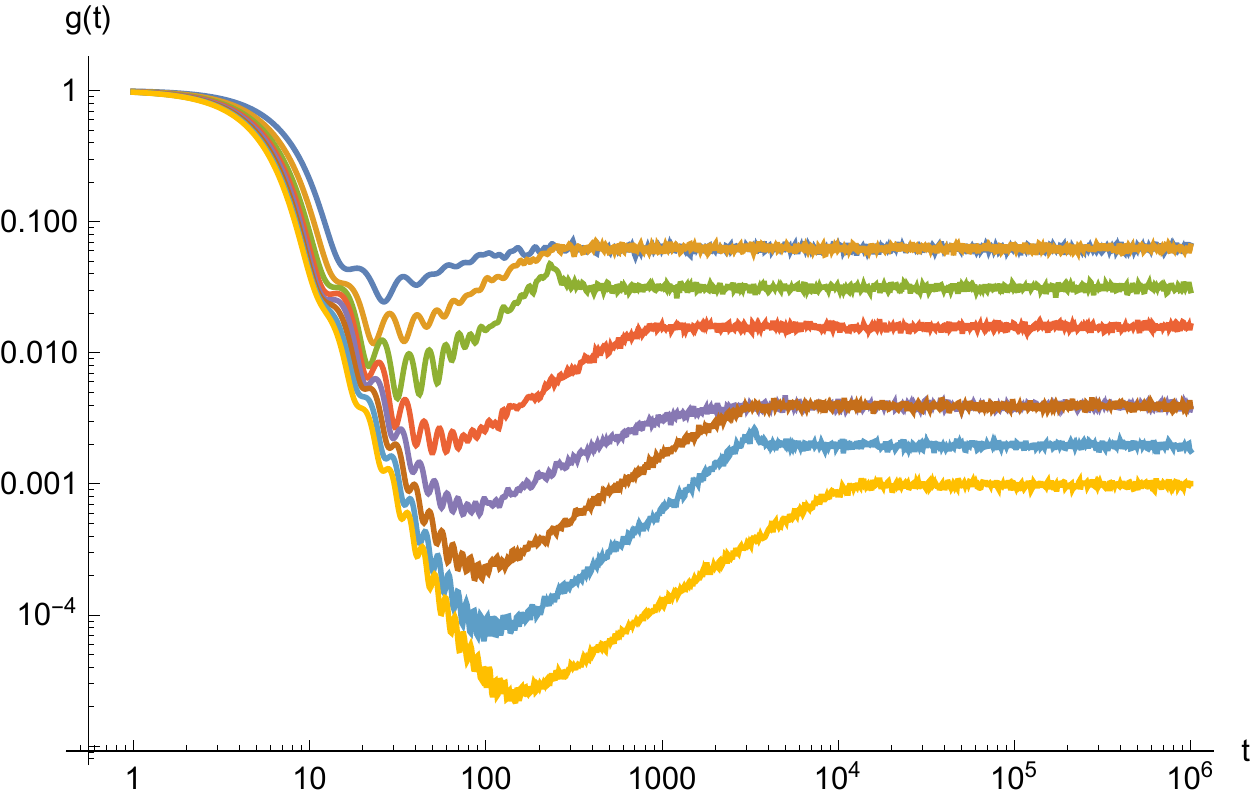}}
	\subfloat[t][\centering{cSFF for the SSYK$_4$}]{\label{fig:ssyk4-SFF-Nscan-L10}
		\includegraphics[height =0.275\linewidth]
		{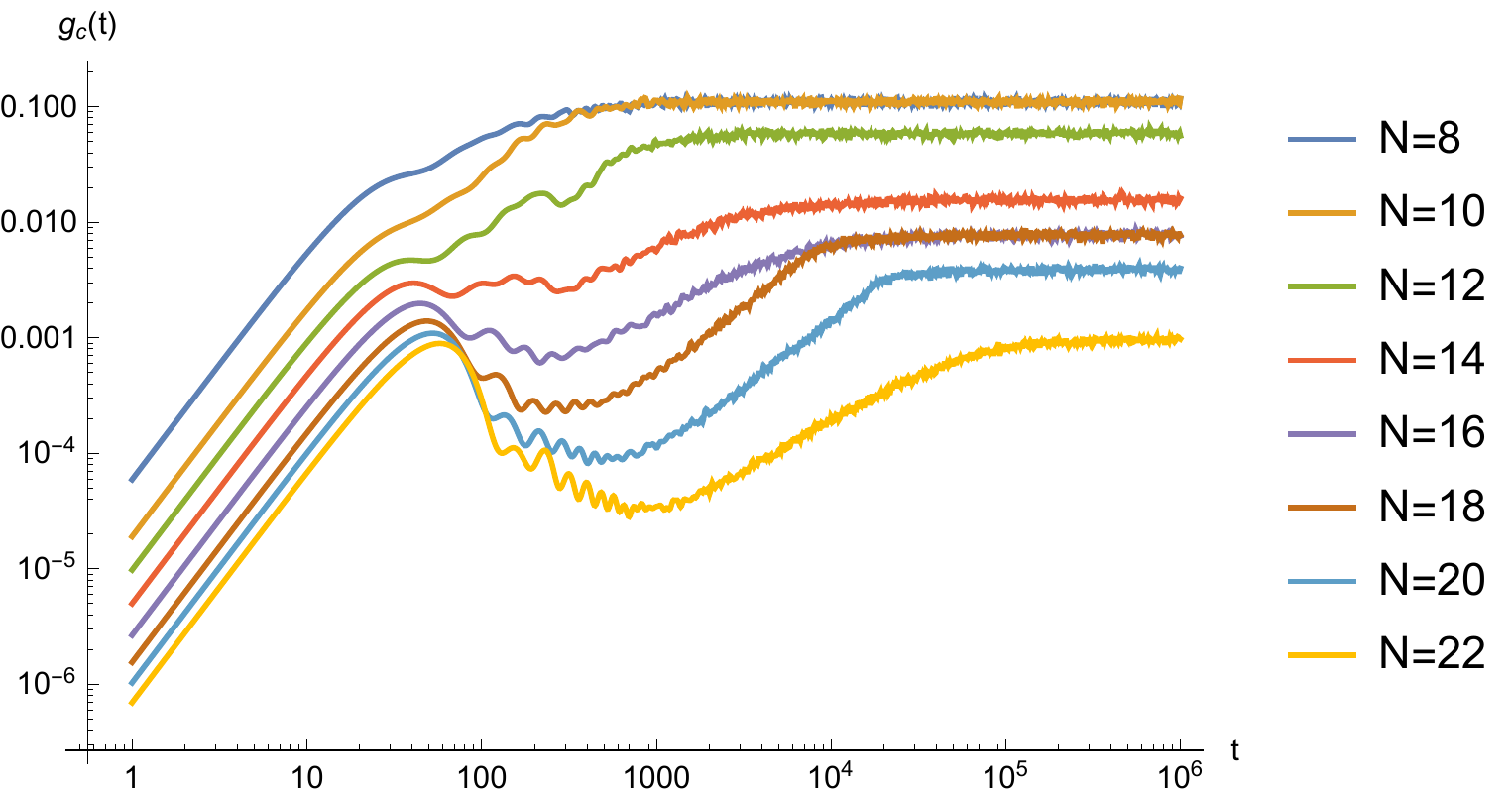}}	
		\caption{Plot of SFF of the (a) deformed SYK$_4$ model and (b) deformed SSYK$_4$ model. We choose $\lambda=-1$, $\beta=0$ and a range of $N$ from 8 to 22. 
		}
	\label{deformed-SFF}
	\vspace{-0.5em}
\end{figure}

In this paragraph, we enumerate the SFFs $g(t,0)$ of the $T\bar{T}$-deformed SYK$_4$ model in Figure \ref{fig:syk4-L10-SFF-Nscan} and its supersymmetric extension version in Figure \ref{fig:ssyk4-SFF-Nscan-L10}. For comparison, we also repeat the result in \cite{Cotler:2016fpe} and display it in Figure \ref{fig:syk4_L0_SFF_Nscan}. As we can see from the plot in Figure \ref{fig:syk4-L10-SFF-Nscan}, the $T\bar{T}$-deformed SFF also similarly exhibits the dip, the ramp, and the plateau regions. The value of $t_d$ and $t_p$ are approximately $O(1) \sim O(10)$ and $O(10)\sim O(10^3)$, respectively, and agree with the time scale mentioned in the last section. Moreover, the shape of the ramp agrees with the RMT prediction for various $N$. In particular, we can find the kink for $N$ mod $8 = 4$, which is the feature of the ramp in the GSE ensemble.

The numerical results of the cSFF of the $T\bar{T}$-deformed SSYK$_4$ model are shown in Figure \ref{fig:ssyk4-SFF-Nscan-L10}. By subtracting the disconnected contribution, the figure shows that the pattern of $g(t,0)$ also exhibits the ramp and the plateau region for large enough $N$. The time scale of $t_d$ and $t_p$ are approximate to the counterpart values of SYK$_4$. Besides, we find the ramp and plateau connect smoothly in the case of $N$ mod $8=0,\,6$ and connect at a kink in the case of $N$ mod $8=2,\,4$. This property indicates that the SFFs for $N$ mod $8=0,\,6$ follow the GOE behavior, and the SFFs for $N$ mod $8=2,\,4$ follow the GSE behavior. This result agrees with the expectation introduced in Sec.~\ref{sec:ssyk}.

To study the effect of the $T\bar{T}$ deformation on the SFF more explicitly, we show the SFF (cSSF) of the deformed SYK$_4$ and SSYK$_4$ models with different $\lambda$ in Figure \ref{SFFSYK4lambda} and \ref{fig:ssyk4_SFF_Lscan_N22}. We find that the image of $g(t,0)$ with different $\lambda$ shifts along the horizon axis by approximately the same amount in the log-log coordinate. As mentioned in Sec.~\ref{QMttbar}, the $T\bar{T}$ deformation acting on large-$N$ SYK is equivalent to replacing $J_0$ with $J_{\rm eff}(\lambda)$. We show the SFF of SYK$_4$ and cSSF of SSYK$_4$ with $T\bar{T}$ deformation in the time units $J_{\rm eff}(\lambda)$ in Figure \ref{SFFSYK4JL} and \ref{PlotSSYK4SFFcJ0L}. We find that the curves of SFF with different $\lambda$ overlap.

\subsection{The SFF in the $T\bar{T}$-deformed SYK$_2$ models}
\label{sec:SFF:SYK2}

\begin{figure}[htbp]
	\centering
	\captionsetup[subfloat]{farskip=10pt,captionskip=1pt}
	\subfloat[t][\centering{SFF with different $N$}]{\label{fig:syk2_L1_SFF_Nscan}
		\includegraphics[width =0.5\linewidth]{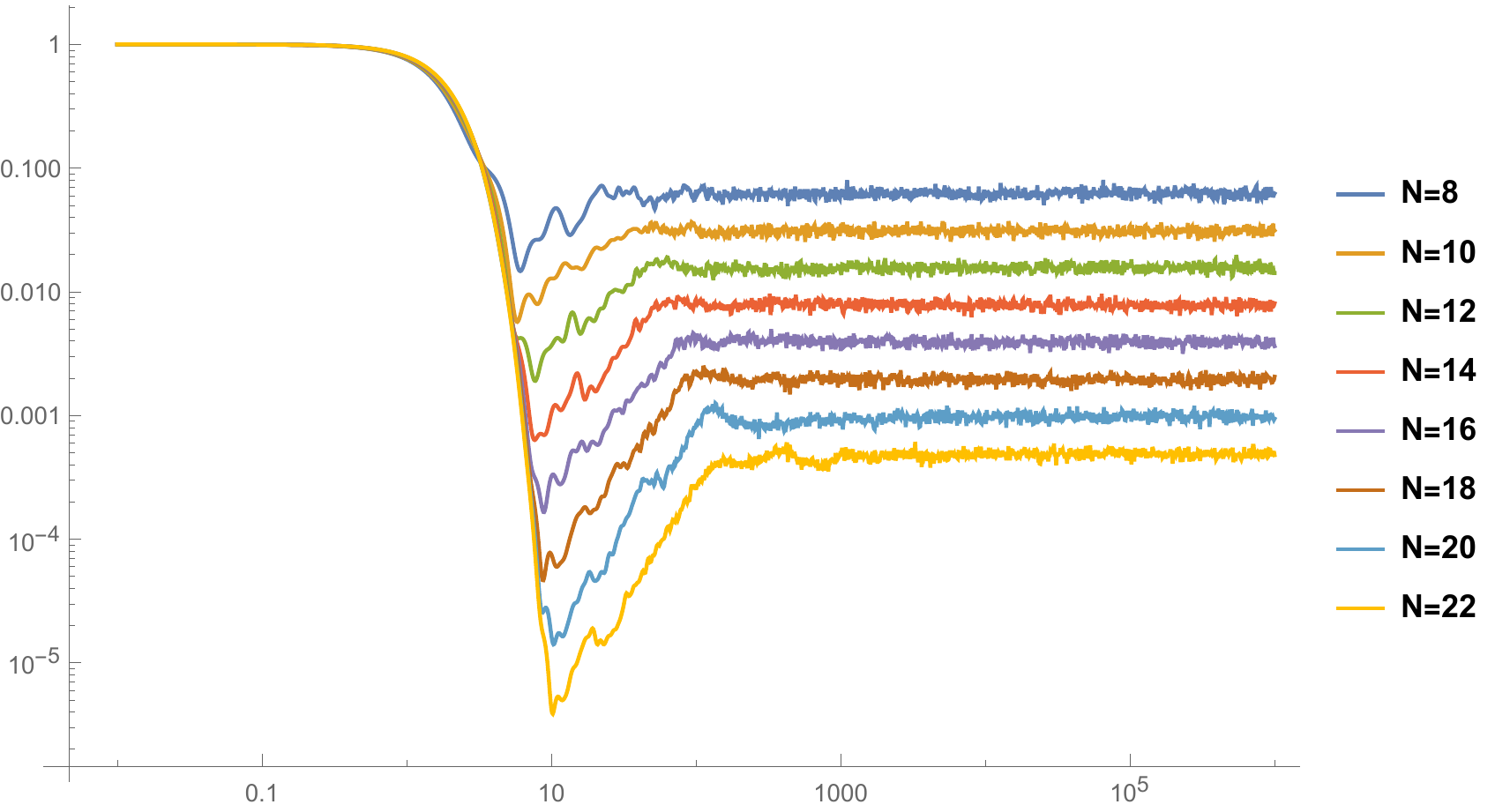}}\\
	\subfloat[t][\centering{SFF with different $\lambda$}]{\label{fig:syk2_N20_SFF_Lscan}
		\includegraphics[height =0.275\linewidth]
		{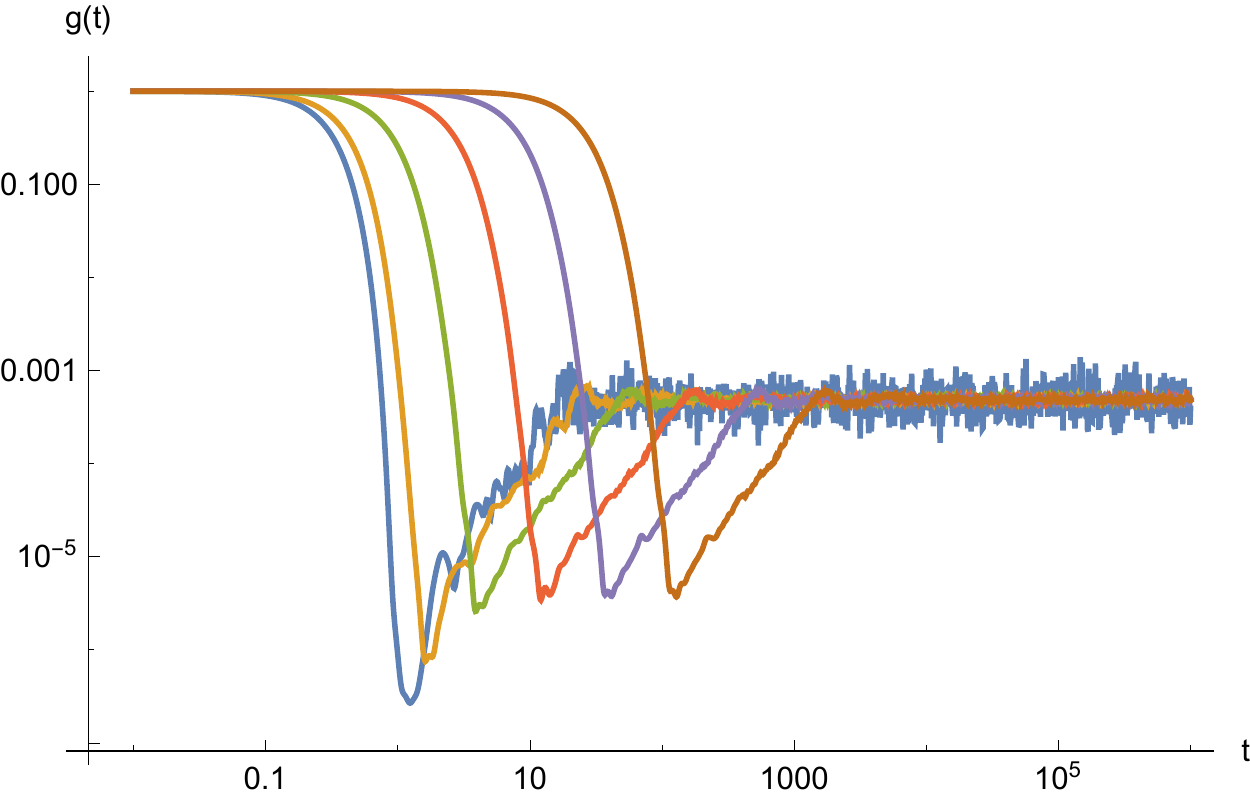}}
	\subfloat[t][\centering{SFF with respect to $J_{\rm eff}t$}]{\label{PlotSFFSYK2JL}
		\includegraphics[height =0.275\linewidth]{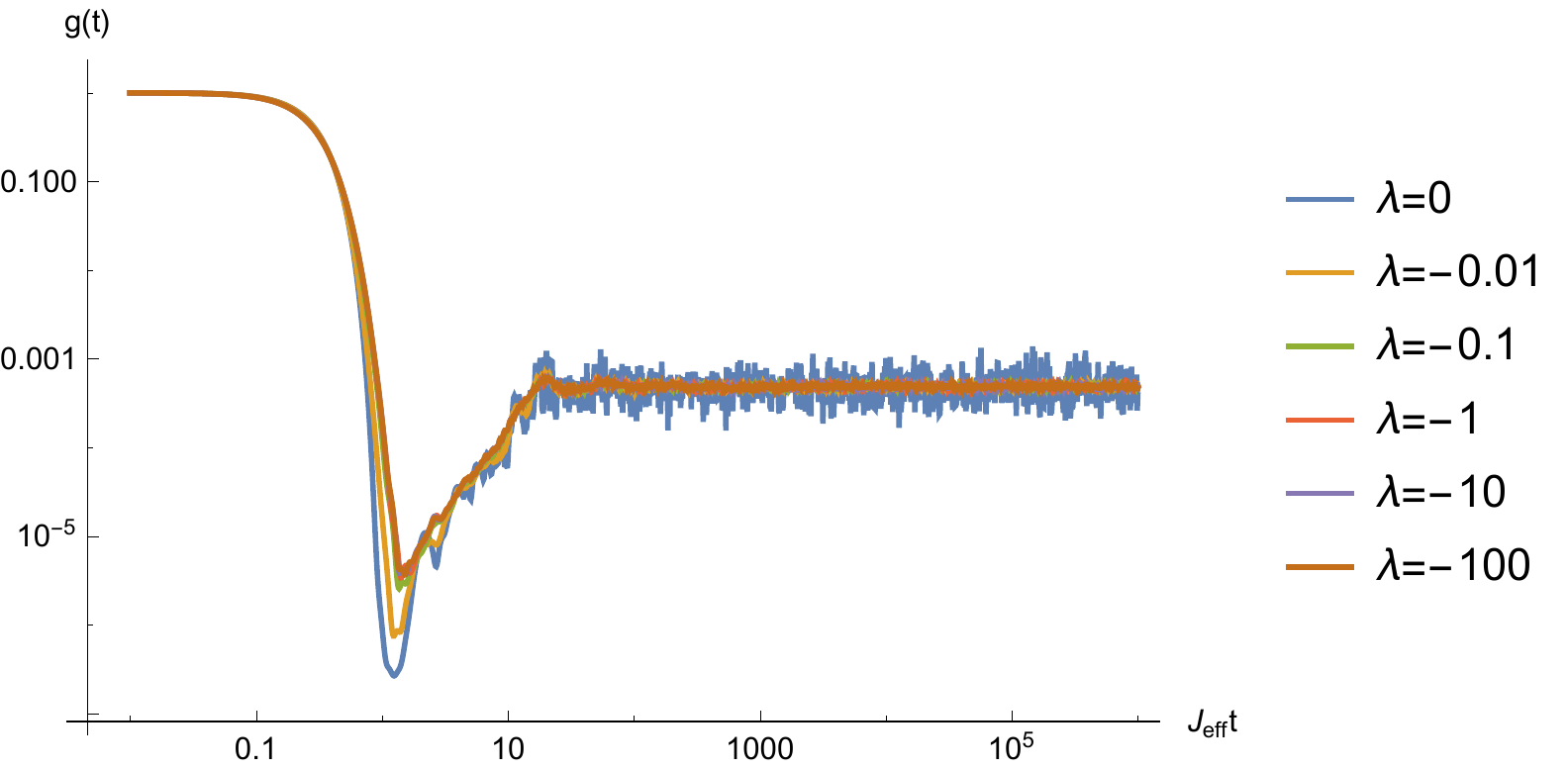}}
	\caption{(a): SSF of the $T\bar{T}$-deformed SYK$_2$ with various value of $N$ and $\lambda=-1$. (b): SSF of the $T\bar{T}$-deformed SYK$_2$ with various value of $\lambda$. (c): SSF of the $T\bar{T}$-deformed SYK$_2$ with respect to $J_{\rm eff}t$.  We choose $N=22$ in (c) and (d). 
	}
	\vspace{-0.5em}
\end{figure}

The SYK$_2$ model is equivalent to a model of free fermions with a random mass matrix \cite{Gross:2016kjj,Magan:2015yoa,Anninos:2016szt}. It can be described with a $N\times N$ random matrix which leads to the ``mini-ramp" and  the ``mini-plateau." The plateau time $t_p\sim N$ (about $O(10)$ in Figure \ref{fig:syk2_N20_SFF_Lscan}) in this model instead of $t_p\sim L$ in the Gaussian random ensemble of RMT. We show $g(t)$ of the deformed SYK$_2$ models with different $N$ and different $\lambda$ in Figure \ref{fig:syk2_L1_SFF_Nscan} and Figure \ref{fig:syk2_N20_SFF_Lscan} respectively. As discussed in the SYK$_4$ model, the major effect of the $T\bar{T}$ deformation on the SFF is to shift the curves along the horizontal axis parallelly. Moreover, the patterns of SFF with different $\lambda$ overlap with each other in the time unit $J_{\rm eff}$, as shown in Figure \ref{PlotSFFSYK2JL}.

As reviewed in Sec. \ref{sec:intro}, we distinguish quantum chaos from other concepts (scrambling and operator growth) and detect quantum chaos signals by SFF. From the results presented in Figure \ref{SFFSYK4JL} and \ref{PlotSFFSYK2JL}, we can conclude that the overall effect of the $T\bar{T}$ deformation can be interpreted as a rescaling of the time parameter and the quantum chaotic behavior remains unchanged under the $T\bar{T}$ deformation.

\section{The OTOC in the $T\bar{T}$-deformed SYK models}
\label{section:OTOC}
In the previous section, we provided strong evidence with the SFF in the $T\bar{T}$-deformed SYK model that the $T\bar{T}$ deformation does not change the late-time behavior of quantum chaos (or non-chaos) in the original theories. Motivated by recent advances in holography, scrambling became a new quantity to detect the feature of the time evolution of a system at the early time. 
In this section, we test the early-time behavior and study the effect of the $T\bar{T}$ deformation on the scrambling in deformed theories in more detail by computing the OTOC.

\subsection{Analytical results}
\label{OTOC}
The out-of-time-ordered correlator is defined by
\begin{align}
	F(t) = 
	-\frac{Z(\beta)\Tr\left(yV(t)yW(0)yV(t)yW(0)\right)}
	{\Tr(y^2Vy^2V)\Tr(y^2Wy^2W)}\,,
	\label{def:OTOC}
\end{align}
where $Z(\beta)=\Tr e^{-\beta (H-E_0)}$ and $y=e^{-\beta(H-E_0)/4}$. In this work, we calculate the OTOC of Majorana fermions $\psi_i$ and $\psi_j$ and then average over the index $i$, $j$.

The effect of the $\TT$ deformation on OTOCs of the SYK$_4$ model is characterized by the effective coupling $J_\text{eff}$. When $\beta=0$, the OTOC of SYK model has been investigated in the large-$q$ limit in \cite{Roberts:2018mnp}. The result is
\begin{align}
    F_{\rm OTOC}=1-\frac{1}{2N}\cosh2\mathcal{J}t\,.
\end{align}
It is reasonable to guess that the $\TT$-deformed results are changed by replacing $\mathcal{J}$ to $\mathcal{J}_{\rm eff}$. So the pattern of $ F_{\rm OTOC}$ is universal as a function of $J_{\rm eff}t$.

In the conformal limit, the $T\bar{T}$-deformed OTOC and Lyapunov exponent have been discussed in \cite{Gross:2019ach,Gross:2019uxi}. In conformal limit, the Lyapunov exponent is extracted from the pole of the two point function $\langle\epsilon(u)\epsilon(0)\rangle$
where $\epsilon(u)$ is the fluctuation of the time reparameterization. Under the $T\bar{T}$ deformation, the $SL(2,\mathbb{R})$ symmetries remain unchanged. As a result, the poles contributed to $\langle\epsilon(u)\epsilon(0)\rangle$ remain $n=0,\pm1$. The OTOC is 
\begin{align}
    F_{\rm OTOC}\sim \frac{\beta}{C}\sqrt{1+\lambda C\frac{16\pi^2}{\beta^2}}e^{\frac{2\pi}{\beta}u}\,,
\end{align}
and the Lyapunov exponent $\lambda_L=2\pi/\beta$ is unchanged.

The OTOC of the $\TT$-deformed SYK$_2$ model could be solved by integrability. The original SYK$_2$ model is free. So the operator $\psi_i(t)$ does not grow and the OTOC does not decay. A nontrivial $\TT$ deformation introduces the interaction while maintains the integrability. The OTOC could decay to small values. We will solve the OTOC on the basis $\ket{\vec n}$. We first calculate the denominator in Eq.~(\ref{def:OTOC}) as
\begin{align}
    G(\beta/2)\equiv&\frac1{NZ(\beta)}\sum_i^{N}\Tr\kd{y^2\psi_iy^2\psi_i}\nn\\
    =&\frac2{NZ(\beta)}\Re\sum_{\vec n}\sum_a^{N/2}\bra{\vec n}y^2\chi_a^\dagger y^2\chi_a\ket{\vec n}\nn\\
    =&\frac{4}{NZ(\beta)}\sum_a\sum_{\vec n}^{n_a=1}\exp\kd{-\frac\beta2(f(E_{\vec n})+f(E_{\vec n\backslash a}))}\nn\\
    \approx&\frac{2}{NZ(\beta)}\sum_n\sum_a \exp\kd{-\beta f(E_{\vec n})}=1\,,
\end{align}
where $E_{\vec n\backslash a}=\sum_{b\neq a}n_b\varepsilon_b$. In the final two steps, we consider large $N$. So we assume $\varepsilon_a\ll E_{\vec n}$ and $f(E_{\vec n})+f(E_{\vec n\backslash a})\approx 2f(E_{\vec n})$. We further drop the constraint $n_a=1$ in the summation of states and compensate for the overcounting by dividing it by $2$. Then the OTOC is
\begin{align}
    F(t)
    \equiv&\frac{-1}{N^2Z(\beta)G(\beta/2)^2}\sum_{ij}^N\Tr\kd{y\psi_i(t)y\psi_jy\psi_i(t)y\psi_j}\nn\\
    =&-\frac{4}{N^2 Z(\beta)G(\beta/2)^2}\Re\sum_{\vec n} \sum_{ab}^{N/2} \bra{\vec n}y\chi_a^\dagger(t)y\chi_b^\dagger y\chi_a(t)y\chi_b\ket{\vec n}\nn\\
    =&\frac{16}{N^2 Z(\beta)G(\beta/2)^2}\sum_{a\neq b}\sum_{\vec n}^{n_a=n_b=1}\cos\kd{t(f(E_{\vec n})-f(E_{\vec n\backslash a})-f(E_{\vec n\backslash b})+f(E_{\vec n\backslash \{a,b\}}))} \nn\\
    &\hspace{3cm} \times \exp\kd{-\frac\beta4\kc{ f(E_{\vec n})+f(E_{\vec n\backslash a})+f(E_{\vec n\backslash b})+f(E_{\vec n\backslash \{a,b\}})}}\nn \\
    \approx&\frac{4}{N^2Z(\beta)}\sum_{\vec n}\sum_{a\neq b}\cos\kc{t\varepsilon_a\varepsilon_bf''(E_{\vec n})}e^{-\beta f(E_{\vec n})}\nn\\
    \approx&\frac{2^{2+N/2}}{N^2Z(\beta)} \int dE p(E+E_0)e^{-\beta f(E)} \sum_{a\neq b} \cos\kc{t \varepsilon_a\varepsilon_b f''(E)}\,.
    \label{OTOCSYK2}
\end{align}
Similarly, in the penultimate step, we assume $\varepsilon_{a,b}\ll E_{\vec n}$, approximate the sum and the difference of $f(E)$ functions, drop the constraint $n_a=n_b=1$ in the summation of states, and compensate the overcounting by dividing it by $4$.
In the last step, we introduce the density of states $p(E)$ of the original SYK$_2$ model. 
We will work at the large $N$ limit to obtain analytical expressions. Then the single-particle spectrum $\ke{J_k}_{k=1}^{N}$ obeys the semi-circle law
\begin{align}
    \rho_\text{sc}(\varepsilon)=\frac{1}{2\pi J_0}\sqrt{1-\kc{\frac{\varepsilon}{4J_0}}^2}
\end{align}
and $E_0=-4J_0N/3\pi$. The density of states $p(E)$ is the Gaussian distribution \cite{Garcia-Garcia:2016mno,Feng:2018zsx}, namely 
\begin{align}
    p(E)=\frac{1}{J_0\sqrt{\pi N}}\exp\kd{-\kc{\frac{E}{J_0\sqrt{N}}}^2}.
\end{align}
In principle, one can calculate the OTOC by integrating over $\varepsilon_{a,b}$ and $E$ in \eqref{OTOCSYK2} with the above distributions.

To get an analytical expression, we will consider a saddle point approximation for $E$ at high-temperature limit $\beta J_0\ll 1$. According to the distribution $p(E+E_0)e^{-\beta f(E)}$, $E$ concentrates near the saddle point $E_c$ determined by 
\begin{align}
    -2\frac{E_c+E_0}{J_0^2N}-\beta f'(E_c)=0.
\end{align}
We get $E_c\sim J_0N$. In \eqref{OTOCSYK2}, we have $t\varepsilon_a\varepsilon_b|f''(E_c)|\leq t J_0^2/(3\sqrt3 E_c)\sim t J_0/N$, which is very small at the early time $t\ll N/J_0$. So the cosine function in \eqref{OTOCSYK2} is a flat function of $E$ near $E_c$. We take the saddle point value $E_c$ in the cosine function and simplify the OTOC as
\begin{align}
    F(t)
    \approx\frac{4}{N^2} \sum_{a\neq b} \cos\kc{t \varepsilon_a\varepsilon_b f''(E_c)}
    = \tilde F\kc{8J_0^2f''(E_c)t}+O\kc{N^{-2}},
    \label{OTOCSYK2A}
\end{align}
where 
\begin{align}
     \tilde F(x)=&2 \kd{\tilde J_0(x){}^2-\frac{\tilde J_1(x) \tilde J_0(x)}{x}+ \tilde J_1(x){}^2-\frac1N\kc{\tilde J_0(x)\cos x+\tilde J_1(x)\sin x}}\nn\\
     =&1-\frac2N-\kc{\frac18-\frac1{2N}}x^2+ O(x^4)
\end{align}
with $\tilde J_i(x)$ the Bessel function of the first kind. 
As expected from the expansion of the Hamiltonian \eqref{TTSYK2H}, the OTOCs of the $\TT$-deformed SYK$_2$ exhibit power law behaviors, which agree with the phenomenon in MBL \cite{Huang:2016knw,Fan:2016ean}.

\subsection{Numerical results}
\label{data-OTOC}
\begin{figure}[h!]
	\centering
	\captionsetup[subfloat]{farskip=10pt,captionskip=1pt}
	\subfloat[t][\centering{OTOC of SYK$_4$
	}]{\label{PlotSYK4LOTOC}
		\includegraphics[width =0.5\linewidth]{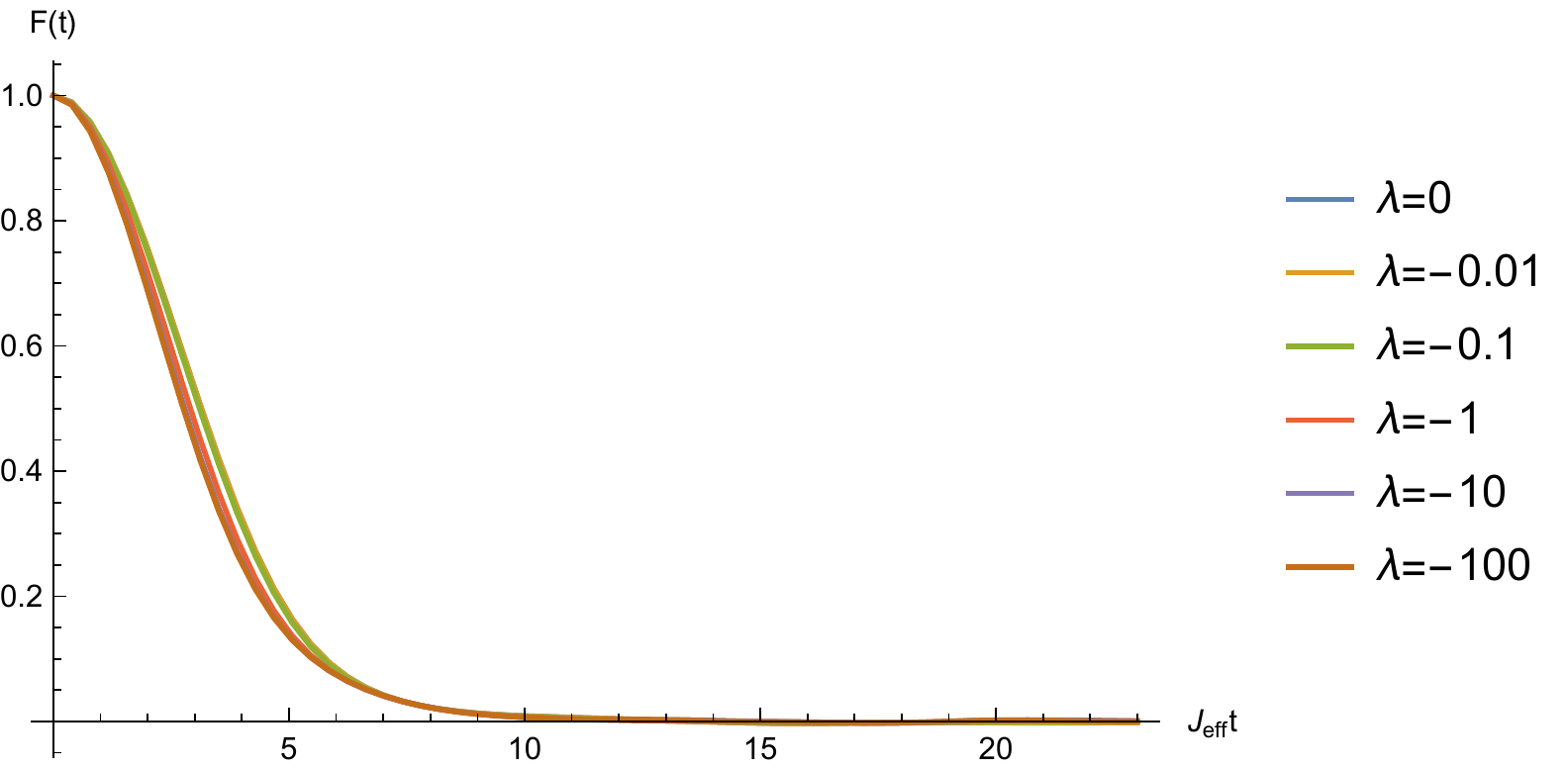}}
	\subfloat[t][\centering{OTOC of SSYK$_4$
	}]{\label{PlotSSYK4J0LOTOC}
		\includegraphics[width =0.5\linewidth]{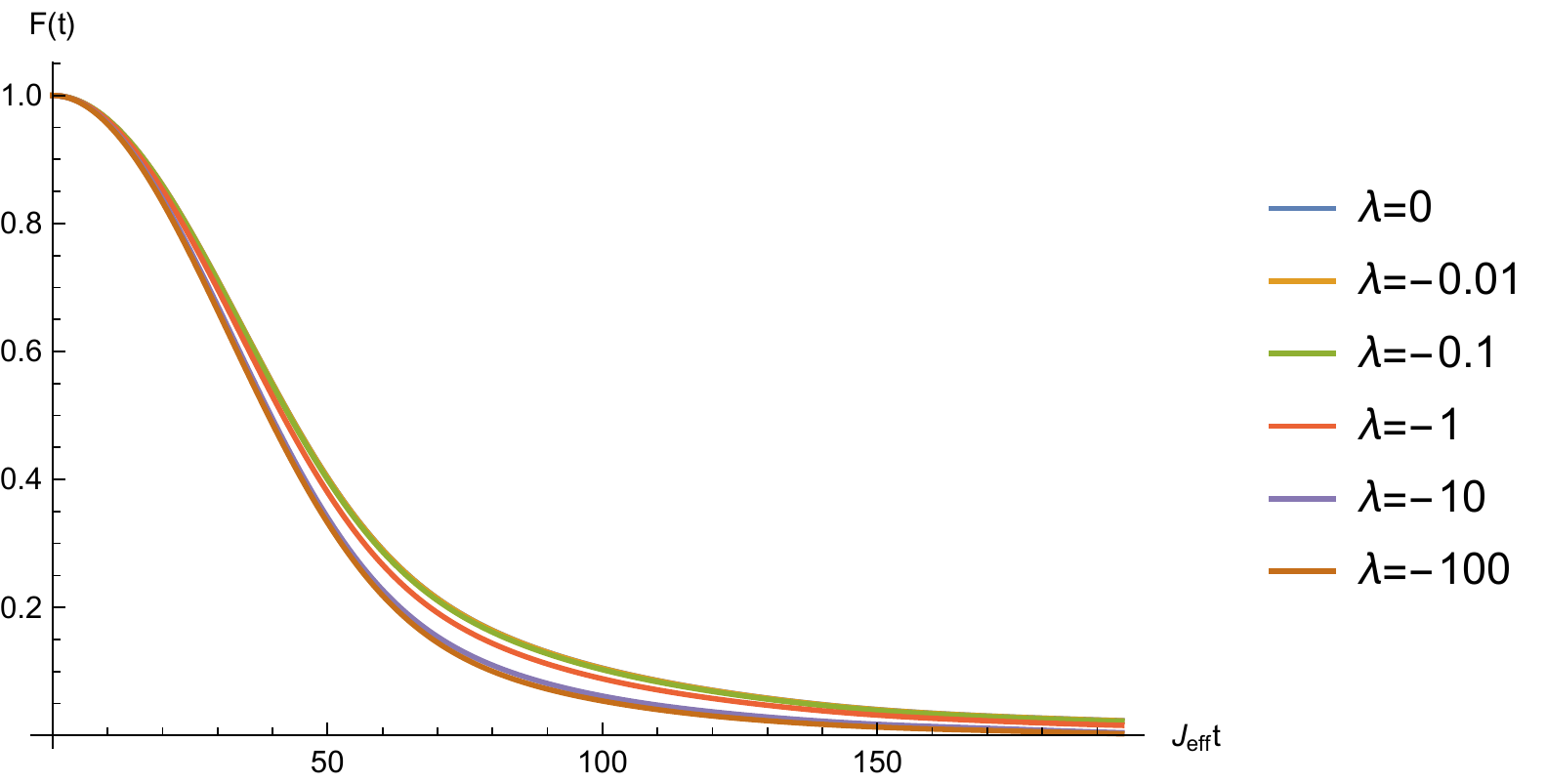}}	
\caption{Plot of OTOC of (a) SYK$_4$ model and (b) SSYK$_4$ model with various value of $\lambda$ at infinite temperature.
}
	\label{fig:syk4_N16_OTOC14_Lscan}
\end{figure}
\begin{figure}[h!]
	\centering
		\includegraphics[width =0.8\linewidth]{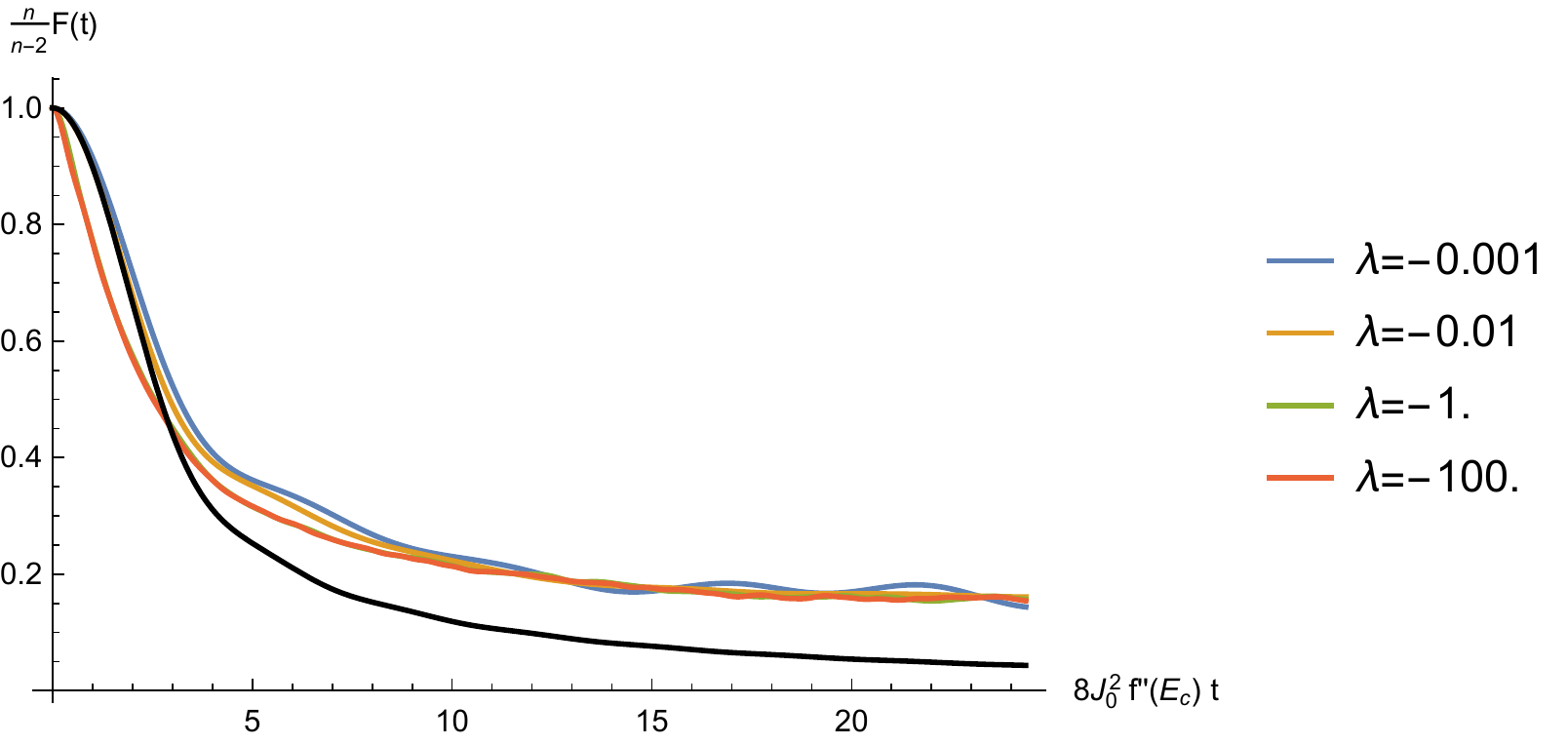}
	\caption{The plot of SYK${}_2$ OTOC with various values of $\lambda$. We choose $N=16$ and the inverse temperature $\beta=0$. The black solid line represents the analytical result calculated by Eq.~(\ref{OTOCSYK2A}). The horizontal axis is $8J_0^2f''(E_c)t$. 
	}
	\label{PlotSYK2LOTOC}
\end{figure}
\begin{figure}[h!]
	\centering
	\captionsetup[subfloat]{farskip=10pt,captionskip=1pt}
	\subfloat[t][\centering{OTOC at high temperature}]{\label{PlotSYK2JLPiLogLinearOTOC}
		\includegraphics[width =0.8\linewidth]
		{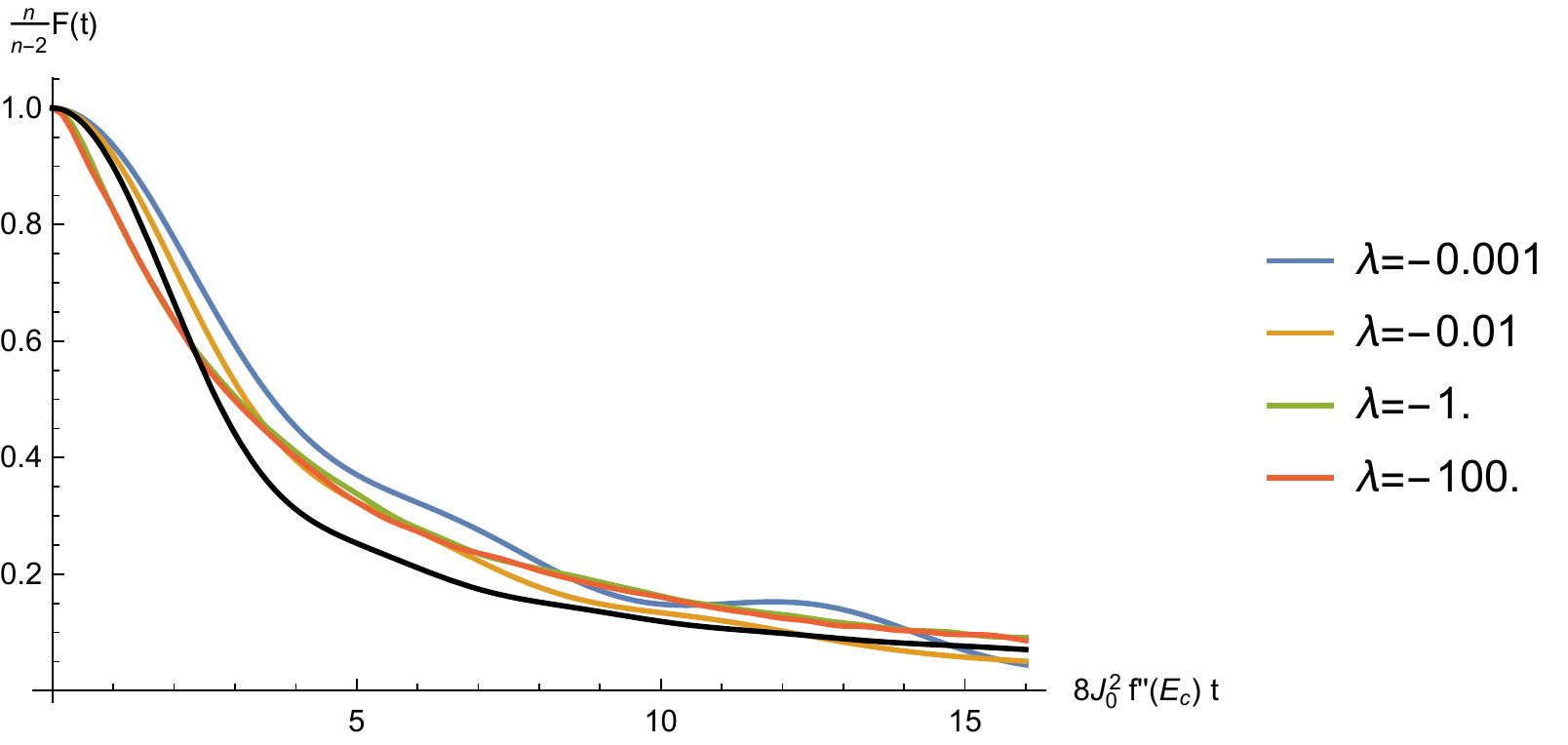}}\\
	\subfloat[t][\centering{OTOC at low temperature}]{\label{PlotSYK2JBLogLinearOTOC}
		\includegraphics[width =0.8\linewidth]
		{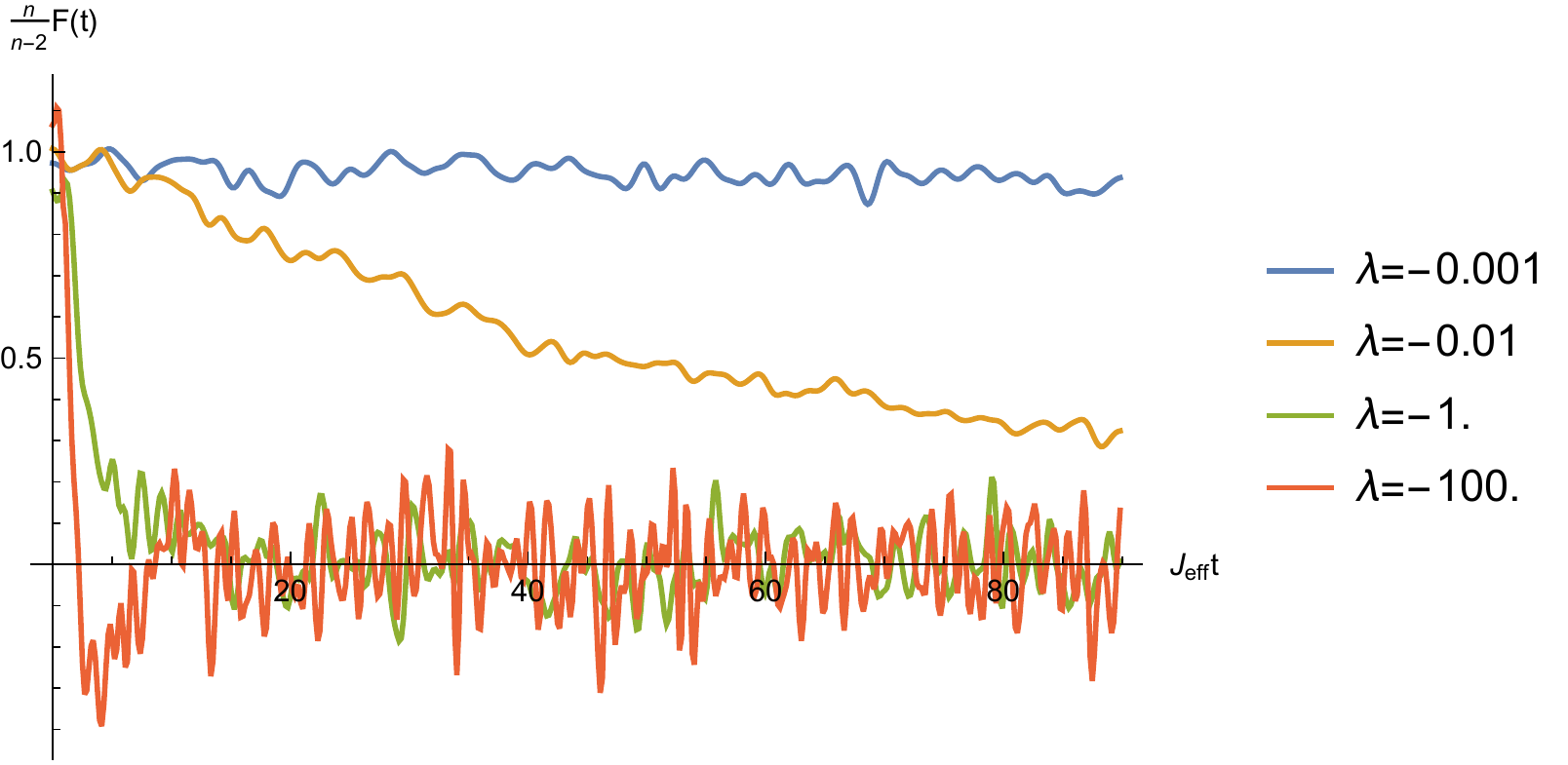}}
		\caption{OTOC of SYK$_2$ with (a) $\beta J_0=\pi/10$ and with (b) $\beta J_{\rm eff}=\pi/2$. In Figure (a), the black solid lines represent the analytical results and the horizontial axis is $8J_0^2f''(E_c)t$. In Figure (b), we plot the OTOC as a function of $J_{\rm eff}t$. 
		}
		\label{fig:OTOC2finiteB}
\end{figure}

In this subsection, we calculate the OTOC numerically. The results of SYK$_4$ and SYK$_2$ with different deformation parameter $\lambda$ are shown in Figure \ref{fig:syk4_N16_OTOC14_Lscan}, \ref{PlotSYK2LOTOC} and \ref{fig:OTOC2finiteB}. In Figure \ref{fig:syk4_N16_OTOC14_Lscan}, we can see the exponential decay behavior of OTOC in both the undeformed SYK$_4$ (SSYK$_4$) model and the $T\bar{T}$-deformed SYK$_4$ (SSYK$_4$) model. The effects of deformation slightly decrease the decay rate. As discussed in the last section, this result can be explained by a rescaling of the coupling constant $J_0$. We also plot the OTOC with respect to $J_{\rm eff}t$ and find that the curves with different $\lambda$ coincide precisely, as shown in Figure \ref{fig:syk4_N16_OTOC14_Lscan}.

We numerically calculate the OTOCs of the SYK$_2$ model with or without $T\bar{T}$ deformation and show them in Figures \ref{PlotSYK2LOTOC} and \ref{fig:OTOC2finiteB}. Based on the Eq.~(\ref{OTOCSYK2A}), we choose the horizontal axis as $J_0f''(E_c)t$. Figure \ref{PlotSYK2LOTOC} shows the OTOC of the SYK$_2$ model at infinite temperature. For small $\lambda$ ($0<|J_0\lambda|\leq0.01$), OTOCs deviate from unity in the early time as a power law. This behavior of OTOC has been observed in MBL systems \cite{Huang:2016knw, Fan:2016ean}. 
For large $\lambda$ ($|J_0\lambda|> 0.1$), our numerical results match the Eq.~(\ref{OTOCSYK2}). We find the saddle-point approximation method (\ref{OTOCSYK2A}) has a small error. However, the patterns of OTOC with large $\lambda$ in Figure \ref{PlotSYK2LOTOC} are universal in time units $8J_0^2f''(E_c)$. The OTOCs at finite temperature is shown in Figure~\ref{fig:OTOC2finiteB}. For the high-temperature case ($\beta J_0\sim \pi/10$), the features are similar to the one at infinite temperature. One can observe a MBL behavior for small $\lambda$ ($0<|J_0\lambda|\leq0.01$) in Figure~\ref{PlotSYK2JLPiLogLinearOTOC}\footnote{In \cite{Garcia-Garcia:2018pwt}, the authors considered a generalized SYK model with two-body and one-body random interactions of finite range. By reducing the range of the two-body interaction, they found a phase transition from the thermal phase to the MBL phase.}. For the low-temperature case ($\beta J_0\geq\beta J_{\rm eff}\sim \pi$), the saddle point approximation method is not applicable so the patterns are not uniform with respect to $8J_0^2f''(E_c)t$. We plot the OTOCs of SYK$_2$ with $\beta J_{\rm eff}=\pi$ ($\beta J_0>\pi$) as the functions of $J_{\rm eff}t$ in Figure~\ref{PlotSYK2JBLogLinearOTOC}.

\section{The K-complexity in the $T\bar{T}$-deformed SYK models}
\label{kcomplexity}
\subsection{Lanczos coefficient and K-complexity}

In this subsection, we will first give a brief overview of K-complexity. The K-complexity is analogous to the circuit complexity, capturing the intrinsic dynamics of the evolution operator \cite{Parker:2018yvk}. Given a Hamiltonian $H$ and a certain simple operator $O$ of a system, the K-complexity of $O$ can be measured in the Krylov basis, which is roughly defined as $[H,[H,\cdots,[H, O]]]$. This definition is motivated by the Taylor expansion of the time evolution of a local operator $O$. The Krylov basis can be generated by the Liouvillian superoperator $\mathcal{L}|O)\equiv|[H, O])$. The exact definition of the Krylov basis is
\begin{align}
	&|A_n)=\mathcal{L}|O_{n-1})-b_{n-1}|O_{n-2}) \,,\nn\\
	&b_n=( A_n|A_n)^{1/2} \,, \nn\\
	&|O_n)=b^{-1}_n|A_n)\,,
	\label{definition-Krylov}
\end{align}
where $|O_0)\equiv|O)$, $b_0=0$ by convention. The inner product of the operator basis are $(A|B)=\Tr[A^\dagger B]$ in the vacuum state and $(A|B)=\Tr[yA^\dagger yB]$ with $y=e^{-\frac{1}{2}\beta \hat{H}}$ in the thermal ensemble. Note that we have regulated the thermal expectation value by splitting the operator of the thermal density matrix. The sequence of positive numbers $\{b_n\}$ is called Lanczos coefficients. The number of Krylov basis corresponds to the dimension of the space spanned by the linear operator acting on a Hilbert space. Consider a system described by a $L$-dimensional Hilbert space. The dimension of the operator space is $L^2$. For the SYK$_4$ model, the Heisenberg evolution of a single Majorana fermion $\psi_i$ is bounded in the subspace of operators consisting of an odd number of fermions. For the SYK$_2$ model, the Heisenberg evolution of $\psi_i$ is bounded in the subspace of a single particle. Thus, the number of Krylov basis of the SYK$_4$ and SYK$_2$ models can be estimated by $d_{\rm Krylov}=(2^{[N/2]})^2/2=2^{N-1}$ and $d_{\rm Krylov}=N$, respectively.

More precise calculations of $d_{\rm Krylov}$ have been discussed in \cite{Barbon:2019wsy}. By (\ref{definition-Krylov}), the Krylov basis is the linear span of the action of the Liouvillian superoperator on $O$, namely
\begin{align}
\left(O,\mathcal{L}O,\mathcal{L}^2O,\cdots,\mathcal{L}^nO,\cdots\right)^T\,.
\label{operatorbasis}
\end{align}
The dimension of this space is the number of the linearly independent vector $\mathcal{L}^n O$. We can expand $\mathcal{L}^n O$ in the energy basis
\begin{align}
\mathcal{L}^n|O)=\delta_{n0}\sum_{a=1}^D O_{aa}|\omega_{aa})
+\sum_{a,b=1,a\neq b}^D O_{ab}\omega_{ab}^n|\omega_{ab})\,,
\label{L-energy}
\end{align}
where $|\omega_{ab})=|E_a\rangle\langle E_b|$ and $\omega_{ab}=E_a-E_b$ is the eigenvalues of the Liouvillian acting on $|\omega_{ab})$. The matrix representation of (\ref{L-energy}) in this basis is
\begin{align}
\begin{pmatrix}
O_{11} & O_{22} & \cdots & O_{DD} & O_{12} & O_{13} & \cdots & O_{D-1,D}\\
0 & 0 &\cdots & 0 & \omega_{12}O_{12} & \omega_{13}O_{13} & \cdots & \omega_{D-1,D}O_{D-1,D}\\
0 & 0 &\cdots & 0  & \omega_{12}^2O_{12} & \omega_{13}^2O_{13} & \cdots & \omega_{D-1,D}^2O_{D-1,D} \\
\vdots & \vdots &\ddots & \vdots  & \vdots & \vdots & \ddots & \vdots\\
0 & 0 &\cdots & 0 & \omega_{12}^{D^2-1}O_{12} & \omega_{13}^{D^2-1}O_{13} & \cdots & \omega_{D-1,D}^{D^2-1}O_{D-1,D}
\end{pmatrix}\,.
\label{matrixL}
\end{align}
The number of linearly independent vectors in Eq.~(\ref{operatorbasis}) is equal to the rank of the matrix (\ref{matrixL}). The rank of the matrix (\ref{matrixL}) can be calculated as follows. First, we omit the columns containing the zero elements of the matrix $O$, and then we omit the columns containing the repeated eigenvalues $\omega_{ab}$ of $\mathcal{L}$. For the SYK$_4$ model with $N \text{ mod } 8=2,4,6$, the energy level is doubly degenerate. As a result, $\omega_{ab}$ are quadruply degenerate. Similarly, for the SSYK$_4$ model with $N \text{ mod }8=0, 6$, the degeneracy of $\omega_{ab}$ is 4 and for the SSYK$_4$ model with $N \text{ mod } 8=2,4$, the degeneracy of $\omega_{ab}$ is 16. Recall that the transformation $H\rightarrow f(H-E_0)$ does not break any symmetries of the original theory. Consequently, the degeneracy of $\omega_{ab}$ and $d_{\rm Krylov}$ remain unchanged under the $T\bar{T}$ deformation. For the SYK$_2$ model, the energy level is non-degenerate but $E_a$ comes in pairs with its opposite value $-E_a$ for any $a$. In this case, $\omega_{ab}$ is doubly degenerate. For example, for any $E_a=-E_{a'}$ and $E_b=-E_{b'}$, we have $\omega_{ab}=E_a-E_b=E_{b'}-E_{a'}=\omega_{b'a'}$. However, the double degeneracy of $\omega_{ab}$ is broken under the $T\bar{T}$ deformation, namely
\begin{align}
\omega_{ab}^\lambda=f(E_a-E_0)-f(E_b-E_0)
= f(-E_{a'}-E_0)-f(-E_{b'}-E_0)\neq\omega_{b'a'}^\lambda\,.
\end{align}
Under the $T\bar{T}$ deformation, the Krylov space enlarges significantly. We list the $d_{\rm Krylov}$ of the SYK$_4$, SSYK$_4$, SYK$_2$ and the $T\bar{T}$-deformed SYK$_2$ model with different $N$ in Table \ref{table: dimension}. 

\begin{table}[]
\centering
\caption{Dimension of the operator space}
\label{table: dimension}
\begin{tabular}{clcccccccc}
\hline
$N$ & 4 & 6 & 8 & 10 & 12 & 14 & 16 & 18 & 20 \\   
\hline
SYK$_4$ & 2 & 12 & 128 & 241 & 512 & 4032 & 32768 & 65281 & 131072 \\
SSYK$_4$ & 1 & 12 & 57 & 57 & 241 & 4032 & 16257 & 16257 & 65295 \\
SYK$_2$ & 4 & 6 & 8 & 10 & 12 & 14 & 16 & 18 & 20 \\
$T\bar{T}$-deformed SYK$_2$ & 8 & 24 & 64 & 160 & 384 & 896 & 2048 & 4608 & 10240 \\
\hline
\end{tabular}
\end{table}

The Heisenberg evolution $O(t)$ of $O$ can be rewritten as 
\begin{align}
	|O(t))=e^{i\mathcal{L}t}|O_0) \,,
\end{align}
and can be expanded on the Krylov basis
\begin{align}
	|O(t))=\sum_ni^n\varphi_n(t)|O_n) \,.
\end{align}
The K-complexity of the Heisenberg operator $O(t)$ is defined as the average of the linear operator $\hat{n}$ in the state $|O(t))$
\begin{align}
	C_K(O)=(O(t)|\hat{n}|O(t))
	=\sum_nn|\varphi_n(t)|^2 \,.
\end{align}
It is obvious that the magnitude of K-complexity is bounded by the number of Krylov basis. Accordingly, the growth of K-complexity is quantitatively captured by the Lanczos coefficients $\{b_n\}$. The asymptotic behavior of the Lanczos coefficients in the large-N limit is $b_n\simeq\alpha n^\delta$ with $0\leq \delta\leq1$. By analogy with the behavior of OTOC, it is generally assumed that for the chaotic system, the growth of $b_n$ depends linearly on $n$. In this case, the K-complexity is $C_K(t)\sim e^{2\alpha t}$ which implies that the Lyapunov exponent is $\lambda_L=2\alpha$. However, based on the discussion in Sec.~\ref{sec:intro} and Sec.~\ref{section:OTOC}, we regard $b_n$ or $C_K$ as another characteristic quantity independent of quantum chaos.

The Lanczos coefficients can be read off from the Hankel determinant $\det(\mu_{2n})$ of the Liouvillian superoperator
\begin{align}
	b_1^{2n}\,b_2^{2n-2}\cdots b_n^2=\det(\mu_{i+j})_{0\leq i ,j\leq n} \,,
	\label{lanczos-mu}
\end{align}
where
\begin{align}
	\mu_{2n}\equiv(O_0|\mathcal{L}^{2n}|O_0)
	\label{momentum}
\end{align}
is the moments of the Liouvillian superoperator $\mathcal{L}$. From Eq.~(\ref{lanczos-mu}), one can show that, for a finite-$N$ system, the asymptotic behavior of the Lanczos coefficients is bounded by \cite{Jian:2020qpp}
\begin{align}
	b_n\leq
	\begin{cases}
		\frac{\lambda_L}{2}n,\quad 1\ll n\ll N/q\\
		\frac{\lambda_C}{2}N,\quad N/q\ll n\ll 2^N
	\end{cases} \,,
	\label{bn-N}
\end{align}
where $\lambda_L$ is the Lyapunov exponent and $\lambda_C$ is a constant independent of $n$. One can estimate $\lambda_C$ by the moments. There is a close relationship between the Green function and the moments
\begin{align}
	G(t)=\frac{Tr[O^\dagger(0)O(t)]}{Tr[O^\dagger O]}
	=(O_0|e^{i\mathcal{L}t}|O_0)
	=\sum_n\frac{(it)^{2n}}{(2n)!}
	(O_0|\mathcal{L}^{2n}|O_0)
	=\sum_n\frac{(it)^{2n}}{(2n)!}\mu_{2n} \,.
	\label{correlator-mumentum}
\end{align}
From Eq.~(\ref{correlator-mumentum}), we find that the momentum $\mu_{2n}$ is determined completely by the 2-point function $G(t)$. Recall that the conformal 2-point function $G_c(t)$ is unchanged under the $T\bar{T}$ deformation. As a result, the Lanczos coefficients, the K-complexity in the conformal limit, and their holographic dual are unchanged under the $T\bar{T}$ deformation.

For a finite $N$ system, it is convenient to express momentum in terms of the energy eigenbasis $|E_a\rangle$. According to the ETH conjecture \footnote{Notice that the SYK$_2$ model doesn't satisfy the ETH conjecture so the analysis below is only applicable to the SYK$_4$ and SSYK$_4$ models.}, the matrix elements $O_{ab}$ in the energy basis can be approximated by $O_{ab}=A(E_a,E_b)\delta_{ab}+A(E_a,E_b)2^{-N/4}R_{ab}$, where $A(E_a,E_b)=A(0,0)F(E_a-E_b)$ for $A(0,0)$ a constant and $F$ some function. $R_{ab}$ is a random matrix with zero mean and the variance can be set to 1 by a rescaling. Substituting this expression into the definition (\ref{momentum}), we obtain
\begin{align}
	\mu_{2n}=2^{-N}\sum_{a,b}(E_a-E_b)^{2n}|F(E_a-E_b)|^2
	\label{mu} \,.
\end{align}
From Eq.~(\ref{lanczos-mu}) and Eq.~(\ref{mu}), it is clear that the Lanczos coefficient $b_n$ is a constant and is dominated by the largest energy difference with $n\gg N/q$ \cite{Jian:2020qpp}. It is worth mentioning that the constant Lanczos coefficients $b_n$ at large-$n$ capture the linear growth of the K-complexity.

\subsection{Numerical results}
\begin{figure}[h!]
	\centering
	\includegraphics[width=0.6\linewidth]{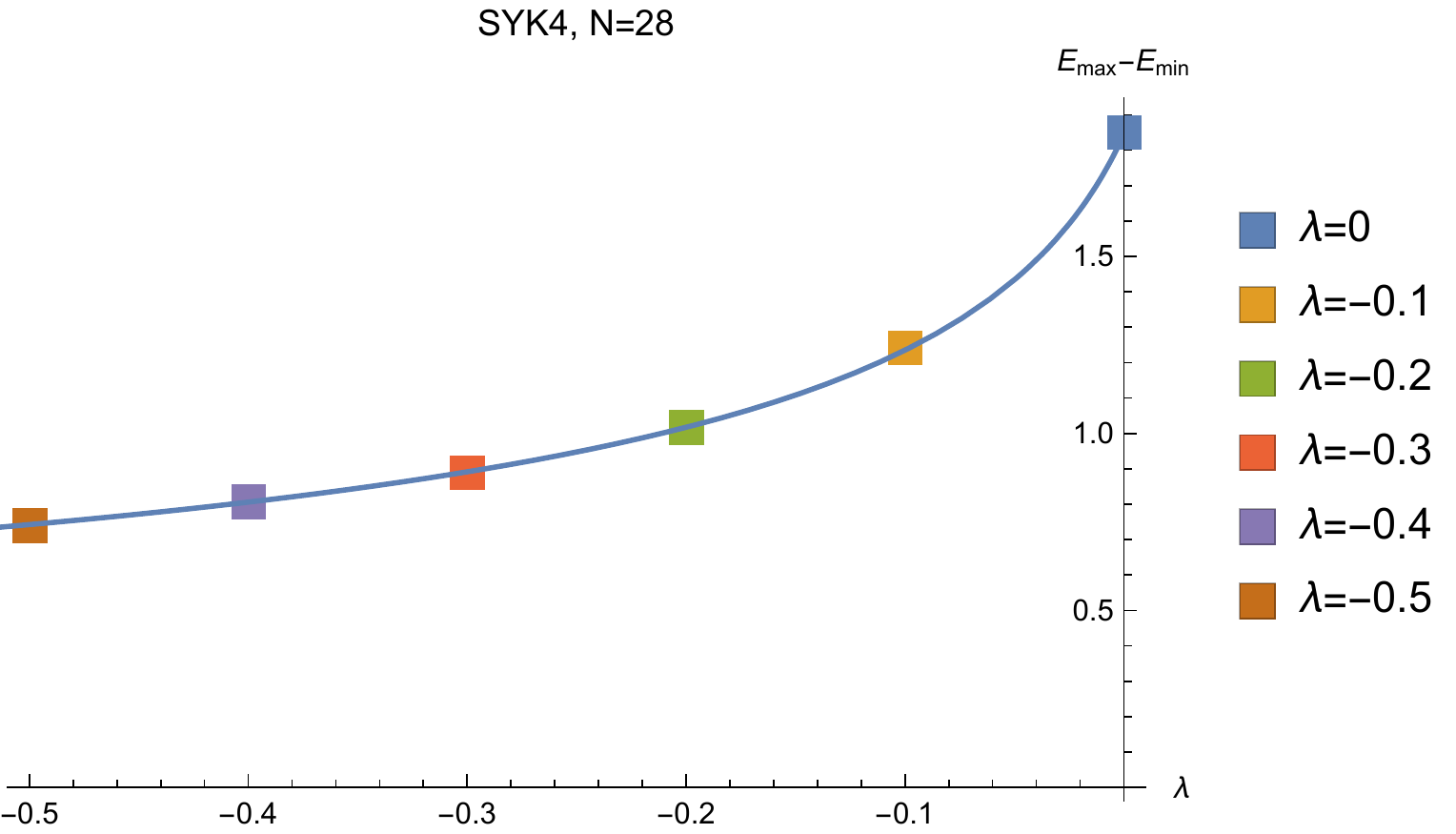}
	\caption{Plot of the maximum energy difference of the SYK$_4$ with respect to the deformation coefficient $\lambda$. The solid squares label the asymptotic values of $b_n$ with different $\lambda$.}
		\label{L-different_energy}
\end{figure}

\begin{figure}[htbp]
	\centering
		\captionsetup[subfloat]{farskip=10pt,captionskip=1pt}
	\subfloat[t][\centering{$b_n/J_{\rm eff}$ for the SYK$_4$}]{\label{syk4_Jeff}
		\includegraphics[width =0.7\linewidth]{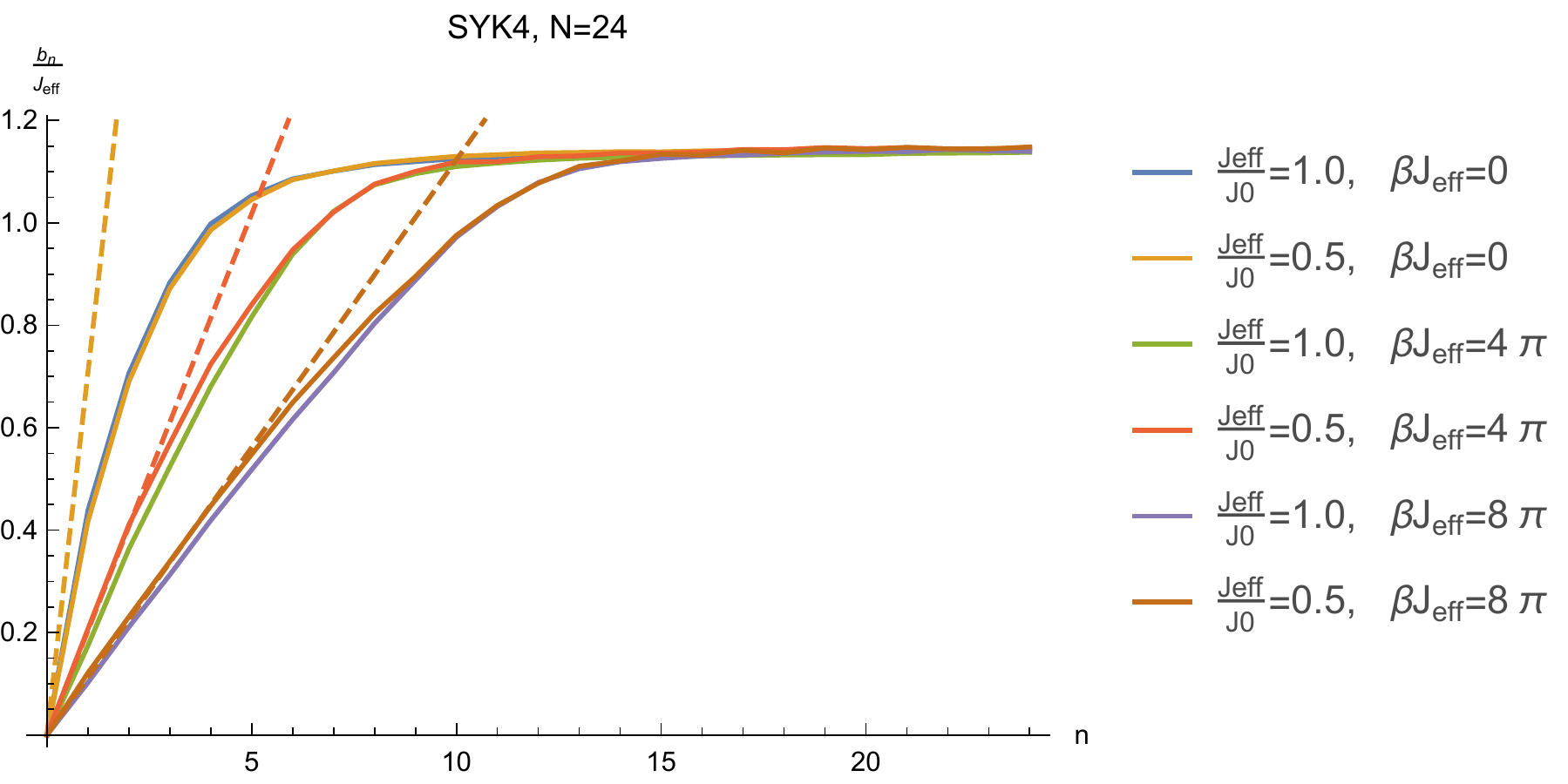}}\\
	\subfloat[t][\centering{$b_n/J_{\rm eff}$ for the SSYK$_4$}]{\label{SPlotb4JL}
		\includegraphics[width =0.7\linewidth]{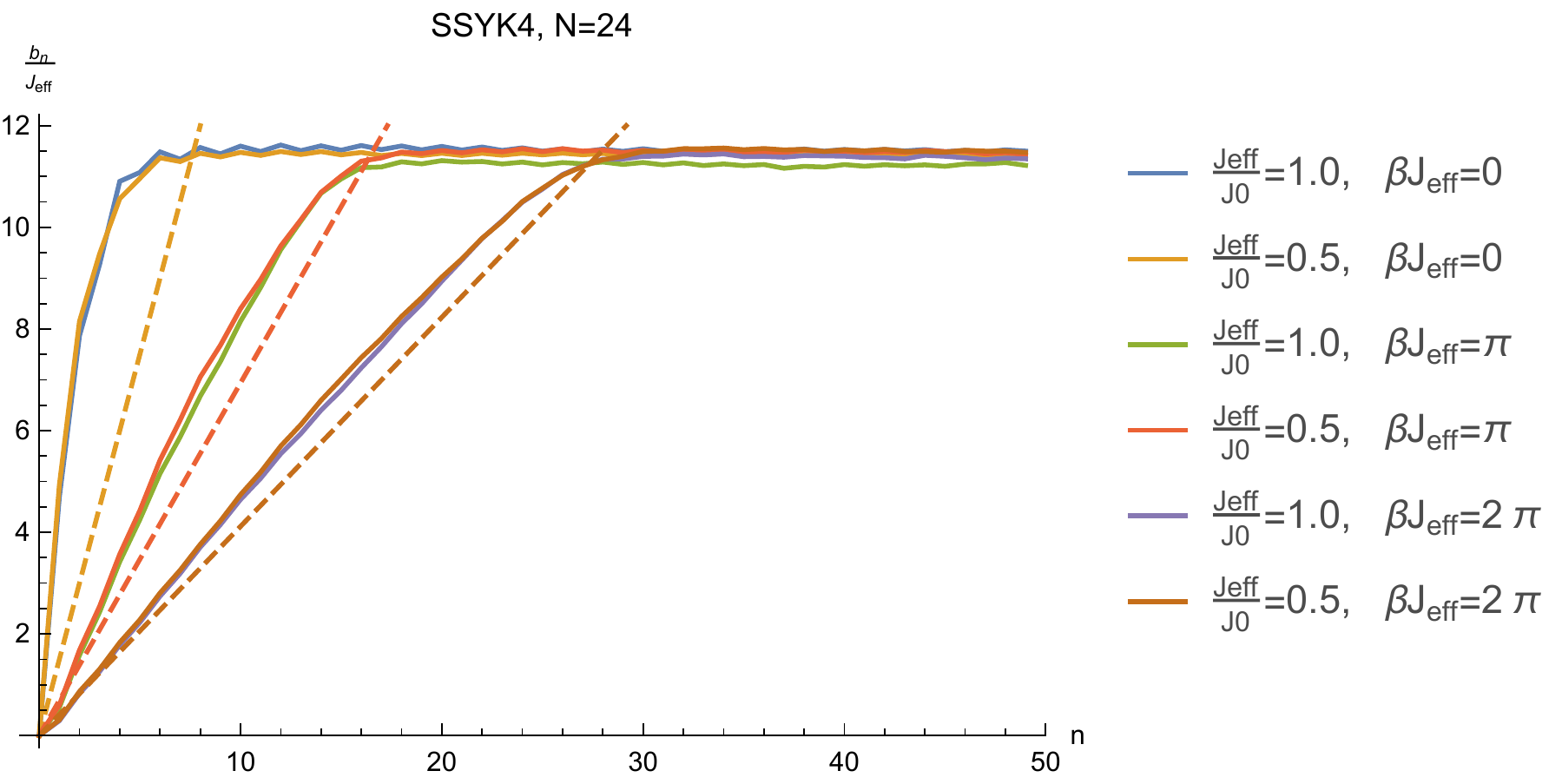}}
	\caption{In Figure (a) and (b), we display the Lanczos coefficients $b_n^\lambda$ of the SYK$_4$ and SSYK$_4$ in the unit $J_{\rm eff}(\lambda)$ with various values of $\lambda$. $b_n^\lambda$ depends on dimensionless parameters $J_{\rm eff}(\lambda)/J_0$ and $\beta J_{\rm eff}$. Dashed lines represent the Lyapunov exponents over $2J_{\rm eff}$ calculated in the large-$q$ limit.}
	\label{bn-SYK4}
	\vspace{-0.5em}
\end{figure}

In this subsection, we present our numerical results on the Lanczos coefficients and the K-complexity of the SYK and SSYK models. We choose the initial simple operator as $O=\psi_1$. The Lanczos coefficients of the deformed SYK$_4$ and SSYK$_4$ models are shown in Figure \ref{bn-SYK4}. The Lanczos coefficients $b_n$ in these figures have the same characteristics: a linearly increasing part and a constant part. Before explaining this result in more detail, we first try to analyze the parameters dependence of $b_n$. In general, $b_n$ should be a function of the inverse temperature $\beta$ and the coupling constant $J$, namely $b_n=b_n(\beta,J)$. In this paragraph, we omit the subscript of $J$ and label both $J_0$ and $J_{\rm eff}$ by $J$. According to Eq.~(\ref{bn-N}), the growth rate in the linearly increasing part of $b_n$ is expected to be half of the Lyapunov exponent $\lambda_L$. The Lyapunov exponent of the SYK$_4$ model can be estimated in the large-$q$ limit \cite{Maldacena:2016hyu} by
\begin{align}
\lambda_L=\frac{2\pi}{\beta}v,\quad \text{where}\quad 
\beta\mathcal{J}=\frac{\pi v}{\cos(\frac{\pi v}{2})}\quad \text{and}\quad
\mathcal{J}=\frac{J}{\sqrt{2}}\,.
\label{LambdaL-large-q}
\end{align}
From Eq.~(\ref{LambdaL-large-q}), we can deduce that $\lambda_L/J$ depends on $\beta J$ alone, and $b_n/J$ is a function of the single variable $\beta J$ in the linearly increasing part. Moreover, the asymptotic value of $b_n$ at large $n$ is dominated by the largest energy difference. We show the deformed largest energy difference as a function of $\lambda$ in Figure \ref{L-different_energy}. Of course, the largest energy difference is proportional to the coupling constant $J$. Thus, the value of $b_n/J$ in this part should be universal. Combined with the previous discussion, we can expect that the dimensionless quantity $b_n/J$ depends only on $\beta J$ and is universal for large $n$. Under the $\TT$ deformation, the largest energy different is $f(E_{\rm max}-E_0)$. In order to match the asymptotic value of $b_n$, we redefine the effective coupling $J_{\rm eff}$ in this section by
\begin{align}
    J_{\rm eff}=J_0\frac{f(E_{\rm max}-E_0)}{E_{\rm max}-E_0}\,.
    \label{JeffsMe}
\end{align}

In Sec. \ref{section-SFF} and \ref{section:OTOC}, we explain the effect of the $T\bar{T}$ deformation as the rescaling of the coupling constant $J_0$ and find that the SFF and OTOC with different $\lambda$ coincide with each other in the time units $J_{\rm eff}(\lambda)$. So it is rational to expect that the effect of $T\bar{T}$ deformation on $b_n$ is also equivalent to the rescaling of the original coupling $J_0$, namely 
\begin{align}
b_n^\lambda(\beta,J_0)=b_n^0(\beta,J_{\rm eff})\,.
\label{bnL}
\end{align}
Note that the right-hand side of Eq.~(\ref{bnL}) is equal to $J_{\rm eff}(\lambda)b_n^0(\beta J_{\rm eff}(\lambda))$ according to the previous analysis. Thus, the dimensionless quantity $b_n^\lambda/J_{\rm eff}$ is only a function of $\beta J_{\rm eff}(\lambda)$ and is universal at large $n$. $J_{\rm eff}(\lambda)$ is determined by $\lambda$ and Eq.~(\ref{JeffsMe}). We also denote $\lambda$ by a dimensionless parameter $J_{\rm eff}(\lambda)/J_0$. Note that $J_{\rm eff}/J_0=1$ means that $\lambda=0$ and we recover the undeformed theory in this case.

Based on the previous analysis, we plotted $b_n^\lambda/J_{\rm eff}(\lambda)$ of the SYK$_4$ and SSYK$_4$ models against different dimensionless parameters $\beta J_{\rm eff}(\lambda)$ and $J_{\rm eff}(\lambda)/J_0$. One can easily notice that the results presented in Figure \ref{bn-SYK4} are consistent with our analysis. The patterns with the same $\beta J_{\rm eff}$ but different $J_{\rm eff}/J_0$ coincide very well, and the asymptotic values at large $n$ are universal. For comparison, we also compute $\lambda_L/2J_{\rm eff}$ of SYK$_4$ and SSYK$_4$ in the large $q$ limit and represent them by the dashed lines in Figure \ref{bn-SYK4}.

\begin{figure}[htbp]
	\centering
	\captionsetup[subfloat]{farskip=10pt,captionskip=1pt}
	\subfloat[t][\centering{$C_K(t)$ for the SYK$_4$}]{\label{PlotCk40BB}
		\includegraphics[width =0.7\linewidth]{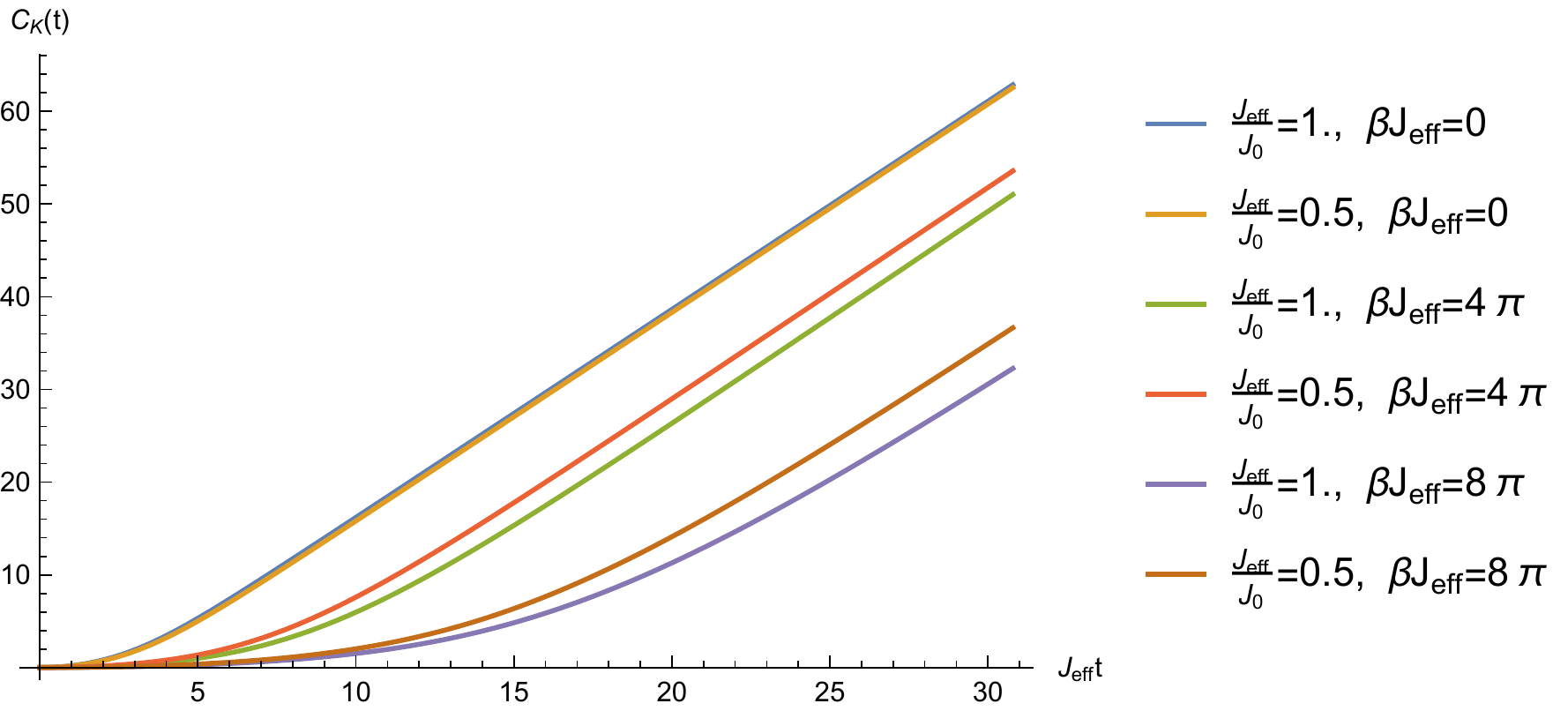}}\\
	\subfloat[t][\centering{$C_K(t)$ for the SSYK$_4$}]{\label{SPlotCk40BB}
		\includegraphics[width =0.7\linewidth]{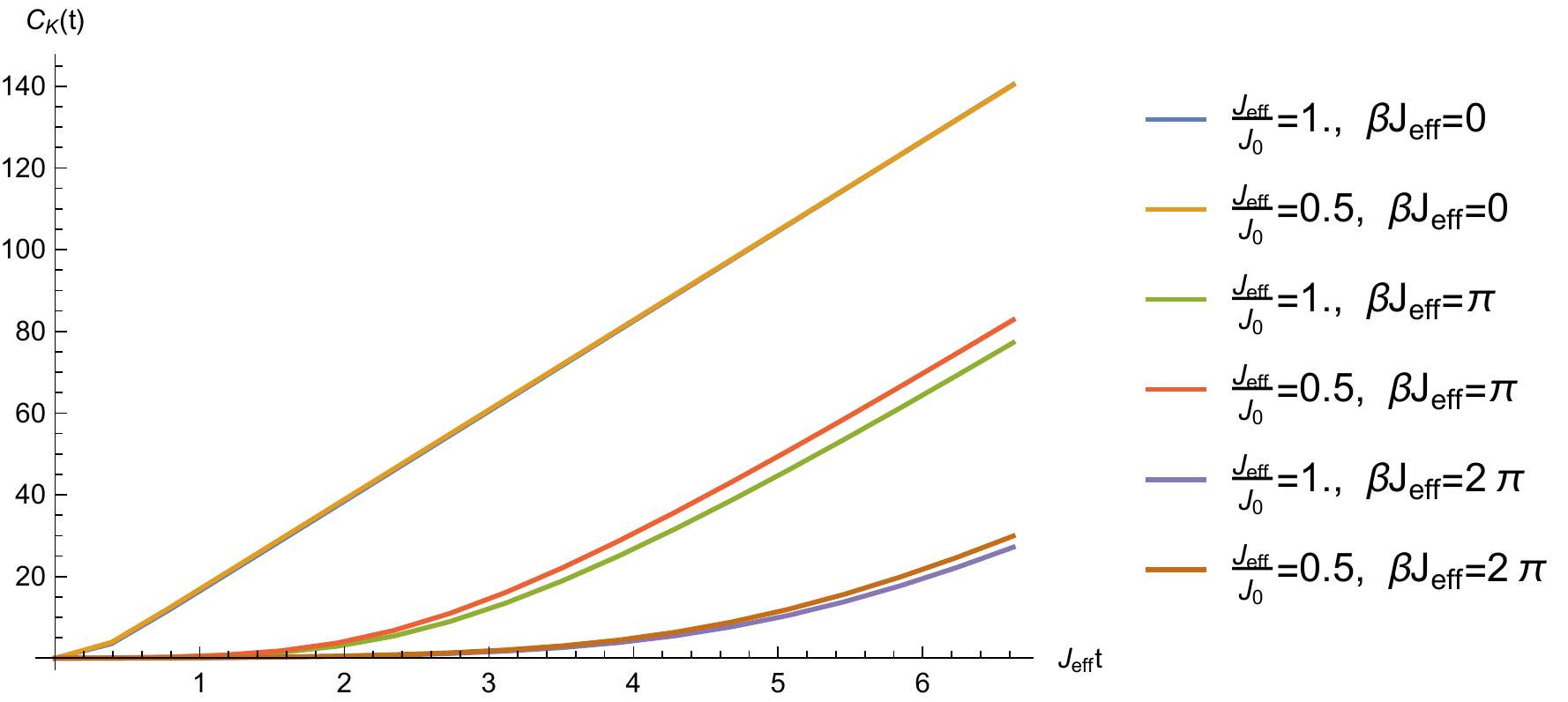}}
	\caption{K-complexity for (a) SYK$_4$ and (b) SSYK$_4$ model with various values of $\beta J_{\rm eff}$ and $J_{\rm eff}/J_0$.}
	\label{figure:CK4}
\end{figure}

In Figure \ref{figure:CK4}, we show the K-complexity of the deformed SYK$_4$ and SSYK$_4$ models. In these two figures, the K-complexity shares the same behavior: exponential and linear growth in the initial time. The exponential growth of K-complexity is not obvious due to the small system size. Similar to Figure. \ref{bn-SYK4}, we show the results with different dimensionless parameters $\beta J_{\rm Jeff}$ and $J_{\rm eff}/J_0$. For the Lanczos coefficients, we have shown that $b_n^\lambda/J_{\rm eff}$ is universal for fixed $\beta J_{\rm eff}$. This result is equivalent to saying that $C_K(t)$ is universal as a function of $J_{\rm eff} t$ for fixed $\beta J_{\rm eff}$. From Figure \ref{figure:CK4}, it is clear that the patterns coincide with the same $\beta J_{\rm eff}$. The small error is due to the small number of ensembles we took. The behavior for different $\beta J_{\rm eff}$ in the early time reflects that the Lyapunov exponents depend on $\beta J_{\rm eff}$ alone. The linear growth rate is universal because the asymptotic values of $b^\lambda_n/J_{\rm eff}$ are universal for arbitrary parameters.
\begin{figure}[htbp]
	\centering
	\captionsetup[subfloat]{farskip=10pt,captionskip=1pt}
	\subfloat[t][\centering{$\beta J_{\rm eff}=4\pi$}]{\label{Plotbn2BN}
		\includegraphics[height =0.25\linewidth]{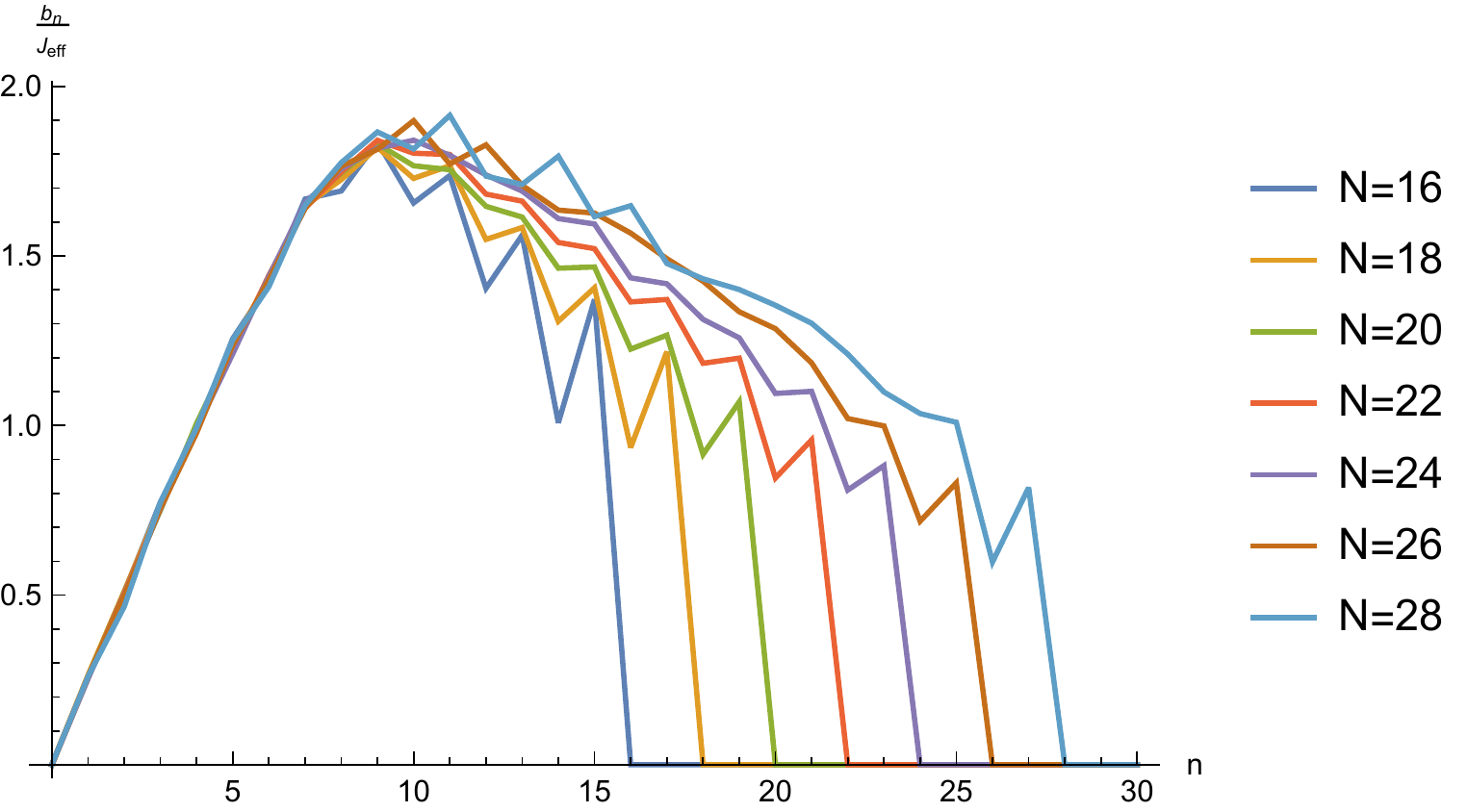}}
	\subfloat[t][\centering{$\beta J_{\rm eff}=0$}]{\label{PlotSYK2J0Lbn}\quad
		\includegraphics[height =0.25\linewidth]{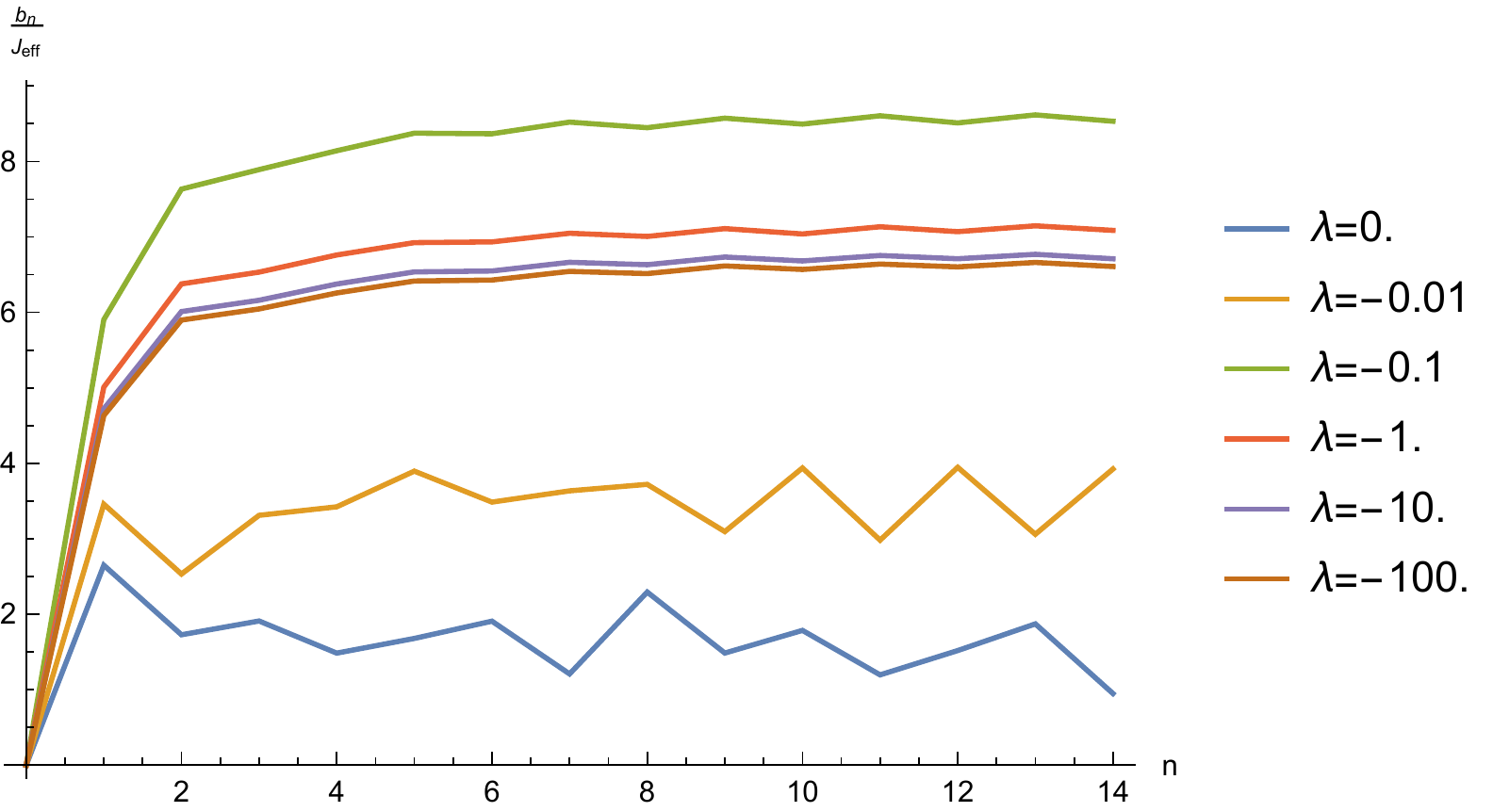}}
        \caption{(a): $b_n$ for the initial SYK$_2$ with different $N$. (b): $b_n$ for the SYK$_2$ with different $\lambda$ at infinite temperature.}
        \label{plot:bn-SYK2}
\end{figure}

\begin{figure}[htbp]
	\centering
	\captionsetup[subfloat]{farskip=10pt,captionskip=1pt}
	\subfloat[t][\centering{$\beta J_{\rm eff}=\pi$}]{\label{PlotSYK2JPiLbn}
		\includegraphics[height =0.27\linewidth]{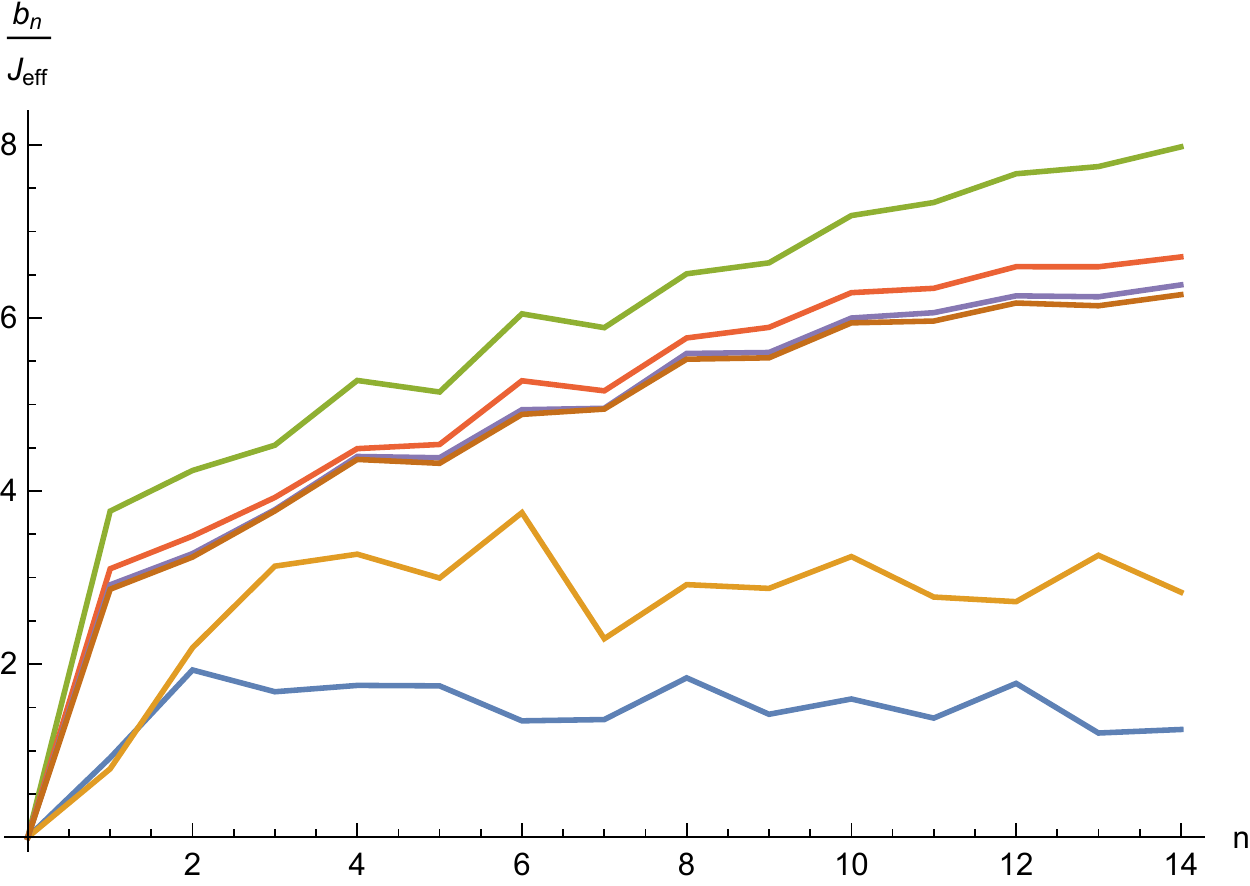}}
	\subfloat[t][\centering{$\beta J_{\rm eff}=2\pi$}]{\label{PlotSYK2J2PiLbn}\quad\quad
		\includegraphics[height =0.27\linewidth]{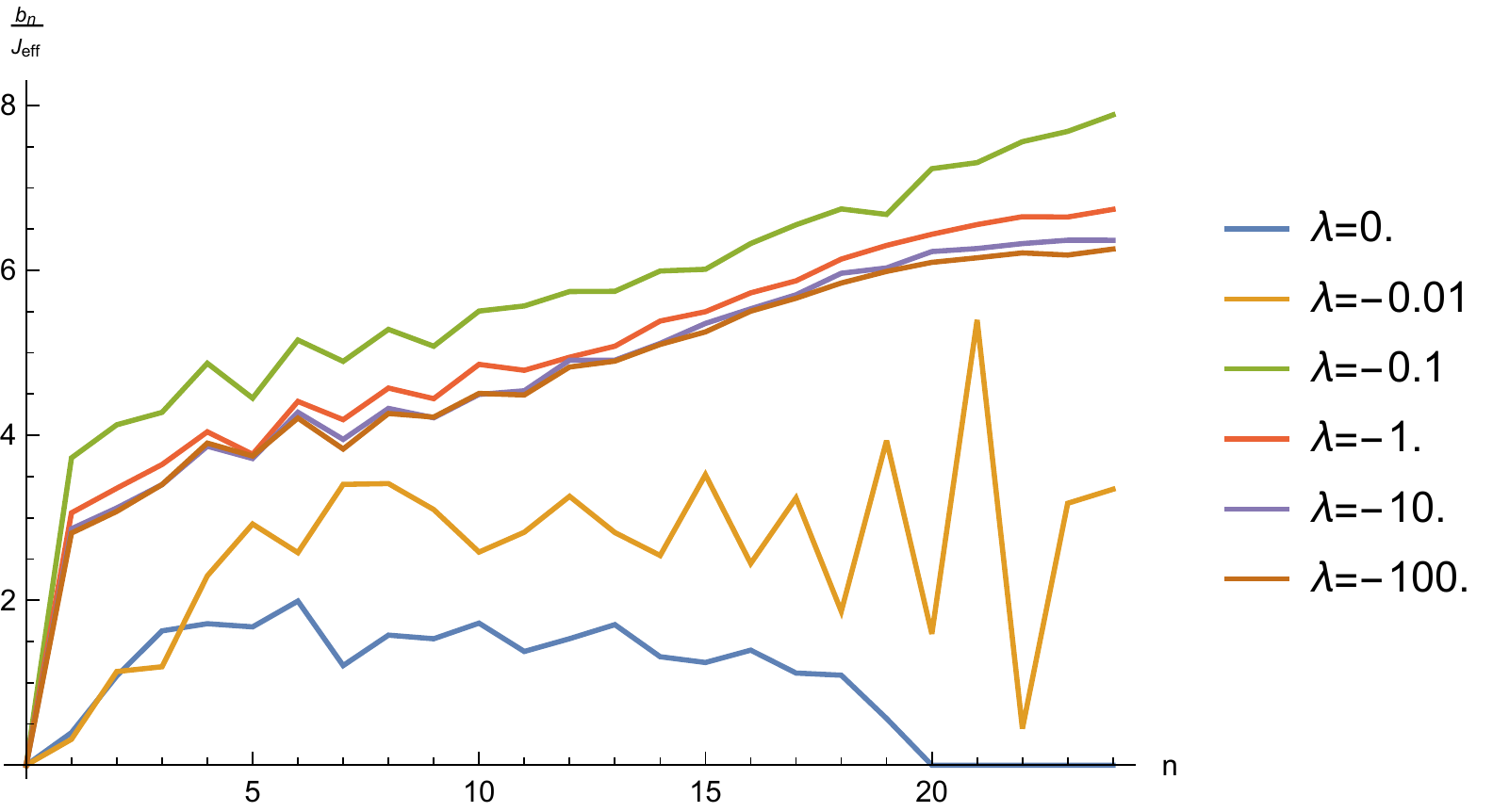}}
        \caption{Dimensionless Lanczos coefficients $b_n^\lambda/J_{\rm eff}$ of the SYK$_2$ model with various values of $\lambda$, where $N=20$, (a) $\beta J_{\rm eff}=\pi$ and (b) $\beta J_{\rm eff}=2\pi$.}
	\label{plot:bn2JL}
\end{figure}

In Sec.~\ref{sec:SFF:SYK2} and \ref{data-OTOC}, the SFF and OTOC results show that the original SYK$_2$ model is a free theory and the $T\bar{T}$-deformed SYK$_2$ model is a theory of interaction by integrability. In this paragraph, we study this property in more detail by using the Lanczos coefficient. The Lanczos coefficients $b_n$ with different $N$ and $\lambda$ are shown in Figure \ref{plot:bn-SYK2} and \ref{plot:bn2JL}, respectively. Note that the original SYK$_2$ model is a free theory and the dimension of the Krylov basis is $N$, so the number of nonzeros $b_n$ is $N-1$. This property is demonstrated in Figure \ref{Plotbn2BN}.

In Figure \ref{PlotSYK2LOTOC} and \ref{PlotSYK2JBLogLinearOTOC} we observe a MBL like behavior of OTOC for small $\lambda$ ($0<|J_0\lambda|\leq 0.01$ in Figure \ref{PlotSYK2LOTOC}) case. However, the deviation from the unity of the OTOCs is always slower than the exponential function for all the $\lambda$. For K-complexity, the chaotic signal is expected to be detected by the early behavior of $b_n$ at infinite temperature \cite{Parker:2018yvk}. The results are plotted in Figure \ref{PlotSYK2J0Lbn}. We find that $b_n/J_{\rm eff}$ for vanishing or small $\lambda$ are non-growth. For large $\lambda$, $b_n/J_{\rm eff}$ grow initially but are slower than the linear function and then reach the asymptotic values. It is also interesting to study the effect of $T\bar{T}$ deformation on $b_n$ at finite temperature. We list our results in Figure \ref{plot:bn2JL}. For large $\lambda$, we find a significant linear growth regions for large $\beta J_{\rm eff}$ in Figure \ref{PlotSYK2JPiLbn} ($1\leq n\lesssim 11$) and \ref{PlotSYK2J2PiLbn} ($1\leq n\lesssim 20$) and we can estimate that the growth rate of $b_n/J_{\rm eff}$ is about $1/\beta J_{\rm eff}$. It's worth noting that the linear growth region still exists even though the deformation vanishes. This feature is exhibited in Figure \ref{Plotbn2BN} more clearly. However, these undeformed results are harmless because the dimension of the operator space is small and $C_K(t)$ oscillates with time rapidly. But for the deformed case, the operator space is extended shown in Table \ref{table: dimension}. In this case, $C_K(t)$ grows exponentially and then linearly which is similar to the results of the SYK$_4$ model. As a result, we suggest that the K-complexity is not a suitable quantity to detect quantum chaos at finite temperatures.

\section{Summary and prospect}
\label{Sec:Summary}

We have investigated the quantum chaotic behaviors of the SYK models with the $T\bar{T}$ deformation. For comparison, we considered the (S)SYK$_4$ model and the SYK$_2$ model. The first two models are chaotic and the last model is integrable. In the literature, there is no mathematical definition of quantum chaos. However, it is commonly believed that the energy level spacing captures the main feature of quantum chaos. In this work, we detected the signals of quantum chaos through the SFF. Up to the time rescaling, we found that the $T\bar{T}$ deformation does not affect the properties of quantum chaos. Based on this assertion, we further investigate the effects of the $T\bar{T}$ deformation on scrambling and operator growth captured by OTOC and K-complexity, respectively. We summarize our main results in Table \ref{table:result}.

\begin{table}[]
\centering
\caption{Summary of the results at infinite temperature}
\label{table:result}
\begin{tabular}{|c|c|c|c|c|}
\hline
&~& SFF & OTOC & Lanczos \\   
\hline
(S)SYK$_4$ & All $\lambda$ & chaotic & $F(t)\sim 2-e^{\lambda_L t}$ &  $b_n\sim\mathcal{J} n$ \\
\hline
\multirow{3}*{SYK$_2$} & $\lambda=0$ & non-chaotic & $F(t)\sim 1$ & non-growth \\
\cline{2-5}
~&$|\lambda|\ll1$ & non-chaotic & $F(t)\sim 1-\alpha t^2$ & non-growth \\
\cline{2-5}
~&$|\lambda|\gg1$ & non-chaotic & $F(t)\sim 1-\alpha t^2$ & $b_n= an^{\delta}$, $\delta<1$\\
\hline
\end{tabular}
\end{table}

\begin{itemize}
	\item For the (S)SYK$_4$ model, the effect of $T\bar{T}$ deformation on SFF, OTOC, and K-complexity is equivalent to a rescaling of the coupling constant $J_0$. The patterns of these quantities as functions of $J_{\rm eff}t$ do not depend on $J_{\rm eff}/J_0$, but only on $\beta J_{\rm eff}$. At the late time, the OTOC decays to near zero. So the scrambling remains unchanged. The K-complexity in the $\TT$-deformed SYK4 also exhibits an exponential-to-linear growth at an early time. The Lanczos coefficient increases linearly and then saturates, whose slope matches the Lyapunov exponent in the $\TT$-deformed OTOC consistently.
\item For the SYK$_2$ model, the effect of $T\bar{T}$ deformation on SFF is also equivalent to a rescaling of the coupling constant $J_0$. For the scrambling, the deviation from the unity of OTOCs with the $\TT$-deformation is slower than the exponential function. For the small $\lambda$ case, we find an MBL behavior for both the infinite temperature case and the finite temperature case. So the $\TT$-deformation here strongly affects the scrambling compared to the original SYK$_2$ model, whose OTOC does not decay. Then we investigated the operator growth, namely Lanczos coefficients. At infinite temperature, the initial growth of $b_n$ is always slower than the linear function. The behaviors of OTOC and operator growth are in agreement with the expectation of the features of the non-chaotic system. At finite temperature, the Lanczos coefficient exhibits a linear growth region in SYK$_2$ (non-chaotic system) with or without the $T\bar{T}$ deformation, so we suggest that the K-complexity is not a suitable quantity to detect the quantum chaos in this case.
	
\end{itemize}

The systems in MBL phases exhibit many interesting features, such as a logarithmic increase of the entanglement entropy \cite{Fan:2016ean} and zero diffusion \cite{Jian:2017unn}. The $\TT$-deformed SYK$_2$ model and its spatial generalizations would be the ideal platforms for studying these phenomena of the MBL. Also, we also expect MBL phases in other integrable models with $\TT$ deformation.

\section*{Acknowledgments}
We would like to thank Shao-Kai Jian, Cheng Peng, Yuan Sun, Jia Tian, and Hua-Jia Wang for valuable discussions related to this work. S.H. also would like to appreciate the financial support from Jilin University and the Max Planck Partner group, as well as the Natural Science Foundation of China Grants No.~12075101, No.~12235016. Z. Y. X. acknowledges support from the National Natural Science Foundation of China under Grants No.~11875053 and No.~12075298, the Deutsche Forschungsgemeinschaft (DFG, German Research Foundation) under Germany's Excellence Strategy through the W\"urzburg-Dresden Cluster of Excellence on Complexity and Topology in Quantum Matter ct. qmat (EXC 2147, project id 390858490), and  the DFG via SFB 1170 ‘Topological and Correlated Electronics at Surfaces and Interfaces’ (project id 258499086). P.H.C.L acknowledges the support from JSPS KAKENHI (Grant No. 20H01902), MEXT KAKENHI (Grant No. 21H05462), and the National Center for Theoretical Sciences during this work.

\end{document}